\documentclass[a4paper,11pt]{article}
\pdfoutput=1 
\usepackage{jheppub} 
\usepackage[utf8x]{inputenc}
\usepackage{fontawesome}
\usepackage[dvipsnames,table]{xcolor}
\usepackage[caption=false]{subfig}
\usepackage{epsfig}
\usepackage{colortbl}
\usepackage[T1]{fontenc} 
\usepackage{adjustbox}
\usepackage{tablefootnote}
\usepackage{mathtools}
\usepackage{multirow}
\usepackage{bbold}
\usepackage{slashed}
\usepackage{bigints}
\usepackage[normalem]{ulem}
\usepackage{amsmath}
\usepackage{textcomp}
\usepackage{float}
\usepackage{graphicx}
\usepackage{booktabs}
\usepackage{academicons}
\usepackage{array}
\usepackage[toc,page]{appendix}
\usepackage[compat=1.1.0]{tikz-feynman} 
\usepackage{silence}
\tikzfeynmanset{warn luatex=false}
\newcommand{\stkout}[1]{\ifmmode\text{\sout{\ensuremath{#1}}}\else\sout{#1}\fi}
\definecolor{calpolypomonagreen}{rgb}{0.12, 0.3, 0.17}
\newcommand{\bea}{\begin{eqnarray}}
\newcommand{\eea}{\end{eqnarray}}
\newcommand{\beq}{\begin{equation}}
\newcommand{\eeq}{\end{equation}}
\usepackage{xcolor}

\title{\boldmath Constraining anomalous $Wtb$ and related SMEFT couplings using low-energy and electroweak precision observables }

\author[a]{Subhajit Kala,}
\author[a]{Lipika Kolay,}
\author[b]{Lopamudra Mukherjee,}
\author[a]{and Soumitra Nandi.}

\affiliation[a]{Department of Physics, Indian Institute of Technology Guwahati,\\North Guwahati, Assam-781039, India.}
\affiliation[b]{School of Physics, Nankai University,\\Tianjin 300071, China.}
\emailAdd{s.kala@iitg.ac.in}
\emailAdd{klipika@iitg.ac.in}
\emailAdd{lopamudra.physics@gmail.com}
\emailAdd{soumitra.nandi@iitg.ac.in}

\abstract{We investigate constraints on couplings of Standard Model effective field theory (SMEFT) operators contributing to $Wtb$ effective vertex at tree level. We study the one-loop level impact of these couplings on the low-energy flavour changing charged and neutral current processes and on the electroweak precision observables. We use the available data on these relevant processes to constrain the associated SMEFT/$Wtb$ couplings. Solving the renormalisation group equations, we connect the SMEFT couplings at different scales and use the bounds at low energy to obtain the relevant bounds at the large scale $\Lambda$. Our findings indicate significantly improved constraints on the couplings compared to existing constraints on $Wtb$ couplings by ATLAS and CMS. Additionally, we predict branching ratios for various top-FCNC processes, which exceed SM expectations by several orders of magnitude but remain within the reach of future colliders. These SMEFT couplings, or anomalous couplings of the effective $Wtb$ vertex, can further constrain different UV-complete and simplified models that generate such interactions at the tree or loop level.
}


\begin{document}
\maketitle
\flushbottom

\section{Introduction}
The top quark is the heaviest known fermion in the standard model (SM) with $\mathcal{O}(1)$ Yukawa coupling. It plays a special role in many theories within the SM and beyond. Due to its large mass and tiny lifetime, its effective couplings to the electroweak gauge bosons are still not strongly constrained. Hence, it is possible that large new physics (NP) effects could still hide there that may result in deviations in top quark properties with respect to the SM. In the SM, this vertex is purely left-handed at the tree level; however, various beyond the standard model (BSM) scenarios could be predicted with additional right-handed vector and tensor-type interactions, which can modify the top quark decay and production processes. 

The most common probe of anomalous structures in the $Wtb$ vertex is the collider studies, like the spin or the angular observables in top pair decays, single top production, study of spin observables of $W$ boson in a top decays, etc. \cite{Boos:1999dd, Aguilar-Saavedra:2006qvv, MohammadiNajafabadi:2008llf, Aguilar-Saavedra:2010ljg, Kolodziej:2011ir, Kolodziej:2012dg, RomeroAguilar:2015plc, Aguilar-Saavedra:2015yza, D0:2011fer}. Additionally, in recent studies, the ATLAS and CMS collaborations have provided constraints on the anomalous couplings, studying the combination of $W$ boson polarisation measurements in the top quark decay \cite{CMS:2020ezf}. To constraint these anomalous couplings, the low energy flavour changing neutral current (FCNC) processes of the $B$ meson such as the rare and radiative decays, $B_q-\bar{B}_q$ (q=d,s) oscillations etc. could be pretty promising since the effective $Wtb$ vertex contributes to these processes at one loop wherein the top quark gives the dominant contribution \cite{Burdman:1999fw, Grzadkowski:2008mf, Drobnak:2011aa, Drobnak:2011wj, Kozachuk:2020mwa, Xiong:2018mkc}. Electroweak precision observables (EWPOs) are also severely impacted by $Wtb$ effective couplings \cite{Frigeni:1991cd, Malkawi:1994tg}. In contrast, various studies have used experimental data and collider analyses to constrain effective operators that modify top quark interactions~\cite{Durieux:2019rbz, Hioki:2015env, Zhang:2010dr, Berger:2009hi, Birman:2016jhg, Bernardi:2022hny, Alioli:2017ces, Cirigliano:2016nyn}.

However, updated data are available from experiments like Belle-II \cite{Belle-II:2018jsg}, LHCb \cite{LHCb:2008vvz}, ATLAS \cite{ATLAS:2008xda}, and CMS \cite{CMS:2008xjf}, a comprehensive analysis that takes into account all possible observables and systematically connects both high-energy and low-energy observables using renormalisation group equations (RGEs) to evolve couplings across different energy scales, followed by a global fit, is still missing.

 The anomalous couplings of the $Wtb$ vertex may arise from various new physics scenarios, either at tree level or at the one-loop level. At tree level, such couplings can be generated through the mixing of new heavy particles with the SM fields, for instance, a new heavy gauge boson $W^{\prime}$\footnote{The new charged vector boson, often referred to as $W'$ bosons, plays an essential role in many extensions of the standard model.} or vector-like quarks $T$ or $B$ etc. The examples of such models include vector-like quark models~\cite{Aguilar-Saavedra:2013qpa, Chen:2017hak}, the existence of right-handed counterparts to the usual left-handed $W$ bosons, arising from a broken $SU(2)_L \times SU(2)_R$ symmetry \cite{Pati:1974yy} and similar kind of other extensions with heavy charged gauge bosons~\cite{Collins:2015wua, Capdevila:2024gki}, $G_{221}$ models \cite{Hsieh:2010zr}, Kaluza Klein model \cite{Garcia-Jimenez:2016cqe, Datta:2000gm} and possibly other models with similar features. At the one-loop level, frameworks like composite Higgs model \cite{Kaplan:1983fs}, Little Higgs model with T-parity (LHT) \cite{Penunuri:2008pb, Belyaev:2006jh}, scalar extensions such as the two-Higgs-doublet model~\cite{ Grzadkowski:1991nj, Gunion:1989we}, extensions with a new spin-0 particles~\cite{Biswas:2021pic,Kolay:2024wns,Kolay:2025jip}, spin-1 particles~\cite{Langacker:2008yv}, axion-like particles~\cite{Bauer:2021mvw}, as well as supersymmetric models \cite{Cao:2003yk}, can also give rise to anomalous $Wtb$ couplings. Many of these models may also modify the interaction vertices of fermion pairs to neutral gauge bosons. However, the effective couplings of the relevant charged and neutral current vertices will differ in each case. Consequently, probing one vertex does not directly constrain the other. This necessitates a careful and independent treatment of each vertex type in any precision analysis. Moreover, the physical manifestations of these modified couplings are distributed across different processes, depending on the nature of the gauge boson involved. 

In an effective theory framework like the Standard Model effective field theory (SMEFT),  higher-dimensional operators incorporate the essence of new heavy degrees of freedom, allowing for a model-independent analysis of NP beyond the standard model. The SMEFT can generate anomalous $Wtb$ couplings at the tree level. In this work, we plan to correlate anomalous $Wtb$ couplings with the SMEFT dimension-6 operator basis with a tree-level matching. We constrain the effective $ Wtb$ couplings of the related SMEFT operators from a combined study of the flavour-changing charged current (FCCC) and FCNC processes and EWPOs. The impact of the anomalous couplings is systematically analysed within the SMEFT framework. First, we match the SMEFT operators contributing to the anomalous \( Wtb \) couplings at the tree level and obtain the respective matched coefficients. We then use RGE to evolve these coefficients from the NP scale $\Lambda$ to the scale $\mu_t = m_t$, the top quark's mass and further down to the electroweak scale $\mu_{EW}$. We can obtain the bounds on these coefficients at the electroweak scale by studying the $W$ and $Z$-pole observables. Afterwards, to connect these coefficients to low-energy data, we matched the contribution of these operators to the effective theory operator bases relevant to different low-energy processes like decays and mixings. We obtained the corresponding Wilson coefficients (WCs) at the electroweak scale. Furthermore, following an RGE, we have obtained the WCs at the scale $\mu_b = m_b$. We have analysed the low-energy data and obtained constraints on these WCs.    


This article is structured as follows: In sec.~\ref{sec:framework}, we discuss the $Wtb$ effective vertex framework and corresponding SMEFT operators, along with our methodology. Secs.~\ref{sec:FCNC_process}, \ref{sec:FCCC_process}, and \ref{sec:EWPOs} explore the impact of these SMEFT operators on the FCCC, FCNC, and electroweak observables, respectively. In sec.~\ref{sec:results_analysis}, we present and analyse our fit results from these observables individually and simultaneously. Finally, we summarize our findings in sec.~\ref{sec:summary}.

\section{Framework and Methodology}\label{sec:framework}
There are different NP scenarios which could impact the anomalous $ Wtb$ vertex at the tree or one-loop levels \cite{Gunion:1989we,Langacker:2008yv,Biswas:2021pic,Bauer:2021mvw,Kolay:2024wns,Kolay:2025jip}. In an effective theory framework, the most general Lorentz-invariant vertex is parametrised as follows \cite{Li:1990qf, Kane:1991bg, delAguila:2002nf, Aguilar-Saavedra:2006qvv} : 
\begin{equation}
\label{eq:Wtb_Lagrangian}
\mathcal{L}_{Wtb} =-\frac{g}{\sqrt{2}} \biggl( \bar{b} \gamma_{\mu} (V_L^{\prime} P_L +V_R P_R) t W^{\mu-}  + \bar{b} \frac{i \sigma_{\mu \nu} q^{\nu}}{m_W} (g_L P_L + g_R P_R) t W^{\mu-} \biggr) + \rm{h.c.}
\end{equation}
Generally in SM, at tree level $  V_L^{\prime} = V_{tb} \approx 1$ and all other couplings are zero i.e $ V_R = g_L = g_R = 0 $. In our case, we can write 
\begin{equation}\label{eq:VL_from_Vtb}
V_{L}^{\prime} = V_{tb} (1 + \delta V_{L})\,,
\end{equation}
where $V_{tb}$ is the SM counterpart of $V_{L}^{\prime}$ and $\delta V_{L}$ is the correction to the left-handed vector current. At our convention, we will write 
\begin{equation}
V_{tb} \, \delta V_{L} \equiv V_{L}\,.
\end{equation}

\noindent However, these couplings can be generated at one-loop level in the SM or at tree or one-loop level in BSM scenarios \cite{Gonzalez-Sprinberg:2011beg, Gonzalez-Sprinberg:2015dea, Bernreuther:2008us, Duarte:2013zfa}. The $Wtb$ vertex has been mostly probed via top-quark production and top decays at the LHC; see \cite{Andrea:2023yap} and the references therein.

\noindent However, flavour-changing charged and neutral current processes involving the top quark should also be considered to obtain proper constraints on the couplings. There are a few studies where bounds on the anomalous couplings were derived from the study of radiative decays \cite{Grzadkowski:2008mf}. Similar studies, in the context of right-handed vector currents involving different quarks, have also attempted to constrain these couplings through other flavour-changing processes \cite{Crivellin:2011ba, Buras:2010pz}.

\paragraph{\underline{\bf SMEFT Correspondence:}}

After a tree-level matching, we can generate the $Wtb$ effective couplings defined in eq.~\eqref{eq:Wtb_Lagrangian} from a few SMEFT operators. The most general SMEFT Lagrangian is represented as
\begin{align}
\mathcal{L}_{\rm eff}=\mathcal{L}_{\rm SM}^{(4)}+\sum_{i}^{n_d} \frac{\mathcal{C}_i^{(d)}}{\Lambda^{d-4}} \mathcal{O}_i^{(d)}\,,
\end{align}
where the first term in the above equation is the 4-dimensional SM Lagrangian, and the second term corresponds to the tower of higher-dimensional operators; for this work, we use the Warsaw basis for dimension-6 operators \cite{Grzadkowski:2010es}, with the Lagrangian 
\begin{equation}\label{eq:dim6_lagrangian}  \mathcal{L}_{\rm eff}^{(6)} = \sum_{i} \frac{\mathcal{C}_{i}}{\Lambda^{2}} \mathcal{O}^{(6)}_{i}\,.
\end{equation}
The SMEFT operators that contribute to the effective $Wtb$ vertices at tree level are:
\begin{subequations}\label{eq:relevant_SMEFT_operators}
	\begin{align}
	\mathcal{O}^{uW}_{pr}&=(\bar{q}_p\sigma_{\mu\nu}u_r)\tau^I\tilde{\phi} \, W^I_{\mu\nu} \label{eq:SMEFT_op1} \,,\\
	\mathcal{O}^{dW}_{pr}&=(\bar{q}_p \sigma_{\mu\nu} d_r)\tau^I \phi \,  W^I_{\mu\nu} \label{eq:SMEFT_op2}\,, \\
	\mathcal{O}^{\phi u d}_{pr}&=i(\tilde{\phi}^{\dagger} D_{\mu} \phi) \, (\bar{u}_p\gamma_{\mu} d_r) \label{eq:SMEFT_op3}\,,\\
	\mathcal{O}^{\phi q(3)}_{pr}&=(\phi^{\dagger} i \overleftrightarrow D_{\mu}^I \phi) \, (\bar{q}_p \tau^I \gamma^{\mu} q_r) \label{eq:SMEFT_op4}\,.
	\end{align}
\end{subequations}
Here $\tau^I$ is the $SU(2)_L$ generator with the flavour indices $p,r$. The derivative of eq.~\eqref{eq:SMEFT_op4} is defined as:
\begin{equation}
\phi^{\dagger} i \overleftrightarrow D_{\mu}^I \phi \equiv i \phi^{\dagger} \left( \tau^I D_{\mu}-\overleftarrow{D}_{\mu}\tau^I\right)\phi\,.
\end{equation}
Expanding the Lagrangian $\mathcal{L}_{\rm eff}^{(6)}$ of eq.~\eqref{eq:dim6_lagrangian} in the mass basis and matching it with eq.~\eqref{eq:Wtb_Lagrangian} at the top scale, gives the form 

\begin{align}
\label{eq:leff}
    \mathcal{L}_{\rm eff}^{(6)}&=-\frac{g}{\sqrt{2}}W_{\mu}^- \bar{d}_i \gamma^{\mu}\left(K_{ij} P_L+\frac{v^2}{2}\mathcal{C}_{ij}^{\phi u d} P_R\right)_{ij} u_j\nonumber\\
    &-i 2v W_{\mu}^-\bar{d}_i \sigma^{\mu \nu} q_{\nu}\left(\mathcal{C}_{j i}^{dW*} P_L+ \mathcal{C}^{uW}_{i j}P_R\right)_{ij} u_j+ \rm h.c.\,,
\end{align}
Where $i,j$ are the flavour indices, and the quark mixing matrix $K$ is written as:
\begin{align}
K_{\rm CKM}\equiv K\equiv U^{\dagger}_{u_L}\left(\mathbb{1}+v^2 \tilde{\mathcal{C}}^{\phi q (3)}\right)U_{d_L}= \left(V_{\rm CKM}\right)_{ij}+v^2 \mathcal{C}^{\phi q (3)}_{ij}\,.
\end{align}
Eq.~\eqref{eq:leff} is expressed in the mass basis. We redefine the Wilson coefficients of the operators involving fermionic currents by absorbing the fermionic flavour rotation matrices used in the transformation from the interaction basis to the mass basis. Here, the Wilson coefficients with a tilde (\(\tilde{\mathcal{C}}\)) correspond to couplings in the interaction basis, while those without a tilde (\(\mathcal{C}\)) represent those in the mass basis, and they are related as:
\begin{subequations}
    \begin{align}
    \mathcal{C}^{uW}&=U^{\dagger}_{d_L}\tilde{\mathcal{C}}^{uW} U_{u_R}\,,\\
\mathcal{C}^{dW}&=U^{\dagger}_{u_L}\tilde{\mathcal{C}}^{dW} U_{d_R} \,, \\
\mathcal{C}^{\phi q (3)}&=U^{\dagger}_{u_L} \tilde{\mathcal{C}}^{\phi q (3)}U_{d_L} \,, \\
\mathcal{C}^{\phi u d}&=U^{\dagger}_{u_R}\tilde{\mathcal{C}}^{\phi u d}U_{d_R} \,.
\end{align}
\end{subequations}
\noindent The contribution to these couplings in terms of effective $Wtb$ couplings of eq.~\eqref{eq:Wtb_Lagrangian} can be written as
\begin{eqnarray}\label{eq:Wtb_SMEFT_correspondence}
& V_L  = \frac{v^2}{\Lambda^2}\mathcal{C}_{33}^{\phi q (3)}\,,  \ \ \ \ \ \  V_R = \frac{1}{2}\frac{v^2}{\Lambda^2}\mathcal{C}^{\phi u d}_{33} \,, \nonumber \\ 
&g_L = \sqrt{2}\frac{v^2}{\Lambda^2}\mathcal{C}^{*\;dW}_{33} \,, \ \ \ \ \ \ g_R =\sqrt{2}\frac{v^2}{\Lambda^2}\mathcal{C}_{33}^{uW} \,.
\end{eqnarray}
Here $v$ is the vacuum expectation value of the Higgs boson, having a value $v=(\sqrt{2}G_F)^{-1/2}\simeq246 \, \rm{ GeV}$. It is noteworthy that among the four dimension-six SMEFT operators contributing to the matching in eq.~\eqref{eq:Wtb_SMEFT_correspondence}, only operators $\mathcal{O}^{\phi q(3)}_{33}$, $\mathcal{O}^{*\,dW}_{33}$ and $\mathcal{O}^{uW}_{33}$ contain $\mathrm{SU}(2)$ structures that couple to both charged and neutral gauge bosons. Therefore, for each of these operators, the Wilson coefficients associated with the charged and neutral gauge boson interaction appear as the same parameters in the SMEFT. However, their physical effects manifest differently across processes, depending on the gauge boson involved. In other words, while the coupling constants are the same, the physical observables they influence differ due to the decomposition of the $\mathrm{SU}(2)$ currents into charged and neutral components. The operator $\mathcal{O}^{\phi ud}_{33}$, on the other hand, takes part only in charged current interactions.\\
\indent{}Nevertheless, in the mass basis of the SM, the effective couplings for charged current and neutral current operators can differ, and the CKM matrix naturally appears in the charged current sector. This structure is preserved in SMEFT also, because the operators are built from gauge-invariant combinations of SM fields, and the CKM matrices arise from the misalignment of the up and down quark mass eigenstates after EWSB. The explicit decompositions of the operators in the mass basis can be found in ref.~\cite{Dedes:2017zog}. We note that the effective couplings may differ upon decomposition in the mass basis, though they may be correlated via the SMEFT WCs. However, this may not be the case in model-dependent studies with different new physics scenarios. Several well-established models are available in the literature \cite{Aguilar-Saavedra:2013qpa,Chen:2017hak,Pati:1974yy,Collins:2015wua, Capdevila:2024gki,Hsieh:2010zr,Garcia-Jimenez:2016cqe, Datta:2000gm,Kaplan:1983fs,Penunuri:2008pb, Belyaev:2006jh,Grzadkowski:1991nj, Gunion:1989we,Biswas:2021pic,Kolay:2024wns,Kolay:2025jip,Langacker:2008yv,Bauer:2021mvw,Cao:2003yk} that contribute to modifications in the $Wtb$ vertex, each with distinct mechanisms and implications. These include frameworks such as extra-dimensional theories, extended gauge symmetries, vector-like quark models, composite Higgs scenarios, supersymmetric models and scalar extensions like the Two-Higgs-Doublet Model. There may be many more such models. A majority of these models contribute to $Wtb$, $Zt\bar t$, $Zb\bar b$, $\gamma t\bar t$ vertices. However, the effective couplings in charged and neutral current interactions are generally uncorrelated due to the distinct gauge structures and mixing mechanisms inherent to each model. As previously noted, while certain observables may receive contributions from both interaction types, these arise through different sets of Feynman diagrams and coupling structures. Consequently, any experimental bound derived from such observables will reflect this lack of correlation and must be treated independently. Besides this, we have observables that are exclusively sensitive to either charged or neutral current interactions, with no simultaneous contributions. These observables are particularly valuable for disentangling the effects of charged and neutral currents, thereby enhancing the discriminatory power of experimental analyses. This will be a crucial point to consider while we study their impact on physical processes. This distinction is necessary for precision fitting in SMEFT and interpreting experimental data correctly.

Additionally, in collider searches, deviations in the $Wtb$ vertex are primarily probed through top quark decay properties and single top production processes~\cite{Boos:1999dd, Aguilar-Saavedra:2006qvv, MohammadiNajafabadi:2008llf, Aguilar-Saavedra:2010ljg, Kolodziej:2011ir, Kolodziej:2012dg, RomeroAguilar:2015plc, Aguilar-Saavedra:2015yza, D0:2011fer}. In contrast, deviations in the $Zt\bar t$ vertex are best constrained through precision measurements of top quark pair production at future colliders and the associated production of top quarks with a $Z$ boson at the LHC. These complementary channels further emphasise the need to treat charged and neutral current interactions as distinct probes of new physics.

In our analysis, we restrict ourselves to the charged-current part of these SMEFT couplings, as this aligns with our objective of constraining the anomalous $Wtb$ couplings using low-energy flavour data, electroweak precision observables, and many other relevant measurements that are affected by the $Wtb$ effective vertex at one loop. However, there are studies in the literature where these SMEFT couplings have been investigated. For instance, in ref.~\cite{Garosi:2023yxg}, the couplings $\mathcal{C}^{uW}_{33}$ and $\mathcal{C}^{\phi q \,(3)}_{33}$ were analyzed, but with emphasis on their neutral-current interactions (i.e.~$\gamma t\bar t,\, Zt\bar t,\, Zb\bar b$). This is evident from their analysis, as a few dominant observables constraining these Wilson coefficients arise from processes such as $h \to \gamma\gamma$, $h \to Z\gamma$, and $Z \to \ell^+\ell^-$, which cannot be influenced by the charged-current counterparts (i.e.~$Wtb$) of the operators at one-loop. Additionally, in their analysis, the contribution in $Z\to b\bar{b}$ decays will be at the tree level, which is not applicable for $Wtb$ effective operators. Furthermore, our analysis includes a larger set of flavour observables related to FCNC processes compared to their study, including the branching fractions of the exclusive radiative decays such as $\mathcal{B}(B \to K^*(\rho)\gamma)$, $\mathcal{B}(B_s \to \phi\gamma)$ and an extensive collection of observables in $B\to K^{(*)}\mu^+\mu^-$ and $B_s\to \phi\mu^+\mu^-$ decays. In the case of flavour-changing charged currents (FCCC), we have incorporated both leptonic and semileptonic decay modes, which do not apply to the analysis in \cite{Garosi:2023yxg} since they considered only the fermionic interactions with the neutral gauge bosons.
	
In contrast, in ref.~\cite{Allwicher:2023shc}, the authors considered a large set of top-philic operators, including those studied in our analysis. However, as mentioned earlier, the three SMEFT operators : $\mathcal{O}^{\phi q (3)}_{33}$, $\mathcal{O}^{*\, dW}_{33}$, and $\mathcal{O}^{uW}_{33}$, contain vertices such as $\gamma t\bar t$, $Zt\bar t$, $\gamma b\bar b$, $Zb\bar b$, and $Wtb$. In ref.~\cite{Allwicher:2023shc}, the contributions of these vertices were treated simultaneously across the various observables under study.  As mentioned above, some of these observables do not receive contributions in our framework. Their analysis of FCNC processes includes contributions from charged and neutral current vertices. In contrast, in our study, the contribution will be only from the charged current interactions. Additionally, the operator $\mathcal{O}^{uW}_{33}$ with neutral-current interactions contributes to the top-quark dipole moment already at tree level, which is not the case in our analysis. The bounds on $\mathcal{C}^{\phi ud}_{33}$ have also been discussed in ref.~\cite{Allwicher:2023shc}. This operator couples only to charged gauge bosons and is similar to ours. However, the authors have considered only the lepton flavour universality (LFU) ratios $R(D)$ and $R(D^*)$ to constrain the WC. However, in our analysis, we have considered a large data set available on $B\to D^{(*)} \mu(e) \nu$ decays, which includes the data on the decay rate distributions in $w$ bins (recoil angle) and different angular bins which adds approximately 40 additional data points on FCCC processes as compared to ref.~\cite{Allwicher:2023shc}.

\paragraph{\underline{\bf Low-energy correspondence:}}\label{sec:methodology}

Various low-energy flavour observables and high-energy EWPOs will be impacted at one loop by the effective vertices of eq.~\eqref{eq:Wtb_Lagrangian} or by their corresponding SMEFT operators defined in eq.~\eqref{eq:relevant_SMEFT_operators}. 
In this study, we have utilised these low-energy observables to constrain the anomalous couplings as well as the respective SMEFT couplings. 

At different energy scales, one could define different operator bases with the appropriate degrees of freedom relevant to that scale. The impact of the operators defined at a high energy scale on the low energy processes/observables can be extracted by following appropriate matching calculations and RGE to relate the coefficients at different scales. For instance, the effective $Wtb$ vertex affects electroweak precision observables, like the $W$ and $Z$-pole observables, and triple gauge boson couplings and $\Gamma(h\to b \bar{b})$, which are particularly important. Hence, we can constrain these couplings from the measured values of the W and Z-pole observables or the Higgs decay rates by studying the one-loop level contributions of these operators to these observables. Describing these bounds at the electroweak scale $\mu_{EW}$ will be appropriate. However, we can obtain the corresponding bounds at the scale $\mu_{t}$ and $\Lambda$ following an appropriate RGE. 

The anomalous couplings of $Wtb$ vertex can modify various FCCC and FCNC low-energy processes. In this study, we focus on the low-energy observables related to the semileptonic, leptonic, rare and radiative decays of the $B$ and $B_s$ mesons and the $B_q-\bar{B}_q$ mixing (with $q=d,s$). For each of these processes, we define the respective dimension-6 low energy effective operator (LEFT) basis $\mathcal{L}_{\rm LEFT}^{(6)}$.

 To study the impact of low-energy flavour observables at the electroweak symmetry breaking (EWSB) scale, we match the contribution of the operators defined in eq.~\eqref{eq:Wtb_Lagrangian} onto the respective LEFT bases by integrating out the electroweak bosons $(W, Z, H)$ and the top quark. In addition, to define the SMEFT couplings at the low energy scale, we need to evolve them to the low energy scale of our interest by solving the RGEs \cite{Jenkins:2013zja, Jenkins:2013wua, Alonso:2013hga}. The detailed methodology of our case is discussed in the latter part of this section. The corresponding matching relations for these processes are discussed in the subsequent sections.

\begin{figure}[t]
	\centering
	\includegraphics[width=1.3\linewidth]{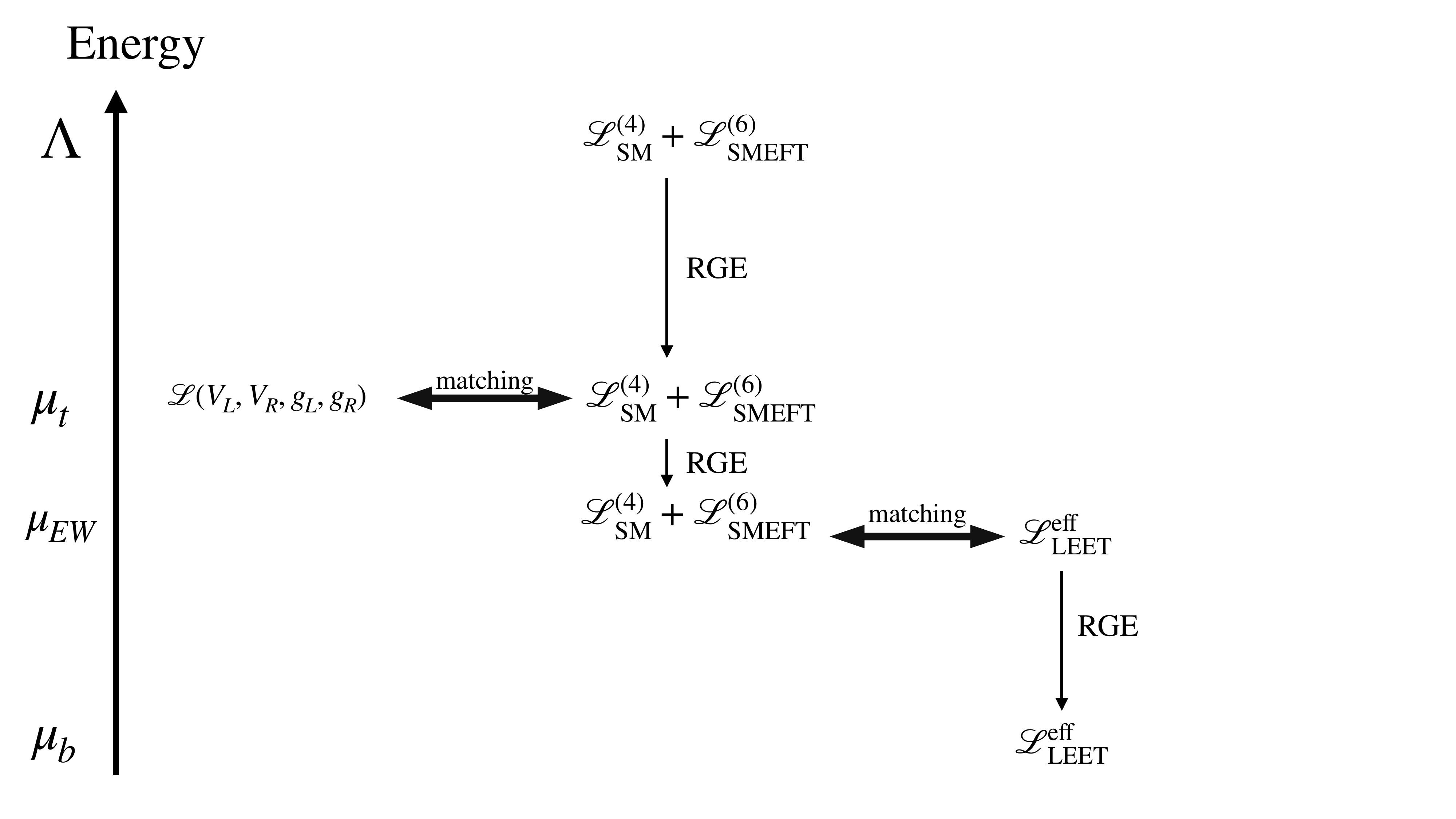}
	\caption{Energy scale hierarchy: effective $Wtb$ vertices are matched with the higher dimension SMEFT Lagrangian at scale $\mu_t$; eventually SMEFT Lagrangian is matched with the effective Hamiltonian governing B physics at the EW symmetry scale ($\mu_{EW}$). Wilson coefficients in different scales are connected via appropriate RGEs. }
	\label{fig:RGE_sketch}
\end{figure}

\paragraph{\underline{\bf Plan of action - Matching and RGE:}}
This work focuses on constraining the anomalous couplings of $Wtb$ vertex and the corresponding SMEFT coefficients. The relevant operators will contribute to various low-energy FCNC and FCCC processes and electroweak observables like EWPOs and Higgs observables at the one-loop level. As we have mentioned, the bounds on the couplings or WCs obtained from different observables will only be relevant to that scale. However, all these coefficients at different scales are linked by an appropriate RGE. Therefore, we need the corresponding Wilson coefficients and the loop contributions at the relevant scales to calculate all these observables at different scales. The methodology we followed is illustrated in the flowchart of fig.~\ref{fig:RGE_sketch}. 

It is customary to acknowledge that the particle content and the underlying symmetry of the SM differ above and below the EW scale. For the SMEFT operator basis, we set a cutoff scale $\Lambda$, which represents the energy scale where the new physics becomes relevant. Furthermore, we define the corresponding SMEFT coefficients $\mathcal{C}_i(\Lambda)$ at this scale. We obtain these coefficients at any other scales following an RGE at leading order as given below:
\begin{align}\label{eq: RGE}
	\frac{d\mathcal{C}_i}{d \log\mu}=\frac{1}{16\pi^2}\sum_j \gamma_{ij} \mathcal{C}_j \equiv\frac{1}{16\pi^2}\beta_i \,.
\end{align}
Here, $\mu$ is the renormalisation scale, $\gamma_{ij}$ is the anomalous dimension matrix (ADM), and $\beta_i$s are the beta functions of SMEFT operators, the detailed expression for $\beta_i$ in our chosen basis is given in Appendix \ref{appndx:RGE_beta}. The solutions of the above differential equation will give us the coefficients $\mathcal{C}_i(\mu)$ at a scale $\mu$ as functions of $\mathcal{C}_j(\Lambda)$. In the above equation, the anomalous dimension matrices $\gamma_{ij}$s are functions of EW gauge couplings. Since the EW couplings remain nearly constant with respect to the scale, the solution of eq.~\eqref{eq: RGE} can be written in the leading-log approximation as:
\begin{align}
   	\mathcal{C}_i(\mu)=\left(1+\frac{\gamma_{ii}}{16\pi^2}\log\left(\frac{\mu}{\Lambda}\right)\right)\mathcal{C}_i(\Lambda)+ \sum_{i\neq j}\frac{\gamma_{ij}}{16\pi^2}\log\left(\frac{\mu}{\Lambda}\right) \mathcal{C}_j(\Lambda)\,.
   \end{align} 
It is important to note that the same RGEs can be applied to the WCs normalised as $\mathcal{C}/\Lambda^2$.  Although we present all  ADMs in terms of $\mathcal{C}$, the same matrices remain valid for $\mathcal{C}/\Lambda^2$.
Following the above equation and the given ADMs, we obtain the following connection between the NP $(1~\rm{ TeV})$ scale and the top-scale $\mu_{t}$ ($\approx 172 \, \rm GeV$):
 \begin{align}
 		\begin{pmatrix}
 			\mathcal{C}^{uW}_{33}\\
 			\mathcal{C}^{dW}_{33}\\
 			\mathcal{C}^{\phi u d}_{33}\\
 			\mathcal{C}^{\phi q (3)}_{33}
 		\end{pmatrix}_{\mu_t}&=\begin{pmatrix}
 			0.92573 & 0.00013 & 0 & 0\\
 			0.00013 & 0.926252 & 0& 0\\
 			0 & 0 & 0.92966 & 0\\
 			0 & 0 & 0 & 0.959319
 		\end{pmatrix}  \begin{pmatrix}
 			\mathcal{C}^{uW}_{33}\\
 			\mathcal{C}^{dW}_{33}\\
 			\mathcal{C}^{\phi u d}_{33}\\
 			\mathcal{C}^{\phi q (3)}_{33}
 		\end{pmatrix}_{\Lambda = \rm 1\, TeV} \,.
 	\end{align}
We note that due to RGE, there is a small mixing between the operators $\mathcal{O}^{uW}_{pr}$ and $\mathcal{O}^{dW}_{pr}$ in eq.~\eqref{eq:SMEFT_op1} while the other two operators do not mix. Furthermore, each of the four operators also mix via RGE with several other dimension-six SMEFT operators not listed here but discussed in refs.~\cite{Jenkins:2013zja, Jenkins:2013wua, Alonso:2013hga}. However, numerical estimates indicate that these mixing effects are subleading. We have explicitly verified that their impact on the running of the four Wilson coefficients is negligible and does not significantly alter the phenomenological conclusions of our analysis. After obtaining the coefficients at the scale $\mu_t$, we have done a tree-level matching of the operators in eq.~\eqref{eq:SMEFT_op1} with the effective anomalous $Wtb$ operators. Afterwards, using the matching relations given in eq.~\eqref{eq:Wtb_SMEFT_correspondence} we obtain the $Wtb$ anomalous couplings in terms of the SMEFT couplings.        

Following a similar method as mentioned above, we obtain the SMEFT coefficients at the EW scale by a RGE evolving from $\mu_t \to \mu_{EW}$. For the EW scale, $ \mu_{\rm EW}$ $( \approx 91.18 \,\rm{ GeV})$ the relations are given as
	\begin{align}\label{eq:RGE_SMEFT}
	\begin{pmatrix}
		\mathcal{C}^{uW}_{33}\\
		\mathcal{C}^{dW}_{33}\\
		\mathcal{C}^{\phi u d}_{33}\\
		\mathcal{C}^{\phi q (3)}_{33}
	\end{pmatrix}_{\mu_{\rm EW}}&=\begin{pmatrix}
		0.96745 & 0.00006 & 0 & 0\\
		0.00006 & 0.96764 & 0 & 0\\
		0 & 0 & 0.96952 & 0\\
		0 & 0 & 0 & 0.98115
	\end{pmatrix}  \begin{pmatrix}
		\mathcal{C}^{uW}_{33}\\
		\mathcal{C}^{dW}_{33}\\
		\mathcal{C}^{\phi u d}_{33}\\
		\mathcal{C}^{\phi q (3)}_{33}
	\end{pmatrix}_{\mu_t}  \,.
\end{align}
The running from $\mu_t \to \mu_{\rm EW}$ is numerically not very significant; however, it is conceptually important, since the LEFT (or WET) basis is valid only up to $\mu_{\rm EW}$. Since $\mu_t$ lies above $\mu_{\rm EW}$, any matching or analysis performed within the LEFT/WET framework must be preceded by a proper renormalization group evolution down to $\mu_{\rm EW}$. Neglecting this step would lead to an inconsistent treatment of the effective theory, even if the numerical impact is minor. Therefore, we include the running from $\mu_t$ to $\mu_{\rm EW}$ to ensure theoretical consistency and completeness in our analysis.
We have pointed out that there are different LEFT operator bases to describe low-energy FCNC and FCCC processes, and the respective observables are expressed in terms of LEFT WCs, which are valid up to the EW scale. To effectively utilise flavour and other low-energy observables, we must match our theory, i.e., the SMEFT, with the LEFT at the EW scale. Furthermore, we must run these LEFT couplings down to the corresponding low-energy scale. The advantage here is that, due to the matching at the EW scale, the information from SMEFT is encapsulated in the LEFT Wilson coefficient.

These LEFT WCs receive contributions from the high energy SMEFT or $Wtb$ effective operators from a one-loop level matching procedure at $\mu_{EW}$. Therefore, for the low-energy observables, we calculate the one-loop Feynman diagrams and match them with the LEFT basis to get the LEFT Wilson coefficients in terms of SMEFT Wilson coefficient at the $\mu_{\rm EW}$ scale. We further run down the LEFT Wilson coefficients to the low energy scale $(\mu_{b})$ to study the impact on the B-meson observables.    

Furthermore, in the low-energy effective theory, we get the correction to the higher dimension operators from QED and QCD, due to the underlying broken gauge symmetry $SU(2)_L\times U(1)_Y$. Analytic solutions of one-loop RGEs of LEFT Wilson Coefficients are discussed in the following literature \cite{Aebischer:2017gaw,Jenkins:2017jig}. 
We can write RGE in terms of the ADM as:
\begin{eqnarray}\label{eq:RGE}
\frac{d\vec{C}_i}{d\log\mu}&=\frac{1}{16\pi^2}\sum_{j}\beta_{ij}\vec{C}_j=\frac{\alpha_s}{4\pi}\sum_j\beta_{ij}^s \vec{C}_j+\frac{\alpha_e}{4\pi}\sum_j \beta_{ij}^e \vec{C}_j\,.
\end{eqnarray}
Here, $\vec{C}_i$ represents the different Wilson coefficients of the LEFT Lagrangian, while $\alpha_s$ and $\alpha_e$ denote the strong and electromagnetic couplings, respectively. The matrices $\beta^s$ and $\beta^e$ are ADMs, arising from QCD and QED corrections to the LEFT operators.
There might be contributions coming from higher order terms $\mathcal{O}(\alpha_s^2,\alpha_e^2, \alpha_s\alpha_e)$, but we will restrict ourselves to the linear order, and that allows us to write  the RGE in a simpler form:
\begin{eqnarray}
\vec{C}_i(\mu)&=U_{ij}(\mu,\mu_0)\vec{C}_j(\mu_0)=\Bigg[U_{ij}^s(\mu,\mu_0)+U_{ij}^e(\mu,\mu_0)\Bigg] \vec{C}_j(\mu_0)\,.
\end{eqnarray}
We have used the evolution matrices for QCD, \( U_{ij}^s \) from \cite{Chetyrkin:1996vx,Buras:1998raa,Gonzalez-Alonso:2017iyc} and for QED, \( U_{ij}^e \) from \cite{Aebischer:2017gaw}, respectively, to perform RGE evolution. Although reference~\cite{Aebischer:2017gaw} contains both \( U_{ij}^s \) and \( U_{ij}^e \) evolution matrix at one-loop, the operator basis used in that work differs from the conventional LEFT basis. As noted by the corresponding author, a rotation matrix is required to transform from their basis to the traditional one. Using this approach, we have verified that $U_{ij}^s$ is consistent with the results available in the existing literature.

We will address the appropriate LEFT operator bases and the relevant WCs when we discuss the low-energy FCNC and FCCC processes. Also, we will provide the numerical solutions of the RGEs of the respective WCs while running them down from the EW scale to the low-energy scale.

\section{Impacts on Different FCNC Processes} \label{sec:FCNC_process}
Different FCNC processes such as neutral meson mixing, $b \to s(d) \ell \ell $ semileptonic decays, rare decays, and invisible decays are particularly important for our study. 
We know that in the SM, the FCNC processes are loop-level suppressed and highly sensitive to new interactions. In the following subsections, we will study the impact on various FCNC observables in detail. The effective vertex of $Wtb$ can contribute to various FCNC vertices at the one-loop level. In this regard, the processes involving the $b$-quark as an external particle are particularly affected. We calculate the one-loop-level corrections to FCNC processes/vertices from the $Wtb$ or the respective SMEFT operators. We use these results and, following a matching procedure, we extract the WCs of the LEFT operator bases relevant to the processes under consideration.

\subsection{Neutral Messon Mixing} \label{sec:meson_mixing}
 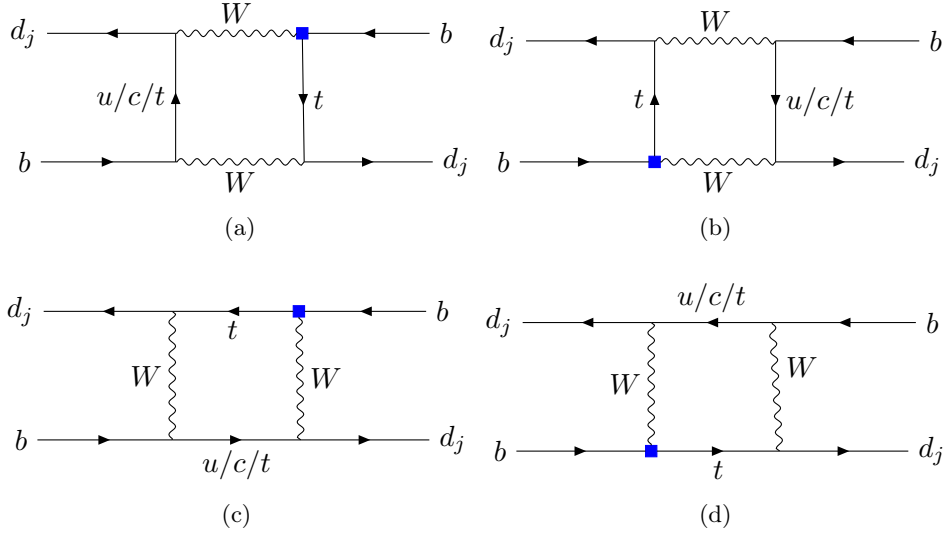
\begin{figure}[t]
\centering
 \subfloat[]{\begin{tikzpicture}
	\begin{feynman}
	\vertex (a1){\(b\)};
	\vertex[square dot,blue,right=2cm of a1](a2);
	\vertex[above=1.7cm of a2](a3);
	\vertex[left=1.7cm of a3](a4){\(d_{j}\)};
	\vertex[right=1.7cm of a2](a5);
	\vertex[square dot,blue,right=1.6cm of a3](a6){};
	\vertex[right=1.9cm of a6](a7){\(b\)};
	\vertex[right=1.7cm of a5](a8){\(d_{j}\)};
	
	\diagram* { 
		(a1) --[fermion, arrow size=1pt](a2) --[fermion, arrow size=1pt,edge label=\(u/c/t\)](a3) --[fermion, arrow size=1pt](a4),
		(a2) --[boson,edge label'=\(W\)](a5),
		(a3) --[boson,edge label={\(W\)}](a6),
		(a7) --[fermion, arrow size=1pt](a6) --[fermion, arrow size=1pt,edge label=\(t\)](a5) --[fermion, arrow size=1pt](a8),
	};	
	\end{feynman}
	\end{tikzpicture}}
    \subfloat[]{\begin{tikzpicture}
	\begin{feynman}
	\vertex (a1){\(b\)};
	\vertex[square dot,blue,right=2cm of a1](a2){};
	\vertex[above=1.6cm of a2](a3);
	\vertex[left=1.7cm of a3](a4){\(d_{j}\)};
	\vertex[right=1.6cm of a2](a5);
	\vertex[square dot,blue,right=1.6cm of a3](a6);
	\vertex[right=1.9cm of a6](a7){\(b\)};
	\vertex[right=1.7cm of a5](a8){\(d_{j}\)};
	
	\diagram* { 
		(a1) --[fermion, arrow size=1pt](a2) --[fermion, arrow size=1pt,edge label=\(t\)](a3) --[fermion, arrow size=1pt](a4),
		(a2) --[boson,edge label'=\(W\)](a5),
		(a3) --[boson,edge label={\(W\)}](a6),
		(a7) --[fermion, arrow size=1pt](a6) --[fermion, arrow size=1pt,edge label=\(u/c/t\)](a5) --[fermion, arrow size=1pt](a8),
	};	
	\end{feynman}
	\end{tikzpicture}}\\
    \subfloat[]{\begin{tikzpicture}
	\begin{feynman}
	\vertex (a1){\(b\)};
	\vertex[square dot,blue,right=2cm of a1](a2);
	\vertex[above=1.7cm of a2](a3);
	\vertex[left=1.7cm of a3](a4){\(d_{j}\)};
	\vertex[right=1.7cm of a2](a5);
	\vertex[square dot,blue,right=1.6cm of a3](a6){};
	\vertex[right=1.9cm of a6](a7){\(b\)};
	\vertex[right=1.7cm of a5](a8){\(d_{j}\)};
        \vertex[below=0.5cm of a8]{};
	
	\diagram* { 
		(a1) --[fermion, arrow size=1pt](a2) --[boson,edge label=\(W\)](a3) --[fermion, arrow size=1pt](a4),
		(a2) --[fermion,arrow size=1pt,edge label'=\(u/c/t\)](a5),
		(a6) --[fermion,arrow size=1pt,edge label={\(t\)}](a3),
		(a7) --[fermion, arrow size=1pt](a6),
            (a6)--[boson, arrow size=1pt,edge label=\(W\)](a5) --[fermion, arrow size=1pt](a8),
	};	
	\end{feynman}
	\end{tikzpicture}}
     \subfloat[]{\begin{tikzpicture}
	\begin{feynman}
	\vertex (a1){\(b\)};
	\vertex[square dot,blue,right=2cm of a1](a2){};
	\vertex[above=1.7cm of a2](a3);
	\vertex[left=1.7cm of a3](a4){\(d_{j}\)};
	\vertex[right=1.7cm of a2](a5);
	\vertex[square dot,blue,right=1.6cm of a3](a6);
	\vertex[right=1.9cm of a6](a7){\(b\)};
	\vertex[right=1.7cm of a5](a8){\(d_{j}\)};
        \vertex[below=0.5cm of a8];
	
	\diagram* { 
		(a1) --[fermion, arrow size=1pt](a2) --[boson,edge label=\(W\)](a3) --[fermion, arrow size=1pt](a4),
		(a2) --[fermion,arrow size=1pt,edge label'=\(t\)](a5),
		(a6) --[fermion,arrow size=1pt,edge label'={\(u/c/t\)}](a3),
		(a7) --[fermion, arrow size=1pt](a6),
            (a6)--[boson, arrow size=1pt,edge label=\(W\)](a5) --[fermion, arrow size=1pt](a8),
	};	
	\end{feynman}
	\end{tikzpicture}}
	\caption{Feynman diagrams contributing to the process of neutral meson-mixing. The blue square dots refer to the $ Wtb $ effective vertex. The quark $d_{j}$ denotes down-type quarks $(s,d)$, which are corresponding to $B_{s}^{0}$ and $B^{0}$ mesons, respectievely. }
    \label{fig:meson_mixing}
\end{figure}

The anomalous $Wtb$ or the relevant SMEFT operators can contribute to the neutral meson mixing processes through the box diagrams of fig.~\ref{fig:meson_mixing}. Here we have $d_{j}=(d,s)$, so the processes that will be modified are: $B_{s}^{0}-\bar{B}_{s}^{0}$ and $B^{0}-\bar{B}^{0}$. Other meson mixing processes, such as $K^{0}-\bar{K}^{0}$ and $D^{0}-\bar{D}^{0}$, will not receive any modification in our case. We have calculated the one-loop contributions by considering one-operator insertion, like the contributions at order $1/\Lambda^2$. We have added the contributions of all the diagrams shown in fig.~\ref{fig:meson_mixing}. Considering two effective vertices at the same time will give us a higher-order effect, which we have neglected in our calculations. 

The effective four-fermion operators relevant to the meson mixing can be written as: 
\begin{equation}\label{eq:mixing_lagrangian}
    \mathcal{L}^{ \bar{b} d_{j} \leftrightarrow \bar{d}_{j} b } = C_{V_{L}}  (\bar{d}_{j} \gamma_{\mu } P_{L} b) (\bar{d}_{j} \gamma_{\mu } P_{L} b) + C_{V_{R}}  (\bar{d}_{j} \gamma_{\mu } P_{R} b) (\bar{d}_{j} \gamma_{\mu } P_{R} b)\,.
\end{equation}
From the diagrams in fig.~\ref{fig:meson_mixing}, we do not have any contribution to the four-fermion operators involving scalar or pseudoscalar quark currents. Furthermore, it is noteworthy that the light quark mass suppresses the contributions from the right-handed quark current operator, and thus, we have neglected it. Following a matching condition, we have extracted the contributions in $C_{V_{L}}$ from the diagrams in fig.~\ref{fig:meson_mixing}. The expression for $C_{V_{L}}$ is given in eq.~\eqref{eq:meson_mixing_loop} of appendix \ref{appndx:mixing_loop}. Note that, for simplicity, we have neglected the masses of the external quarks. Hence, apart from the CKM factors, the expressions for $C_{V_{L}}$ will be same for both $B_{s}^{0}-\bar{B}_{s}^{0}$ and $B^{0}-\bar{B}^{0}$ mixing processes. The corresponding coefficients at the $\mu_b \approx (4.8 \,\, \text{GeV})$ scale can be obtained using the equation below, derived from the RGE: 
\begin{align}\label{eq:Inv_EW_to_mb}
	\begin{pmatrix}
		C_{V_L}\\
		C_{V_R}
	\end{pmatrix}_{{\mu_b}}&=\begin{pmatrix}
		0.856037 & 0\\
		0 & 0.856037
	\end{pmatrix}    \begin{pmatrix}
		C_{V_L}\\
		C_{V_R}
	\end{pmatrix}_{\mu_{\rm{EW}}}\,.
\end{align}
For obtaining the numerical results of the aforementioned evolution matrices, we have used $\alpha_{\rm em} = \frac{1}{137}$, $\alpha_s(\mu_b) = 0.212$, and $\alpha_s(\mu_{\rm EW}) = 0.118$, respectively.

In the absence of CP-violating new physics, the observable associated with the meson oscillations is the mass difference, i.e. $\Delta M$, which is expressed as : 
\begin{equation}\label{eq:mixing_deltaM}
    \Delta M = 2 |M_{12}| = 2 \frac{|\mathcal{M}|}{2 m_{B}}\,,
\end{equation}
Where $m_{B}$ is the mass of the mesons $B_{s}^{0}$ and $B^{0}$. The $|\mathcal{M}|$ is the mixing amplitude we obtain from the sum of the dispersive parts of the meson mixing diagrams in fig.~\ref{fig:meson_mixing}. We can express $\Delta M$ as the sum of contributions from the SM and BSM scenarios:
\begin{equation}
    \Delta M_{\rm tot} = \Delta M_{\rm SM} + \Delta M_{\rm NP}\,.
\end{equation}
$\Delta M_{\rm SM}$ the contribution coming from SM. In SM, $\Delta M_{\rm SM}$ can be expressed as: 
\begin{equation}
	\Delta M_{\rm SM}^{q} =  \frac{G_{F}^2}{6 \pi ^2}  M_W^2 m_{B_{q}} f_{B_{q}}^2 B_{B_{q}} \eta_{B_{q}} |V_{tq}^{*} V_{tb} |^2 S_{0}(x_t) \,,
\end{equation}
where, the loop factor $S_{0}(x_t)$ is defined by: 
\begin{equation}\label{eq:mixing_SM}
	S_{0}(x_t) = x_{t} \left( \frac{1}{4} + \frac{9}{4} \frac{1}{1-x_{t}} - \frac{3}{2} \frac{1}{(1-x_{t})^2 }\right)-\frac{3}{2}\frac{x_t^3\log x_t}{(1-x_t)^3}\,,
\end{equation}
with $x_{t} = \frac{m_{t}^2}{M_W^2}$. $f_{B_{q}} $ is the decay constant defined as below: 
\begin{equation}
	\langle 0 | \bar{q}_{d} \gamma_{\mu} b | B (p) \rangle = i f_{B_{q}} P_{\mu}\,.
\end{equation}
$B_{B_{q}}$ is the bag factor, defined in \cite{Gay:2000utx}. $\eta_{B}$ is the QCD correction at next-to-leading order. The major sources of uncertainty in the predictions of the mixing amplitudes are the decay constants and the bag factors. In NP, the expression of $\Delta M_{\rm NP}^{q}$ is given by: 
\begin{equation}
	\Delta M_{\rm NP}^{q} = \frac{2}{3} m_{B_{q}} B_{B_{q}} f_{B_{q}}^2  C_{V_{L}} \,.
\end{equation}
$C_{V_{L}}$ is the loop factor mentioned in eq.~\eqref{eq:mixing_lagrangian}. The contribution from the NP is also sensitive to the decay constant and the bag factor.   

For the numerical analysis, we have defined the ratio of the NP and the SM contributions in $\Delta M$:  
\begin{equation}\label{eq:mixing_observable}
    \Delta_{q} = \frac{\Delta M_{\rm NP}^{q}}{\Delta M_{SM}^{q}} = \left( \frac{\Delta M_{\rm exp}^{q}}{\Delta M_{SM}^{q}}  - 1 \right)\,.
\end{equation}
In the above equation, $q = (s,d)$ for the mesons $B^{0}_{s}$ and $B^{0}$, respectively. In $\Delta_{q}$s, the bag parameters and the decay constants cancel; hence, they will be less uncertain. Using the available data and the respective SM predictions, we obtain estimates of the $\Delta_d$ and $\Delta_s$, which are useful to constrain the new WCs. The current theoretical and experimental status of the mixing parameters are \cite{Albrecht:2024oyn, LHCb:2023sim, Belle-II:2023bps} : 
	\begin{subequations}
		\begin{eqnarray}
		 \Delta M^{d}_{\rm Exp}  =  0.5065\pm0.0019\, \rm ps^{-1}\,, &\quad& \Delta M^{d}_{\rm SM} = 0.535 \pm 0.021 \, \rm ps^{-1} \,,\\
		\Delta M^{s}_{\rm Exp} = 17.765 \pm 0.006\, \rm ps^{-1}\,,  &\quad& \Delta M^{s}_{\rm SM} = 18.23\pm 0.63 \rm \, ps^{-1}\,. 
 		\end{eqnarray}
	\end{subequations} 
Hence, the allowed value of the parameter $ \Delta_{q} $, calculated using updated SM and experimental results,  is given as follows: 
	\begin{subequations}\label{eq:mixing_delta_Exp}
		\begin{eqnarray}
		\Delta_{d} &= & -0.05327 \pm 0.03733 \,, \\
		\Delta_{s} &= & -0.02551 \pm 0.03368\,.
		\end{eqnarray}
	\end{subequations}
We use these inputs to constrain the SMEFT coefficients, which we will present in the results and analysis section.

\subsection{Radiative Decays} \label{sec:radiative}

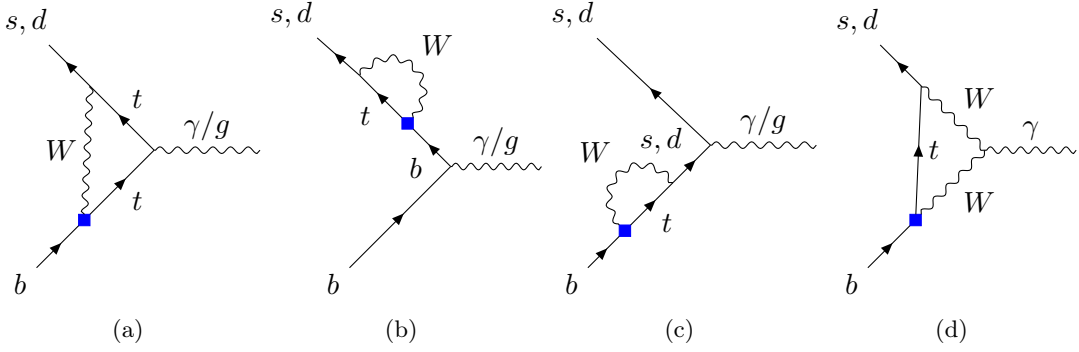
\begin{figure}[t]
	\centering
	\subfloat[]{\begin{tikzpicture}
			\begin{feynman}
				\vertex (a1){\( s,d\)};
				\vertex [below right=1.2cm of a1](a2);
				\vertex [large,below right=1.2cm of a2](a3);
				\vertex [square dot,blue,below left=1.2cm of a3](a4){ };
				\vertex [below left=1.2cm of a4](a5){\( b\)};
				\vertex [right=1.4cm of a3](a6);
				
				\diagram* {
					(a3) -- [fermion, arrow size=1pt,edge label'={\( t\)}] (a2) -- [fermion, arrow size=1pt] (a1),
					(a5) --[fermion, arrow size=1pt] (a4) -- [fermion, arrow size=1pt,edge label'={\( t\)}] (a3),
					(a2) --[boson, edge label'={\(W\)}] (a4),
					(a3) --[boson,edge label={\(\gamma/g\)}](a6),
					
				};
			\end{feynman}
	\end{tikzpicture}}
	\subfloat[]{\begin{tikzpicture}
			\begin{feynman}
				\vertex (a1){\( b\)};
				\vertex [above right=2.2cm of a1](a2);
				\vertex [square dot, blue, above left=0.7cm of a2](a3){};
				\vertex [above left=0.9cm of a3](a4);
				\vertex [above left=0.7cm of a4](a5){\( s,d\)};
				\vertex [right=1.2cm of a2](a6);

				\diagram* {
					(a1) --[fermion, arrow size=1pt](a2) --[fermion, arrow size=1pt, edge label={\(b\)}] (a3) --[fermion, arrow size=1pt, edge label={\(t \)}] (a4) --[fermion, arrow size=1pt] (a5),
					(a4) --[boson, half left, looseness=2, edge label = \(W\)] (a3),
					(a2) --[boson, edge label={\( \gamma/g\)}] (a6),
				};
			\end{feynman}
	\end{tikzpicture}}
	\subfloat[]{\begin{tikzpicture}
			\begin{feynman}
				\vertex (a1){\( b\)};
				\vertex [square dot, blue,above right=1cm of a1](a2){};
				\vertex [above right=0.9cm of a2](a3);
				\vertex [above right=0.7cm of a3](a5);
				\vertex [right=1.4cm of a5](a6);
				\vertex [above left=2.0cm of a5](a9){\(s,d\)};

				\diagram* {
					(a1) --[fermion, arrow size=1pt](a2) --[fermion, arrow size=1pt, edge label'={\(t\)}] (a3) --[fermion, arrow size=1pt, edge label={\(s,d \)}] (a5) --[fermion, arrow size=1pt] (a9),
					(a2) --[boson, half left, looseness=2, edge label = \(W\)] (a3),
					(a5) --[boson, edge label={\( \gamma/g\)}] (a6),
				};
			\end{feynman}
	\end{tikzpicture}}\hspace{0.001cm}
	\subfloat[]{\begin{tikzpicture}
			\begin{feynman}
				\vertex (a1){\( s,d\)};
				\vertex [below right=1.2cm of a1](a2);
				\vertex [large,below right=1.2cm of a2](a3);
				\vertex [square dot,blue,below left=1.2cm of a3](a4){};
				\vertex [below left=1.2cm of a4](a5){\( b\)};
				\vertex [right=1.2cm of a3](a6);
				
				\diagram* {
					(a5) --[fermion, arrow size=1pt] (a4) -- [fermion, arrow size=1pt,edge label'={\( t\)}] (a2) -- [fermion, arrow size=1pt] (a1),
					(a3) --[boson,edge label={\(\gamma\)}](a6) ,
					(a2) -- [boson, edge label={\( W\)}] (a3) -- [boson, edge label={\( W\)}] (a4),		
				};
			\end{feynman}
	\end{tikzpicture}}
	\caption{Feynman diagrams contributing to radiative process $b \to d_{j} \gamma(g)$, with blue  squares being effective $Wtb$ vertex.}
	\label{fig:Radiative_feynman}
\end{figure}

The radiative decays are important probes of NP, as the experimental and SM predictions mostly agree with each other within the $1\sigma$ error bar. The SM prediction and experimental values for different channels related to $b \to s(d) \gamma $ transitions are given in table~\ref{tab:radiative_data}. The LEFT operator bases for these radiative decays $b\to d_j\gamma$ are given as  
\begin{equation} \label{eq:Heff_bsgamma}
	{\cal H}^{b\to d_j\gamma}_{eff} = - \frac{4\,G_F}{\sqrt{2}} V_{tb}V_{td_j}^\ast
   \sum_{i=7,8} \biggl(C_i (\mu)  O_i + C'_i (\mu) 
	O'_i\biggr) \,,
\end{equation}
 \begin{align}\label{eq:opro7o8}
	O_{7} &= \frac{e}{16 \pi^2} m_b
	(\bar{d_j} \sigma_{\mu \nu} P_R b) F^{\mu \nu} ,&
	O_{7}^\prime &= \frac{e}{16 \pi^2} m_b
	(\bar{d_j} \sigma_{\mu \nu} P_L b) F^{\mu \nu} , \\
	O_{8} &= \frac{g_s}{16\pi^2} m_b
	(\bar{d_j} \sigma_{\mu \nu} T^a P_R b) G^{\mu \nu \, a} ,&
	O_{8}^\prime &= \frac{g_s}{16 \pi^2} m_b
	(\bar{d_j} \sigma_{\mu \nu} T^a P_L b) G^{\mu \nu \, a}.
	\end{align}

Note that a contribution from $O_8$ to the $b \to s(d)\gamma$ process arises through its mixing with $O_7$ via the RGE~\cite{Chetyrkin:1996vx,Buras:1998raa}, and the relevant RGEs are given by:

\begin{align}\label{eq:RGEC78}
\begin{pmatrix}
	        C_7\\
	        C_8
	     \end{pmatrix}_{\mu_b}=\begin{pmatrix}
	 0.66301 & 0.09259 \\
	 0.00326 & 0.69877
	 \end{pmatrix}\begin{pmatrix}
	         C_7\\
	         C_8
	     \end{pmatrix}_{\mu_{\rm{EW}}} \,.
 \end{align}
Similarly, the operator $\mathcal{O}'_7$ mixes with $\mathcal{O}'_8$ during RGE from the scale $\mu_{EW} \to \mu_b$, however, they do not mix with $\mathcal{O}_7$ or $\mathcal{O}_8$. 

The SMEFT or the $Wtb$ effective operators will modify the $b\to d_j\gamma/g$. The corresponding Feynman diagrams are given in fig.~\ref{fig:Radiative_feynman}.
The sum of the contributions of all these diagrams can be written as
  
\begin{subequations}\label{eq:bsgamma_1loop}
	\begin{align}
		&\mathcal{M}^{1-loop}_{d_{i} \to d_{j} \gamma} =\sum_{I=\{L,R\}} C_{T_{I}}(\bar{d}_j i \sigma_{\mu \nu} P_{I} b)F^{\mu \nu}\,, \\
		&\mathcal{M}^{1-loop}_{d_{i} \to d_{j} g} =\sum_{I=\{L,R\}} C_{T_{I}}(\bar{d}_j i T^A\sigma_{\mu \nu} P_{I} b)G^{A \,\mu \nu}\,.
	\end{align}
\end{subequations}
Here $P_{L,R}=(\mathbb{1}\mp \gamma^5)/2$ corresponds to the chirality projection operators. In the above equations, the coefficients are expressed as  
\begin{subequations}
	\begin{eqnarray}
		C_{T_{I}}^{\gamma} &=&C_{T_{I}}^{\gamma \,\rm vertex} + C_{T_{I}}^{\gamma\, \rm self} \,,\\
		C_{T_{I}}^{g} &=&C_{T_{I}}^{g \,\rm vertex} + C_{T_{I}}^{g\, \rm self} \,,
	\end{eqnarray}
\end{subequations}
and the detailed expressions of the coefficients $C_{i}$'s are given in eqs.~\eqref{eq:radiative_gamma_CTR} and \eqref{eq:radiative_gluon_CTR} of appendix \ref{appndx:Loop_b2sll}. In the SM, both the $C_7$ and $C_8$ contribute to the rates of the processes related to $b \to s(d) \gamma$ decays. In such cases, the WCs $C_i(\mu)$, can be written as follows:
\begin{equation}\label{eq:wcssmnp}
	C_i(\mu)=C_i^{\rm SM} (\mu) +\Delta C_i(\mu) \,.
\end{equation}
Here, $C_i^{\rm SM}$ and $\Delta C_i(\mu)$ are the SM and the new physics contributions respectively. The different $C_i^{\rm SM}$ values at $\mu_b$ scale can be found in \cite{Mahmoudi:2024zna}. From the matching conditions, we obtain from eq.~\eqref{eq:bsgamma_1loop} and \eqref{eq:Heff_bsgamma} 

\begin{align}\label{eq:C7C8NP}
\Delta C_{7}(\mu_{EW}) &= -\frac{v^2}{2 \lambda_t}\frac{16\pi^2}{e m_b} C_{T_{R}}^{\gamma}(\mu_{EW}) , &
C_{7}^\prime(\mu_{EW}) &= -\frac{v^2}{2 \lambda_t}\frac{16\pi^2}{e m_b} C_{T_{L}}^{\gamma}(\mu_{EW}) , \\
\Delta C_{8}(\mu_{EW}) &= -\frac{v^2}{2 \lambda_t}\frac{16\pi^2}{g_s m_b} C_{T_{R}}^g(\mu_{EW}) , &
C_{8}^\prime(\mu_{EW}) &= -\frac{v^2}{2 \lambda_t}\frac{16\pi^2}{g_s m_b} C_{T_{L}}^g(\mu_{EW}) .
\end{align}
Note that the WCs $C_{7/8}^{\prime}$ depend on light quark masses and are therefore highly suppressed compared to their unprimed Wilson coefficients. To give an estimate, tensor type couplings $C_{7,8}^{\prime}$ are $\sim \mathcal{O}(10^3)$ times smaller than $\Delta C_{7,8}$. Hence, in the numerical analysis, we will neglect any contribution from $C_{7,8}^{\prime}$ as compared to $\Delta C_{7,8}$. Also, we obtain the $\Delta C_{7}(\mu_b)$ and $\Delta C_{8}(\mu_b)$ using the RGEs given in eq.~\eqref{eq:RGEC78}.   

For the inclusive radiative decay process, $ B \to X_{s} \gamma $ the branching ratio in the presence of new WCs is given by \cite{Misiak:2020vlo}: 
\begin{equation}
	\mathcal{B}(B \to X_{s} \gamma) \times 10^{4} = (3.40 \pm 0.17) - 8.25 \Delta C_{7} - 2.10 \Delta C_{8} \,. 
\end{equation}
Here, $\Delta C_8$ and $\Delta C_7$ are the new WCs contributing in the rates. 
The branching ratios for the exclusive decay to vector meson can be written as \cite{Paul:2016urs, Beneke:2001at}:
\begin{equation}
	\mathcal{B}(B_q \to V\gamma) = \tau_{B_q} \frac{G_F^2 \alpha_{\text{em}} m_{B_q}^3 m_b^2}{32 \pi^3} 
	\left( 1 - \frac{m_V^2}{m_{B_q}^2} \right)^3 
	|\lambda_t|^2 \left( |C_7(\mu_b)|^2 + |C_7'(\mu_b)|^2 \right) T_1(0),
\end{equation}
with $\lambda_{t} = V_{tq}^{\ast} V_{tb} $ and $T_{1}(0)$ is the tensor Form Factor (FF) can be found from \cite{Bharucha:2015bzk}. The data given in table~\ref{tab:radiative_data} are useful to constrain the new WCs, hence, the SMEFT coefficients. 
\begin{table}[t]
	\centering
	\rowcolors{1}{cyan!10}{blue!20!green!5}
	\renewcommand{\arraystretch}{1.5}
	\setlength{\tabcolsep}{15pt}
	\resizebox{\textwidth}{!}{%
		\begin{tabular}{|c|c|c|}
			\hline
			\rowcolor{cyan!20}
			\multicolumn{3}{|c|}{Radiative Decays} \\ 
			\hline
			$\text{Decay}$  & SM Value & Exp Value  \\
			\hline
			\hline
			$\mathcal{B} (B^{+} \to K^{*+} \gamma)$     & $(4.60 \pm 1.28) \times 10^{-5}$ \cite{Ali:2007sj} &$ (3.76 \pm 0.16 ) \times  10^{-5}$ \cite{Belle:2017hum} \\
			$\mathcal{B} (B^{0} \to K^{*} \gamma)$      & $(4.30 \pm 1.19 ) \times 10^{-5}$  \cite{Ali:2007sj} & $(3.96 \pm 0.16 ) \times  10^{-5}$ \cite{Belle:2017hum} \\
			$\mathcal{B} (B^{+} \to \rho^{+} \gamma)$   & $(0.85 \pm 0.32 ) \times 10^{-6}$  \cite{Ali:2001ez} &$(1.31 \pm 0.23 ) \times 10^{-6} $ \cite{Belle-II:2024tru}   \\ 
			$\mathcal{B} (B^{0} \to \rho^{0} \gamma)$   & $(0.49 \pm 0.17 ) \times 10^{-6}$  \cite{Ali:2001ez} & $(0.75 \pm 0.16 ) \times 10^{-6} $\cite{Belle-II:2024tru} \\
			$\mathcal{B} (B^{0}_{s} \to \phi \gamma)$   & $(4.30 \pm 1.18) \times 10^{-5}$ \cite{Ali:2007sj}& $(3.60 \pm 0.86 ) \times  10^{-5} $  \cite{Belle:2014sac}  \\
			$\mathcal{B} (B \to X_{s} \gamma)_{E_{\gamma}>1.6 \rm \, GeV } $          & $(3.40 \pm 0.17) \times 10^{-4}$ \cite{Misiak:2020vlo} & $ (3.49 \pm 0.19 ) \times 10^{-4} $ \cite{HeavyFlavorAveragingGroupHFLAV:2024ctg}  \\
			\hline
		\end{tabular}%
	}
	\caption{SM and Experimental branching ratios of the available radiative processes which we have taken into account. }
	\label{tab:radiative_data}
\end{table}

\subsection{\texorpdfstring{Semileptonic $b \to d_j \ell \ell$ decays}{b2sll} } \label{sec:b2sll}

The general low-energy effective Hamiltonian describing the $b\to d_j \ell^+\ell^-$ transitions is expressed as:
\begin{equation} \label{eq:Heff}
	{\cal H}^{b\to d_j}_{eff} = - \frac{4\,G_F}{\sqrt{2}} V_{tb}V_{td_j}^\ast
	\left[  \sum_{i=1}^{6} C_i (\mu)
	O_i(\mu) + \sum_{i=7,8,9,10,P,S} \biggl(C_i (\mu)  O_i + C'_i (\mu) 
	O'_i\biggr)\right] \,,
\end{equation}
where the twice Cabibbo suppressed contributions ($\propto  V_{ub}V^*_{us} $) have been neglected. We are following the operator basis as presented in refs.~\cite{Buras:1995iy, Becirevic:2012fy, Bobeth:1999mk, Altmannshofer:2008dz}.
The operators $O^{(')}_{7,8}$ are defined in eq.~\eqref{eq:opro7o8}, the rest of the operators are as given below
\begin{subequations}\label{eq:basisOps}
	\begin{align}
		O_{9} &= \frac{e^2}{16 \pi^2} 
		(\bar{d_j} \gamma_{\mu} P_L b)(\bar{\ell} \gamma^\mu \ell) ,&
		O_{9}^\prime &= \frac{e^2}{16 \pi^2} 
		(\bar{d_j} \gamma_{\mu} P_R b)(\bar{\ell} \gamma^\mu \ell) , \\
		O_{10} &=\frac{e^2}{16 \pi^2}
		(\bar{d_j}  \gamma_{\mu} P_L b)(  \bar{\ell} \gamma^\mu \gamma_5 \ell) ,&
		O_{10}^\prime &=\frac{e^2}{16 \pi^2}
		(\bar{d_j}  \gamma_{\mu} P_R b)(  \bar{\ell} \gamma^\mu \gamma_5 \ell) , \\
		O_{S} &=\frac{e^2}{16\pi^2} m_b
		(\bar{d_j} P_R b)(  \bar{\ell} \ell) ,&
		O_{S}^\prime &=\frac{e^2}{16\pi^2} m_b
		(\bar{d_j} P_L b)(  \bar{\ell} \ell) , \\
		O_{P} &=\frac{e^2}{16\pi^2} m_b 
		(\bar{d_j} P_R b)(  \bar{\ell} \gamma_5 \ell) ,&
		O_{P}^\prime &=\frac{e^2}{16\pi^2} m_b
		(\bar{d_j} P_L b)(  \bar{\ell} \gamma_5 \ell), 
	\end{align}
\end{subequations}
where $\ell = e$ or $\mu$, and the explicit expressions for the QCD penguin operators $\mathcal O_{1-6}$ can be found in the ref.~\cite{Bobeth:1999mk}. Here, $ C_i $ and $C'_i$s are the Wilson coefficients corresponding to the operators $ O_i $ and $ O'_i$, respectively. The operators $O_{7,8,9,10}$ are present in the SM, and most of them also appear in our analysis. Hence, the respective WCs will be expressed as given in eq.~\eqref{eq:wcssmnp}. The rest of the operators will appear only in NP scenarios.
The values of the WCs in the SM at $\mu_b$ scale can be found in \cite{Mahmoudi:2024zna}\,.     

\begin{figure}[t]
	\centering
\subfloat[]{\begin{tikzpicture}
		\begin{feynman}
		\vertex (a1){\( s,d\)};
		\vertex [below right=1.2cm of a1](a2);
		\vertex [large,below right=1.2cm of a2](a3);
		\vertex [square dot,blue,below left=1.2cm of a3](a4){ };
		\vertex [below left=1.2cm of a4](a5){\( b\)};
		\vertex [right=1.2cm of a3](a6);
		\vertex [above right=1.8cm of a6](a7){\(\ell\)};
		\vertex [below right=1.8cm of a6](a8){\(\ell\)};
		
		\diagram* {
			(a3) -- [fermion, arrow size=1pt,edge label'={\( t\)}] (a2) -- [fermion, arrow size=1pt] (a1),
			(a5) --[fermion, arrow size=1pt] (a4) -- [fermion, arrow size=1pt,edge label'={\( t\)}] (a3),
			(a2) --[boson, edge label'={\(W\)}] (a4),
			(a3) --[boson,edge label={\(Z/\gamma\)}](a6),
			(a8) --[fermion, arrow size=1pt](a6) --[fermion, arrow size=1pt](a7),
			
		};
		\end{feynman}
		\end{tikzpicture}}
\subfloat[]{\begin{tikzpicture}
	\begin{feynman}
	\vertex (a1){\( b\)};
	\vertex [above right=2.2cm of a1](a2);
	\vertex [square dot, blue, above left=0.7cm of a2](a3){};
	\vertex [above left=0.7cm of a3](a4);
	\vertex [above left=0.7cm of a4](a5){\( s,d\)};
	\vertex [right=1.2cm of a2](a6);
	\vertex [above right=1.8cm of a6](a7){\(\ell\)};
	\vertex [below right=1.8cm of a6](a8){\(\ell\)};

	\diagram* {
		(a1) --[fermion, arrow size=1pt](a2) --[fermion, arrow size=1pt, edge label={\(b\)}] (a3) --[fermion, arrow size=1pt, edge label={\(t \)}] (a4) --[fermion, arrow size=1pt] (a5),
		(a4) --[boson, half left, looseness=2, edge label = \(W\)] (a3),
		(a2) --[boson, edge label={\( Z/\gamma\)}] (a6),
		(a8) --[fermion, arrow size=1pt](a6) --[fermion, arrow size=1pt](a7),
	};
	\end{feynman}
	\end{tikzpicture}}
\subfloat[]{\begin{tikzpicture}
	\begin{feynman}
	\vertex (a1){\( b\)};
	\vertex [square dot, blue,above right=1cm of a1](a2){};
	\vertex [above right=0.9cm of a2](a3);
	\vertex [above right=0.7cm of a3](a5);
	\vertex [right=1.2cm of a5](a6);
	\vertex [above left=2.2cm of a5](a9){\(s,d\)};
	\vertex [above right=1.8cm of a6](a7){\(\ell\)};
	\vertex [below right=1.8cm of a6](a8){\(\ell\)};

	\diagram* {
		(a1) --[fermion, arrow size=1pt](a2) --[fermion, arrow size=1pt, edge label'={\(t\)}] (a3) --[fermion, arrow size=1pt, edge label={\(s,d \)}] (a5) --[fermion, arrow size=1pt] (a9),
		(a2) --[boson, half left, looseness=2, edge label = \(W\)] (a3),
		(a5) --[boson, edge label={\( Z/\gamma\)}] (a6),
		(a8) --[fermion, arrow size=1pt](a6) --[fermion, arrow size=1pt](a7),
	};
	\end{feynman}
	\end{tikzpicture}}

\hspace{0.001cm}
\subfloat[]{\begin{tikzpicture}
	\begin{feynman}
	\vertex (a1){\( s,d\)};
	\vertex [below right=1.2cm of a1](a2);
	\vertex [large,below right=1.2cm of a2](a3);
	\vertex [square dot,blue,below left=1.2cm of a3](a4){};
	\vertex [below left=1.2cm of a4](a5){\( b\)};
	\vertex [right=1.2cm of a3](a6);
	\vertex [above right=1.8cm of a6](a7){\(\ell\)};
	\vertex [below right=1.8cm of a6](a8){\(\ell\)};
	
	\diagram* {
		(a3) -- [fermion, arrow size=1pt,edge label'={\( t\)}] (a2) -- [fermion, arrow size=1pt] (a1),
		(a5) --[fermion, arrow size=1pt] (a4) -- [fermion, arrow size=1pt,edge label'={\( t\)}] (a3),
		(a2) --[boson, edge label'={\(W\)}] (a4),
		(a3) --[scalar,edge label={\(H\)}](a6),
		(a8) --[fermion, arrow size=1pt](a6) --[fermion, arrow size=1pt](a7),
		
	};
	\end{feynman}
	\end{tikzpicture}}
\subfloat[]{\begin{tikzpicture}
	\begin{feynman}
	\vertex (a1){\( b\)};
	\vertex [above right=2.2cm of a1](a2);
	\vertex [square dot, blue, above left=0.7cm of a2](a3){};
	\vertex [above left=0.7cm of a3](a4);
	\vertex [above left=0.7cm of a4](a5){\( s,d\)};
	\vertex [right=1.2cm of a2](a6);
	\vertex [above right=1.8cm of a6](a7){\(\ell\)};
	\vertex [below right=1.8cm of a6](a8){\(\ell\)};

	\diagram* {
		(a1) --[fermion, arrow size=1pt] (a2) --[fermion, arrow size=1pt, edge label={\(b\)}] (a3) --[fermion, arrow size=1pt, edge label={\(t \)}] (a4) --[fermion, arrow size=1pt] (a5),
		(a4) --[boson, half left, looseness=2, edge label = \(W\)] (a3),
		(a2) --[scalar, edge label={\(H\)}] (a6),
		(a8) --[fermion, arrow size=1pt](a6) --[fermion, arrow size=1pt](a7),
	};
	\end{feynman}
	\end{tikzpicture}}
\subfloat[]{\begin{tikzpicture}
	\begin{feynman}
	\vertex (a1){\( b\)};
	\vertex [square dot, blue,above right=1cm of a1](a2){};
	\vertex [above right=0.9cm of a2](a3);
	\vertex [above right=0.7cm of a3](a5);
	\vertex [right=1.2cm of a5](a6);
	\vertex [above left=2.2cm of a5](a9){\(s,d\)};
	\vertex [above right=1.8cm of a6](a7){\(\ell\)};
	\vertex [below right=1.8cm of a6](a8){\(\ell\)};	
	\diagram* {
		(a1) --[fermion, arrow size=1pt](a2) --[fermion, arrow size=1pt, edge label'={\(t\)}] (a3) --[fermion, arrow size=1pt, edge label={\(s,d \)}] (a5) --[fermion, arrow size=1pt] (a9),
		(a2) --[boson, half left, looseness=2, edge label = \(W\)] (a3),
		(a5) --[scalar, edge label={\(H\)}] (a6),
		(a8) --[fermion, arrow size=1pt](a6) --[fermion, arrow size=1pt](a7),
	};
	\end{feynman}
	\end{tikzpicture}}\hspace{0.001cm}
    \subfloat[]{\begin{tikzpicture}
		\begin{feynman}
		\vertex (a1){\( s,d\)};
		\vertex [below right=1.2cm of a1](a2);
		\vertex [large,below right=1.2cm of a2](a3);
		\vertex [square dot,blue,below left=1.2cm of a3](a4){};
		\vertex [below left=1.2cm of a4](a5){\( b\)};
		\vertex [right=1.2cm of a3](a6);
		\vertex [above right=1.8cm of a6](a7){\(\ell\)};
		\vertex [below right=1.8cm of a6](a8){\(\ell\)};
		
		\diagram* {
			(a5) --[fermion, arrow size=1pt] (a4) -- [fermion, arrow size=1pt,edge label'={\( t\)}] (a2) -- [fermion, arrow size=1pt] (a1),
                (a3) --[boson,edge label={\(Z/\gamma\)}](a6) ,
                 (a2) -- [boson, edge label={\( W\)}] (a3) -- [boson, edge label={\( W\)}] (a4),
                (a8) --[fermion, arrow size=1pt](a6) --[fermion, arrow size=1pt](a7),		
		};
		\end{feynman}
		\end{tikzpicture}}
    \subfloat[]{\begin{tikzpicture}
		\begin{feynman}
		\vertex (a1){\( s,d\)};
		\vertex [below right=1.2cm of a1](a2);
		\vertex [large,below right=1.2cm of a2](a3);
		\vertex [square dot,blue,below left=1.2cm of a3](a4){};
		\vertex [below left=1.2cm of a4](a5){\( b\)};
		\vertex [right=1.2cm of a3](a6);
		\vertex [above right=1.8cm of a6](a7){\(\ell\)};
		\vertex [below right=1.8cm of a6](a8){\(\ell\)};
		
		\diagram* {
			(a5) --[fermion, arrow size=1pt] (a4) -- [fermion, arrow size=1pt,edge label'={\( t\)}] (a2) -- [fermion, arrow size=1pt] (a1),
                (a3) --[scalar,edge label={\(H\)}](a6) ,
                 (a2) -- [boson, edge label={\( W\)}] (a3) -- [boson, edge label={\( W\)}] (a4),
                (a8) --[fermion, arrow size=1pt](a6) --[fermion, arrow size=1pt](a7),		
		};
		\end{feynman}
		\end{tikzpicture}}
    \subfloat[]{\begin{tikzpicture}
	\begin{feynman}
	\vertex (a1){\(b\)};
	\vertex[square dot,blue,right=1.7cm of a1](a2){};
	\vertex[above=1.5cm of a2](a3);
	\vertex[left=1.5cm of a3](a4){\(s,d\)};
	\vertex[right=1.5cm of a2](a5);
	\vertex[right=1.5cm of a3](a6);
	\vertex[right=1.5cm of a6](a7){\(\ell\)};
	\vertex[right=1.5cm of a5](a8){\(\ell\)};
        \vertex[below=1cm of a8](a9){};
	
	\diagram* { 
		(a1) --[fermion, arrow size=1pt](a2) --[fermion, arrow size=1pt,edge label=\(t\)](a3) --[fermion, arrow size=1pt](a4),
		(a2) --[boson,edge label'=\(W\)](a5),
		(a3) --[boson,edge label={\(W\)}](a6),
		(a7) --[fermion, arrow size=1pt](a6) --[fermion, arrow size=1pt,edge label=\(\nu_{\ell}\)](a5) --[fermion, arrow size=1pt](a8),
	};	
	\end{feynman}
	\end{tikzpicture}}
	\caption{Feynman diagrams contributing to FCNC process $b \to d_{j} (=s,d) \ell \ell$. The blue squares refer to the $ Wtb $ effective vertex.}
    \label{fig:FCNC_feynman}
\end{figure}
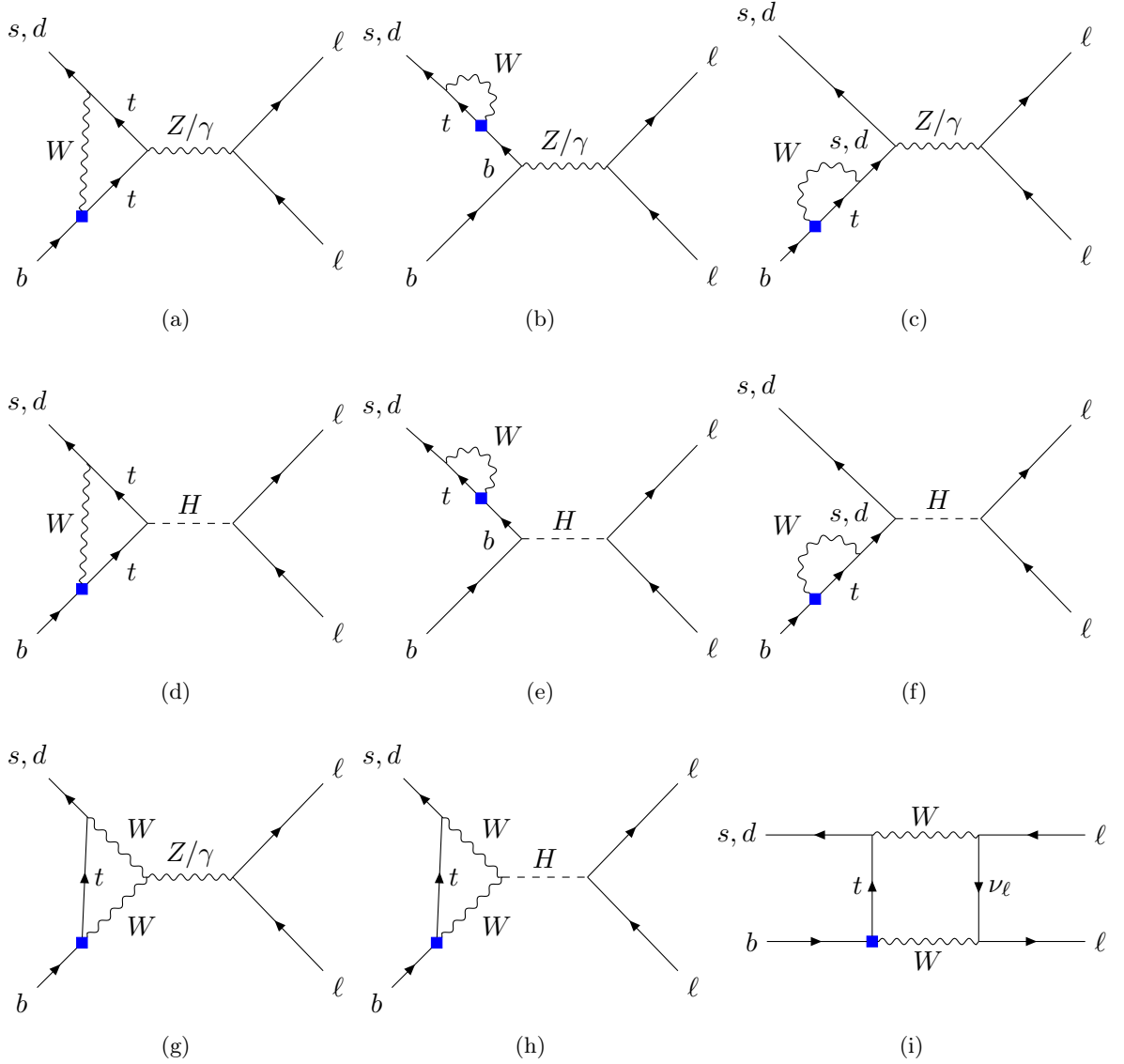

Like $b\to d_j\gamma$ vertex the SMEFT or the $Wtb$ effective operators will modify the $b\to d_j Z$ or $b\to d_j \, H$ vertices at the one-loop level. In fig.~\ref{fig:FCNC_feynman}, we have shown the diagrams that contribute to the flavour-changing neutral current $b \to d_j \ell \ell$ transitions via the contributions in $b\to d_j \gamma(Z/H)$ vertices. The contributions come from the one-loop box and penguin diagrams as well as from corrections to the external legs. The diagrams in fig.~\ref{fig:FCNC_feynman} mediated via $Z$ boson and photon will contribute to the vector current, whereas the diagrams mediated via Higgs boson will contribute to the scalar current. After calculating these loop diagrams, we obtain the matching contributions to the low-energy WCs in eq.~\eqref{eq:Heff} which are given as 
\begin{subequations}\label{eq:NP_wilsons}
	\begin{align}
		\Delta C_{9} &= -\frac{v^2}{2\lambda_t}\frac{16 \pi^2}{e^2} (C_{V_{LL}} + C_{V_{LR}}) , &
		 C_{9}^\prime &= -\frac{v^2}{2\lambda_t}\frac{16 \pi^2}{e^2} (C_{V_{RL}}+ C_{V_{RR}}) , \\
		\Delta C_{10} &= -\frac{v^2}{2\lambda_t}\frac{16 \pi^2}{e^2} (C_{V_{LR}} - C_{V_{LL}}) , &
		C_{10}^\prime &= -\frac{v^2}{2\lambda_t}\frac{16 \pi^2}{e^2} \left(C_{V_{RR}} - C_{V_{RL}} \right) , \\
		 C_{S} \,  &= -\frac{v^2}{2\lambda_t}\frac{16 \pi^2 m_b}{e^2}(C_{S_{RL}}+C_{S_{RR}}) , &
		 C_{S}^\prime \ &= -\frac{v^2}{2\lambda_t}\frac{16 \pi^2 m_b}{e^2}(C_{S_{LL}}+C_{S_{LR}}). 
	\end{align}
\end{subequations}
Here, $\lambda_t=V_{tb} V_{ts}^*$ and the Wilson coefficients are written at the EW scale. The matching contributions for $\Delta C_{7,8}$ and $C'_{7,8}$ are given in eq.~\eqref{eq:C7C8NP}. In eq.~\eqref{eq:NP_wilsons}, the various loop factors are expressed as 
\begin{subequations}
	\begin{eqnarray}
		C_{V_{IJ}} &=& C_{V_{IJ}}^{\rm vertex} + C_{V_{IJ}}^{\rm self} + C_{V_{IJ}}^{\rm box } \,,\\
		C_{S_{IJ}} &=& C_{S_{IJ}}^{\rm vertex} + C_{S_{IJ}}^{\rm self} \,.
	\end{eqnarray}
\end{subequations}
The expressions of the coefficients $C_{i}$'s are given in eqs.~\eqref{eq:loop_b2sll_gamma_med}-\eqref{eq:loop_b2sll_scalar_right} in appendix \ref{appndx:Loop_b2sll}. Note that like $C'_{7,8}$, the WCs $ C_9^{\prime}$, $ C_{10}^{\prime}$ depend on light quark masses and are therefore highly suppressed compared to their unprimed Wilson coefficients. For example, $C^{\prime}_{9,10}$ are $\sim \mathcal{O}(10^2)$ times smaller than $\Delta C_{9,10}$, approximately. The lepton sector will not receive any correction as NP is coming through the effective $Wtb$ vertex. So, the operators such as $ C_{P}$ and $C_{p}^{\prime}$ are zero. Consequently, our analysis focuses on the dominant coefficients $\Delta C_7$, $\Delta C_9$, $\Delta C_{10}$, $ C_S$ and $ C_S^{\prime}$.

Note that the Hamiltonian in eq.~\eqref{eq:Heff} is valid below the EW scale. 
To study the processes involving B-meson, we need the Wilson coefficients at the scale $\mu_{b}$. In our approach, we first use RGE to evolve the NP parameters ($Wtb$ effective couplings as well as SMEFT couplings $(\mathcal{C}_i$)) to the EW scale. Next, we have calculated the loop contributions $(b\to s(d)\ell \ell )$ at the EW scale, which has been matched to the basis of eq.~\eqref{eq:Heff}. Finally, we run the LEFT Wilson coefficients $(C_i)$ down to the $\mu_b$ scale using the RGEs provided below.

\begin{align}\label{eq:FCNC_EW_to_mb}
	\begin{pmatrix}
		C_9\\
		C_{10}\\
		C_S
	\end{pmatrix}_{{\mu_b}}&=\begin{pmatrix}
		 0.99522 & 0.00716 & 0 \\
		 0.00716 & 1.0& 0 \\
	 0& 0 &1.37433 
	\end{pmatrix}   \begin{pmatrix}
		C_9\\
		C_{10}\\
		C_S
\end{pmatrix}_{\mu_{\rm{EW}}}\,,
\end{align}
    
For the numerical analysis, we have considered all the observables corresponding to the quark process $b \to s \mu^{+} \mu^{-}$ including leptonic and semileptonic decays such as $B \to K^{(*)} \mu^{+} \mu^{-}$ and $B \to \phi \mu^{+} \mu^{-}$. The vast amount of data available on differential branching fractions, CP asymmetries, and various angular observables measured for these modes by LHCb, Belle, CDF, ATLAS, and CMS \cite{CDF:2011tds, LHCb:2013lvw, LHCb:2014cxe, LHCb:2014vgu, LHCb:2015svh, Belle:2016fev, CMS:2017rzx, ATLAS:2018gqc, LHCb:2020gog,LHCb:2021zwz,LHCb:2022qnv}.
We also have considered the updated value of the Lepton Flavour Universality Violation (LFUV) observables $R_{K^{(*)}}$, which agree with the SM within a $1\sigma$ error, which is crucial for constraining any NP scenarios. A detail on the observables from $b \to s \mu^{+} \mu^{-} $ channels that are considered in the analysis are discussed in \cite{Biswas:2020uaq}.


\subsection{Rare Decays}
Rare dileptonic decays of neutral mesons are very important probes to new physics since the rates of these decays are suppressed in the SM. For our cases, we will mainly get contributions to the process $B_{s}^{0} \to \ell \ell $ and $B^{0} \to \ell \ell $ $(\ell = e, \mu, \tau)$. Among these, we only have data for the muon channels, and an upper limit exists for electron and tau channels. 
 Following the Hamiltonian given in eq.~\eqref{eq:Heff}, the model independent expression of the branching ratios of $ B_q \rightarrow \mu^+ \mu^- $ decay is given by \cite{Becirevic:2012fy}: 
\begin{equation}
\begin{split} \label{eq:BR_formula}
\mathcal{B}(B_q \rightarrow \mu^+ \mu^-) = & \tau_{B_q} f_{B_q}^2  m_{B_q}^3 \frac{G_F^2 \alpha^2}{64 \pi^3} |V^*_{tq}V_{tb}|^2 \beta_{\mu}(m_{B_q}^2) \left[   \frac{m_{B_q}^2}{m_b^2} |C_s - C'_s|^2 \left(1-\frac{4m_{\mu}^2}{m_{B_q}^2}\right) \right.\\& \left.  + \bigg|\frac{m_{B_q}}{m_b}(C_p - C'_p) + 2\frac{m_{\mu}}{m_{B_q}} (C_{10} - C'_{10})\bigg|^2 \right]\,,
\end{split}
\end{equation}
where, $\beta_\ell(q^2)=\sqrt{  1- { 4 m_\ell^2/q^2}   }$ and the $ B  $ meson decay constant is defined via the matrix element: $ \langle 0| \bar{q}  \gamma_{\mu} P_L b  |B_q(p)\rangle =\frac{i}{2} f_{B_q} p_\mu $, where $q=(d,s)$. We have used the values of decay constants, $f_{B} = (190.0\pm1.3) \text{ MeV}$ and $f_{B_{s}} = (230.3 \pm1.3)$ MeV \cite{FlavourLatticeAveragingGroupFLAG:2024oxs}. In this case, NP contribution comes through the Wilson coefficients $C_{10}$, $C_{S}$ and $C_S^{\prime}$. The expressions of the Wilson coefficients are given in eq.~\eqref{eq:NP_wilsons} in terms of the anomalous couplings of $Wtb$.

The most recent experimental results are as follows \cite{ParticleDataGroup:2024cfk}:
\begin{subequations}\label{eq:rare_exp}
    \begin{eqnarray}
    \mathcal{B}(B^{0} \to \mu^{+} \mu^{-})^{\rm exp} &=& \left( 1.2 ^{+0.8}_{-0.7} \, \, \pm 0.1 \right) \times 10^{-10} \,, \\
    \mathcal{B}(B^{0}_{s} \to \mu^{+} \mu^{-})^{\rm exp} &=& \left(3.83 ^{+0.38 \, \, +0.24}_{-0.36\,\, -0.21}  \right) \times 10^{-9} \,.
\end{eqnarray}\end{subequations}
 The SM contribution will be dependent only on $C_{10}$, including the QED corrections the SM predictions are given as \cite{Beneke:2019slt}:
\begin{subequations}
\begin{align}
\mathcal{B}(B_{s}^{0} \rightarrow \mu^+ \mu^-)^{\rm SM} = (3.66 \pm 0.14)\times 10^{-9}\,,\\ 
\mathcal{B}(B^{0} \rightarrow \mu^+ \mu^-)^{\rm SM} = (1.03 \pm 0.05)\times 10^{-10}\,.
\end{align}
\end{subequations}
\noindent Note that these SM predictions are fully consistent with the respective measurements shown in eq.~\eqref{eq:rare_exp}.

\subsection{Invisible Decays}\label{sec:invisible_decays}
\begin{figure}[t]
	\centering
\subfloat[]{\begin{tikzpicture}
		\begin{feynman}
		\vertex (a1){\( s,d\)};
		\vertex [below right=1.2cm of a1](a2);
		\vertex [large,below right=1.2cm of a2](a3);
		\vertex [square dot,blue,below left=1.2cm of a3](a4){ };
		\vertex [below left=1.2cm of a4](a5){\( b\)};
		\vertex [right=1.2cm of a3](a6);
		\vertex [above right=1.8cm of a6](a7){\(\nu\)};
		\vertex [below right=1.8cm of a6](a8){\(\nu\)};
		
		\diagram* {
			(a3) -- [fermion, arrow size=1pt,edge label'={\( t\)}] (a2) -- [fermion, arrow size=1pt] (a1),
			(a5) --[fermion, arrow size=1pt] (a4) -- [fermion, arrow size=1pt,edge label'={\( t\)}] (a3),
			(a2) --[boson, edge label'={\(W\)}] (a4),
			(a3) --[boson,edge label={\(Z\)}](a6),
			(a8) --[fermion, arrow size=1pt](a6) --[fermion, arrow size=1pt](a7),
			
		};
		\end{feynman}
		\end{tikzpicture}}
\subfloat[]{\begin{tikzpicture}
	\begin{feynman}
	\vertex (a1){\( b\)};
	\vertex [above right=2.2cm of a1](a2);
	\vertex [square dot, blue, above left=0.7cm of a2](a3){};
	\vertex [above left=0.7cm of a3](a4);
	\vertex [above left=0.7cm of a4](a5){\( s,d\)};
	\vertex [right=1.2cm of a2](a6);
	\vertex [above right=2cm of a6](a7){\(\nu\)};
	\vertex [below right=2cm of a6](a8){\(\nu\)};

	\diagram* {
		(a1) --[fermion, arrow size=1pt](a2) --[fermion, arrow size=1pt, edge label={\(b\)}] (a3) --[fermion, arrow size=1pt, edge label={\(t \)}] (a4) --[fermion, arrow size=1pt] (a5),
		(a4) --[boson, half left, looseness=2, edge label = \(W\)] (a3),
		(a2) --[boson, edge label={\( Z\)}] (a6),
		(a8) --[fermion, arrow size=1pt](a6) --[fermion, arrow size=1pt](a7),
	};
	\end{feynman}
	\end{tikzpicture}}
\subfloat[]{\begin{tikzpicture}
	\begin{feynman}
	\vertex (a1){\( b\)};
	\vertex [square dot, blue,above right=1cm of a1](a2){};
	\vertex [above right=0.9cm of a2](a3);
	\vertex [above right=0.7cm of a3](a5);
	\vertex [right=1.2cm of a5](a6);
	\vertex [above left=2.2cm of a5](a9){\(s,d\)};
	\vertex [above right=1.8cm of a6](a7){\(\nu\)};
	\vertex [below right=1.8cm of a6](a8){\(\nu\)};

	\diagram* {
		(a1) --[fermion, arrow size=1pt](a2) --[fermion, arrow size=1pt, edge label'={\(s,d\)}] (a3) --[fermion, arrow size=1pt, edge label={\(t \)}] (a5) --[fermion, arrow size=1pt] (a9),
		(a2) --[boson, half left, looseness=2, edge label = \(W\)] (a3),
		(a5) --[boson, edge label={\( Z\)}] (a6),
		(a8) --[fermion, arrow size=1pt](a6) --[fermion, arrow size=1pt](a7),
	};
	\end{feynman}
	\end{tikzpicture}}\hspace{0.001cm}
    \subfloat[]{\begin{tikzpicture}
		\begin{feynman}
		\vertex (a1){\( s,d\)};
		\vertex [below right=1.2cm of a1](a2);
		\vertex [large,below right=1.2cm of a2](a3);
		\vertex [square dot,blue,below left=1.2cm of a3](a4){};
		\vertex [below left=1.2cm of a4](a5){\( b\)};
		\vertex [right=1.2cm of a3](a6);
		\vertex [above right=1.8cm of a6](a7){\(\nu\)};
		\vertex [below right=1.8cm of a6](a8){\(\nu\)};
		
		\diagram* {
			(a5) --[fermion, arrow size=1pt] (a4) -- [fermion, arrow size=1pt,edge label'={\( t\)}] (a2) -- [fermion, arrow size=1pt] (a1),
                (a3) --[boson,edge label={\(Z\)}](a6) ,
                 (a2) -- [boson, edge label={\( W\)}] (a3) -- [boson, edge label={\( W\)}] (a4),
                (a8) --[fermion, arrow size=1pt](a6) --[fermion, arrow size=1pt](a7),		
		};
		\end{feynman}
		\end{tikzpicture}}\hspace{1cm}
        \subfloat[]{\begin{tikzpicture}
	\begin{feynman}
	\vertex (a1){\(b\)};
	\vertex[square dot,blue,right=2.2cm of a1](a2){};
	\vertex[above=2cm of a2](a3);
	\vertex[left=2cm of a3](a4){\(s,d\)};
	\vertex[right=2cm of a2](a5);
	\vertex[right=2cm of a3](a6);
	\vertex[right=2cm of a6](a7){\(\nu\)};
	\vertex[right=2cm of a5](a8){\(\nu\)};
        \vertex[below=0.8cm of a8](a9){};
	
	\diagram* { 
		(a1) --[fermion, arrow size=1pt](a2) --[fermion, arrow size=1pt,edge label=\(t\)](a3) --[fermion, arrow size=1pt](a4),
		(a2) --[boson,edge label'=\(W\)](a5),
		(a3) --[boson,edge label={\(W\)}](a6),
		(a7) --[fermion, arrow size=1pt](a6) --[fermion, arrow size=1pt,edge label=\(\ell\)](a5) --[fermion,arrow size=1pt](a8),
	};	
	\end{feynman}
    \end{tikzpicture}}
    \caption{Feynman diagrams contributing to invisible decay process $b  \to s (d) \nu \bar{\nu}$. The blue square dots refer to the $ Wtb $ effective vertex.}
    \label{fig:invisible}
    \end{figure}
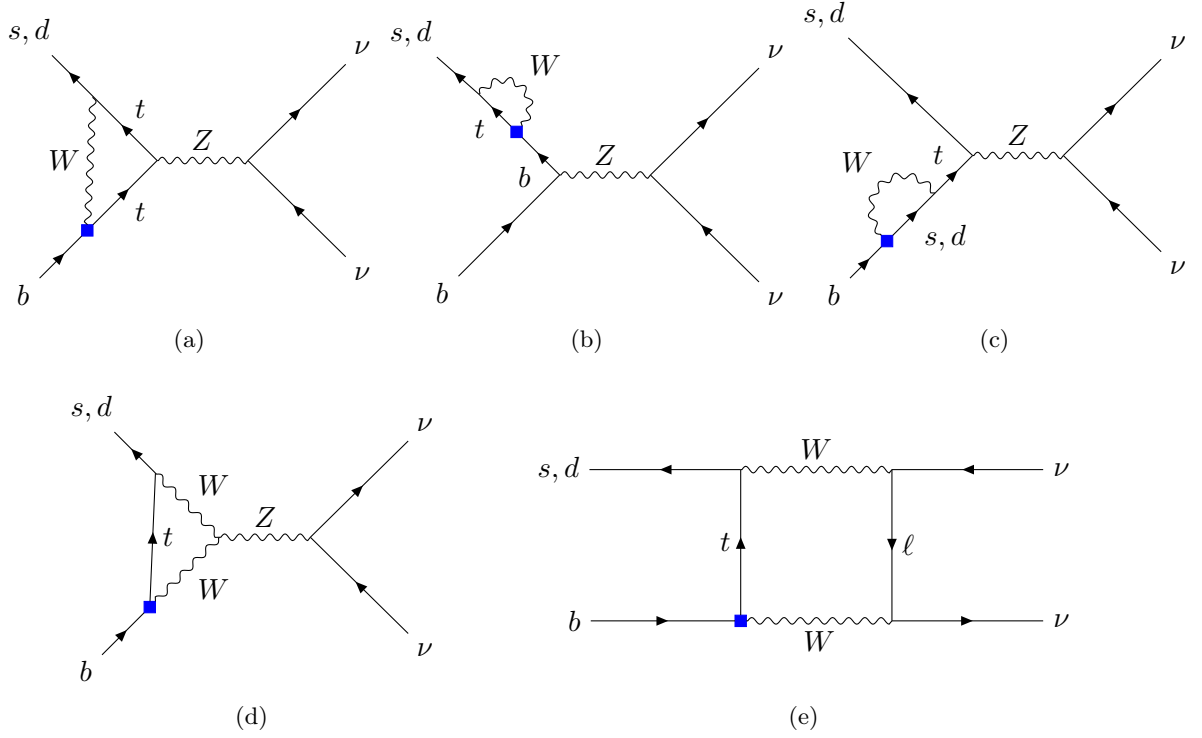

Invisible decays are an important probe for studying the new physics effect. In SM, the decays like $P \to P' \bar{\nu} \nu$, with $P$ and $P'$ being pseudoscalar (vector mesons), are considered as invisible decay as neutrinos cannot be detected in the detector and treated as missing energy. Recently, Belle has provided for the first time the result on the branching ratio of the invisible decay $B^{+} \to K^{+} \bar{\nu} \nu$ decay \cite{Belle-II:2023esi}. The value of the branching ratio of $B^{0} \to K^{*} \bar{\nu} \nu$ decay is provided by Babar collaboration \cite{BaBar:2013npw}. 
\begin{subequations}
    \begin{eqnarray}
        \mathcal{B} (B^{+} \to K^{+} \bar{\nu} \nu)^{\rm exp} &=& (2.30 \pm 0.50 \pm 0.45 ) \times 10^{-5}\,, \\
        \mathcal{B} (B^{0} \to K^{*} \bar{\nu} \nu)^{\rm exp} &=& (3.80 \pm 2.75 ) \times 10^{-5}\,.
    \end{eqnarray}
\end{subequations}


The weak effective or the LEFT Hamiltonian for $b \to s  \bar{\nu} \nu$ process is given as  
\begin{equation}\label{eq:eff_hamilton_b2snunubar}
	\mathcal{H}_{\rm eff}^{b \to d_{j} \bar{\nu} \nu}= \frac{4 G_{F}}{\sqrt{2}} V_{td_{j}}^{*} V_{tb} \sum_{i} C_{i} \mathcal{O}_{i}\,,
\end{equation}
with the operator structure: 
\begin{eqnarray}
	\mathcal{O}_{L}^{\nu} = \frac{e^2}{16 \pi^2}  \left( \bar{d}_{j} \gamma_{\mu} P_{L}b \right) \left(\bar{\nu} \gamma^{\mu}P_{L}\nu\right) , & \ \ \ \ &  \mathcal{O}_{R}^{\nu}=  \frac{e^2}{16 \pi^2}  \left( \bar{d}_{j} \gamma_{\mu} P_{R}b \right) \left(\bar{\nu} \gamma^{\mu}P_{L}\nu\right)\,,
\end{eqnarray}
with $d_{j} = s(d)$ for the decay of $b \to s(d)  \bar{\nu} \nu$ process.

\noindent The $q^2$ distribution of the branching fraction of $B^{+} \to K^{+} \nu \bar{\nu}$ in terms of the above Wilson coefficients is given by 
	\begin{equation}\label{eq:br_B2Knunubar}
	\frac{d \Gamma (B^{+} \to K^{+} \bar{\nu}  \nu  )}{d q^2} =  \frac{G_{F}^2 \alpha^2 }{ 2^{10}  \pi^5 m_{B}^3} |V_{ts}^{*} V_{tb}|^2 \lambda^{3/2} ( m_{B}^2 , m_{K}^2, q^2) \left[f_{+}^{K}(q^2)\right]^2 |C_{L}^{\nu} + C_{R}^{\nu}|^2 \,.
	\end{equation}
Here, $f_{+}$ is the form factor for the vector current of $B \to K$. The branching ratio of the process $B \to K^{*}  \bar{\nu} \nu$ will depend on the polarization of $K^{*}$ vector meson, which can be written as follows:
	
	{\footnotesize \begin{subequations}
		\begin{eqnarray}
		\frac{d\Gamma}{dq^2}(B \to K^{*} \bar{\nu}\nu)_{0}=&& P_{C} \frac{4 (C_{L}^{\nu} - C_{R}^{\nu})^2 \left( 
			A_1(q^2) (m_B + m_{K^{*}})^2 \left(m_B^2 - m_{K^{*}}^2 - q^2 \right) 
			- A_2(q^2) \lambda(m_B^2, m_{K^{*}}^2, q^2) \right)^2 }{3 m_{K^{*}}^2 (m_B + m_{K^{*}})^2} \,,\nonumber \\
		\\
		\frac{d\Gamma}{dq^2}(B \to K^{*} \bar{\nu}\nu)_{\pm}= && P_{C} \frac{16 q^2 \left( A_1(q^2) (C_{L}^{\nu} - C_{R}^{\nu})(m_B + m_{K^{*}})^2 
			\mp (C_{L}^{\nu} + C_{R}^{\nu}) V(q^2) \sqrt{\lambda(m_B^2, m_{K^{*}}^2, q^2)} \right)^2 }{3 (m_B + m_{K^{*}})^2}\,, \nonumber \\
		\end{eqnarray}\end{subequations}}
	
where the index after parentheses on the left-hand side denotes the polarisation of $K^{*}$. $A_{1}, A_{2}, V$ are the FFs of $B \to K^{*}$ transition. $P_{C} $ being the common factor, given as: 
	
	\begin{eqnarray}\label{eq:common_factor_invisible}
	P_{C} = && \frac{3 G_F^2  |V_{ts}^{*}V_{tb} |^2 \alpha^2 
		\sqrt{\lambda(m_B^2, m_{K^{*}}^2, q^2)}} { 16 \pi^5 2^{10}  m_B^3 }.
	\end{eqnarray}
Therefore, the total branching ratio of the process is given as the sum of the branching ratios of all the polarisations.
\begin{eqnarray}
    \frac{d\Gamma (B \to K^{*} \bar{\nu} \nu )}{dq^2} = \sum_{i= \{ 0,+,-\}} \frac{d\Gamma}{dq^2}(B \to K^{*} \bar{\nu}\nu)_{i}\,.
\end{eqnarray}
The SM predictions of the respective $q^2$ integrated branching ratios are given as \cite{Becirevic:2023aov}:
\begin{subequations}
	\begin{eqnarray}
		\mathcal{B} (B^{+} \to K^{+} \bar{\nu} \nu)^{\rm SM} &=& (5.06 \pm  0.14 \pm  0.27) \times 10^{-6}\,, \\
		\mathcal{B} (B^{0} \to K^{*} \bar{\nu} \nu)^{\rm SM} &=& (9.05 \pm 1.25 \pm 0.55 ) \times 10^{-6}\,.
	\end{eqnarray}
\end{subequations}
These predictions are based on the $B \to K$ and $B \to K^{*}$ form factors estimated using the lattice QCD (HPQCD) \cite{Parrott:2022rgu} and LCSR \cite{Bharucha:2015bzk}, respectively. 
    
When we write the decay width for the invisible channel $ B \to K^{(*)} \bar{\nu} \nu $, all three flavours of neutrino are considered in the calculation. 
In our effective operator framework, the contributions in these invisible decay channels will be via the one-loop diagrams shown in fig.~\ref{fig:invisible}. From the matching condition, we get contributions in both $C_{L}^{\nu}$ and $C_{R}^{\nu}$. We represent these contributions as: 
\begin{eqnarray}
    C_{L}^{\nu} = C_{L}^{\nu, \rm SM} + C_{L}^{\nu, \rm NP} , \quad  \quad C_{R}^{\nu} = C_{R}^{\nu, \rm NP} \,.
\end{eqnarray}
The value of the Wilson coefficients in the SM is given by \cite{Chen:2024jlj, Brod:2021hsj, Buras:2014fpa}: \begin{equation}
    C_{L}^{\nu, \rm SM} = - 12.64 \pm 0.14; \ \ \ \ C_{R}^{\nu, \rm SM} =0.
\end{equation}   
The expression for the NP loop contributions are written in eq.~\eqref{eq:loop_invisible} of appendix \ref{appndx:invisible}. Note that since this is a low-energy process similar to the previous case, we assumed that the NP originates at the scale $\Lambda$ (for SMEFT or at $\mu_{t}$ for $Wtb$ effective vertex). We calculated the loop factors and matched them with LEFT of eq.~\eqref{eq:eff_hamilton_b2snunubar} and obtained the WCs at the EW scale. Note that the coefficients $C_{L,R}^{\nu}$ do not change during RGE, hence, scale $C_{L,R}^{\nu, \rm NP}(\mu_{b}) \approx C_{L,R}^{\nu, \rm NP}(\mu_{\rm EW})$. Using the data shown above, we obtain the bounds on $C_{L,R}^{\nu, \rm NP}(\mu_{b})$, which we will discuss later.  

\section{Impacts on Different FCCC Processes}\label{sec:FCCC_process}
In the previous section, we studied the impact of anomalous $Wtb$ couplings on different FCNC processes. In this section, we will study how these couplings modify different FCCC processes. In particular, we focus on the phenomenology of the semileptonic and leptonic decays. A wealth of data is available on various observables related to these decays, which we use to constrain the CKM elements. The strategy for calculating the loop is the same as that of previous FCNC processes. 
The NP correction to the charged current vertex will impact the semileptonic and leptonic decays of the $B$ meson. The most general effective Hamiltonian containing all possible four-fermion operators for the $ b \to c \ell \nu $ transition, taking the neutrinos to be like SM (left-handed), can be written as 
\begin{equation}\label{eq:four_ferm_eff_b2c}
	\mathcal{H}_{eff}^{b \to c \ell \nu} = \frac{4G_F }{\sqrt{2}} V_{cb} \left[( 1 + C^{\ell}_{V_1}) \mathcal{O}_{V_1}^{\ell} +  C^{\ell}_{V_2} \mathcal{O}_{V_2}^{\ell} +  C^{\ell}_{S_1} \mathcal{O}_{S_1}^{\ell}+  C^{\ell}_{S_2} \mathcal{O}_{S_2}^{\ell}+  C^{\ell}_{T} \mathcal{O}_{T}^{\ell} \right] \,,
\end{equation}
With the operators' structure  \cite{Sakaki:2013bfa}:
\begin{eqnarray}	
	\mathcal{O}_{V_1}^{\ell} = (\bar{c}_L \gamma^{\mu}b_{L} )(\bar{\ell}_L \gamma_{\mu} \nu_{\ell L })\,, & \ \ \ & 	\mathcal{O}_{V_2}^{\ell} = (\bar{c}_R \gamma^{\mu}b_{R} )(\bar{\ell}_L \gamma_{\mu} \nu_{\ell L })\,, \nonumber\\
	\mathcal{O}_{S_1}^{\ell} = (\bar{c}_L b_{R} )(\bar{\ell}_R  \nu_{\ell L }) \,, &\ \ \ &
	\mathcal{O}_{S_2}^{\ell} = (\bar{c}_R b_{L} )(\bar{\ell}_R \nu_{\ell R }) \,, \\
	\mathcal{O}_{T}^{\ell} = (\bar{c}_R \sigma^{\mu \nu } b_{L} )(\bar{\ell}_R  \sigma_{\mu \nu } \nu_{\ell L })\,. & \nonumber
\end{eqnarray}
In the SM, we only have the $\mathcal{O}_{V_1}^{\ell}$ operator with a Wilson coefficient of 1. Note that the semileptonic and leptonic decay rates are used to extract the CKM elements. From the effective Hamiltonian given in eq.~\eqref{eq:four_ferm_eff_b2c}, we obtain the differential rate for $P \to M \ell \nu_\ell$ (P and M are pseudoscalar mesons) decay corresponding to the quark level transition $d_j\to u_i \ell\nu_{\ell}$ as

{\footnotesize
	\begin{equation}\label{eq:br_semileptonic}
		\frac{d\Gamma (P \to M \ell \nu_\ell)}{dq^2} = \frac{G_F^2 |V_{u_id_j}|^2}{192 \pi^3 m_P^3} q^2 \sqrt{\lambda_M (q^2)} \left(1-\frac{m_\ell^2}{q^2}\right) |1+C_{V_{1}}^{\ell} + C_{V_{2}}^{\ell}|^2 \left \{ \left(1+\frac{m_\ell^2}{2q^2}\right) {H_{V,0}^{s}}^2 + \frac{3}{2}\frac{m_\ell^2}{q^2} {H_{V,t}^s}^2 \right \}.
\end{equation}}
\noindent All the parameters and helicity amplitudes are defined in \cite{Kolay:2024wns, Biswas:2021pic}. 
Also, the branching fraction for $P\to\ell\nu_{\ell}$ corresponding to the same Hamiltonian is:
\begin{equation}\label{eq:brlep}
	\begin{split}
		\mathcal{B}(P\to\ell\nu_{\ell}) = & \frac{\tau_P}{8\pi}\, m_P \, m_\ell^2\,  f_P^2\,  G_F^2 \left(1-\frac{m_\ell^2}{m_P^2}\right)^2 \left|V_{u_i d_j}(1+C_{V_{1}}^{\ell}-C_{V_{2}}^{\ell}) \right|^2.
	\end{split}
\end{equation}
\begin{figure}[t]
	\centering 
    \subfloat[]{\label{fig:FCCC_a}
	\begin{tikzpicture}
	\begin{feynman}
	\vertex (a1){\( c,u\)};
	\vertex [below right=1.cm of a1](a2);
	\vertex [square dot,blue,large,below right=1.cm of a2](a3){};
	\vertex [square dot,blue,below left=1.cm of a3](a4);
	\vertex [below left=0.75cm of a4](a5){\( b\)};
	\vertex [right=1.cm of a3](a6);
	\vertex [above right=1.5cm of a6](a7){\(\ell\)};
	\vertex [below right=1.5cm of a6](a8){\(\nu\)};
	
	\diagram* {
		(a3) -- [fermion,, arrow size=1.2pt,edge label'={\( b\)}] (a2) -- [fermion, arrow size=1pt] (a1),
		(a5) --[fermion, arrow size=1pt] (a4) -- [fermion,, arrow size=1pt,edge label'={\( t\)}] (a3),
		(a2) --[boson, edge label'={\(W\)}] (a4),
		(a3) --[boson,edge label'={\(W\)}](a6),
		(a8) --[fermion, arrow size=1pt](a6) --[fermion, arrow size=1pt](a7),
		
	};
	\end{feynman}
	\end{tikzpicture} }
          \subfloat[]{\label{fig:FCCC_b}
	\begin{tikzpicture}
	\begin{feynman}
	\vertex (a1){\( c,u\)};
	\vertex [below right=1.cm of a1](a2);
	\vertex [square dot,blue,large,below right=1.cm of a2](a3);
	\vertex [square dot,blue,below left=0.9cm of a3](a4){};
	\vertex [below left=1.cm of a4](a5){\( b\)};
	\vertex [right=1.cm of a3](a6);
	\vertex [above right=1.5cm of a6](a7){\(\ell\)};
	\vertex [below right=1.5cm of a6](a8){\(\nu\)};
	
	\diagram* {
		(a3) -- [fermion,, arrow size=1.2pt,edge label'={\( b/s/d\)}] (a2) -- [fermion, arrow size=1pt] (a1),
		(a5) --[fermion, arrow size=1pt] (a4) -- [fermion,, arrow size=1pt,edge label'={\( t\)}] (a3),
		(a2) --[boson, edge label'={\(W\)}] (a4),
		(a3) --[boson,edge label'={\(W\)}](a6),
		(a8) --[fermion, arrow size=1pt](a6) --[fermion, arrow size=1pt](a7),
		
	};
	\end{feynman}
	\end{tikzpicture} }\\
 \subfloat[]{\label{fig:FCCC_c}
	\begin{tikzpicture}
	\begin{feynman}
	\vertex (a1){\( b\)};
	\vertex [square dot,blue,above right=1.6 cm of a1](a2);
	\vertex [square dot, blue, above left=0.5cm of a2](a3){};
	\vertex [above left=0.8cm of a3](a4);
	\vertex [above left=0.5cm of a4](a5){\( c,u\)};
	\vertex [right=1.cm of a2](a6);
	\vertex [above right=1.5cm of a6](a7){\(\ell\)};
	\vertex [below right=1.5cm of a6](a8){\(\nu\)};

	\diagram* {
		(a1) --[fermion, arrow size=1pt] (a2),
		(a2) --[fermion, , arrow size=1pt,edge label={\(t\)}] (a3) --[fermion, , arrow size=1pt,edge label={\(b \)}] (a4) --[fermion, arrow size=1pt] (a5),
		(a4) --[boson, half left, looseness=2, edge label = \(W\)] (a3),
		(a2) --[boson, edge label=\( W\)] (a6),
		(a8) --[fermion, arrow size=1pt](a6) --[fermion, arrow size=1pt](a7),
	};
	\end{feynman}
	\end{tikzpicture}}  
 \subfloat[]{\label{fig:FCCC_d}
	\begin{tikzpicture}
	\begin{feynman}
	\vertex (a1){\( b\)};
	\vertex [square dot,blue,above right=1.6 cm of a1](a2){};
	\vertex [square dot, blue, above left=0.5cm of a2](a3);
	\vertex [above left=0.8cm of a3](a4);
	\vertex [above left=0.5cm of a4](a5){\( c,u\)};
	\vertex [right=1.cm of a2](a6);
	\vertex [above right=1.5cm of a6](a7){\(\ell\)};
	\vertex [below right=1.5cm of a6](a8){\(\nu\)};

	\diagram* {
		(a1) --[fermion, arrow size=1pt] (a2),
		(a2) --[fermion, , arrow size=1pt,edge label={\(t\)}] (a3) --[fermion, , arrow size=1pt,edge label={\(b/s/d \)}] (a4) --[fermion, arrow size=1pt] (a5),
		(a4) --[boson, half left, looseness=2, edge label = \(W\)] (a3),
		(a2) --[boson, edge label=\( W\)] (a6),
		(a8) --[fermion, arrow size=1pt](a6) --[fermion, arrow size=1pt](a7),
	};
	\end{feynman}
	\end{tikzpicture}}  
\subfloat[]{\label{fig:FCCC_e}
\begin{tikzpicture}
	\begin{feynman}
	\vertex (a1){\( b\)};
	\vertex [square dot, blue, above right=0.8cm of a1](a2){};
	\vertex [ above right=0.7cm of a2](a3);
	\vertex [ above right=0.5cm of a3](a4);
	\vertex [above left=1.5cm of a4](a5){\( c,u\)};
	\vertex [right=1.0cm of a4](a6);
	\vertex [above right=1.5cm of a6](a7){\(\ell\)};
	\vertex [below right=1.5cm of a6](a8){\(\nu\)};

	\diagram* {
		(a1) --[fermion, arrow size=1pt] (a2),
		(a2) --[fermion, arrow size=1pt,edge label'={\(t\)}] (a3) --[fermion, arrow size=1pt,edge label={\(s/d \)}] (a4) --[fermion, arrow size=1pt] (a5),
		(a2) --[boson, half left, looseness=2, edge label = \(W\)] (a3),
		(a4) --[boson, edge label=\( W\)] (a6),
		(a8) --[fermion, arrow size=1pt](a6) --[fermion, arrow size=1pt](a7),
	};
	\end{feynman}
	\end{tikzpicture}}\\
\subfloat[]{\label{fig:FCCC_f}
\begin{tikzpicture}
	\begin{feynman}
	\vertex (a1){\( s,d\)};
	\vertex [below right=1.cm of a1](a2);
	\vertex [square dot,blue,large,below right=1.cm of a2](a3){};
	\vertex [below left=1.cm of a3](a4);
	\vertex [below left=0.6 cm of a4](a5){\( c,u\)};
	\vertex [right=1.cm of a3](a6);
	\vertex [above right=1.5cm of a6](a7){\(\ell\)};
	\vertex [below right=1.5cm of a6](a8){\(\nu\)};
	
	\diagram* {
		(a3) -- [fermion,, arrow size=1pt,edge label'={\( t\)}] (a2) -- [fermion, arrow size=1pt] (a1),
		(a5) --[fermion, arrow size=1pt] (a4) -- [fermion,, arrow size=1pt,edge label'={\( b\)}] (a3),
		(a2) --[boson, edge label'={\(W\)}] (a4),
		(a3) --[boson,edge label={\(W\)}](a6),
		(a8) --[fermion, arrow size=1pt](a6) --[fermion, arrow size=1pt](a7),
	};
	\end{feynman}
	\end{tikzpicture}}
	\subfloat[]{\label{fig:FCCC_g}
    \begin{tikzpicture}
	\begin{feynman}
	\vertex (a1){\( c,u\)};
	\vertex [square dot,blue,above right=1.6 cm of a1](a2);
	\vertex [square dot, blue, above left=0.5cm of a2](a3){};
	\vertex [above left=0.7cm of a3](a4);
	\vertex [above left=0.5cm of a4](a5){\( s, d\)};
	\vertex [right=1.0cm of a2](a6);
	\vertex [above right=1.5cm of a6](a7){\(\ell\)};
	\vertex [below right=1.5cm of a6](a8){\(\nu\)};

	\diagram* {
		(a1) --[fermion, arrow size=1pt] (a2),
		(a2) --[fermion, arrow size=1pt,edge label={\(b\)}] (a3) --[fermion, arrow size=1pt,edge label={\(t \)}] (a4) --[fermion, arrow size=1pt] (a5),
		(a4) --[boson, half left, looseness=2, edge label = \(W\)] (a3),
		(a2) --[boson, edge label=\( W\)] (a6),
		(a8) --[fermion, arrow size=1pt](a6) --[fermion, arrow size=1pt](a7),
	};
	\end{feynman}
	\end{tikzpicture}}
\subfloat[]{\label{fig:FCCC_h}
\begin{tikzpicture}
	\begin{feynman}
	\vertex (a1){\( c,u\)};
	\vertex [above right=0.8cm of a1](a2);
	\vertex [square dot, blue, above right=0.7cm of a2](a3){};
	\vertex [above right=0.5cm of a3](a4);
	\vertex [above left=1.5cm of a4](a5){\( s, d\)};
	\vertex [right=1.0cm of a4](a6);
	\vertex [above right=1.5cm of a6](a7){\(\ell\)};
	\vertex [below right=1.5cm of a6](a8){\(\nu\)};

	\diagram* {
		(a1) --[fermion, arrow size=1pt] (a2),
		(a2) --[fermion, arrow size=1pt,edge label'={\(b\)}] (a3) --[fermion, arrow size=1pt,edge label={\(t \)}] (a4) --[fermion, arrow size=1pt] (a5),
		(a2) --[boson, half left, looseness=2, edge label = \(W\)] (a3),
		(a4) --[boson, edge label=\( W\)] (a6),
		(a8) --[fermion, arrow size=1pt](a6) --[fermion, arrow size=1pt](a7),
	};
	\end{feynman}
	\end{tikzpicture}}
	\caption{Feynman diagrams contributing to the processes $ d_{i} (u_{i}) \to u_{j} (d_{j}) \ell \nu $.  The blue square dots denote the $ Wtb  $ effective vertex.}
	\label{fig:FCCC_diagrams}
	\end{figure}
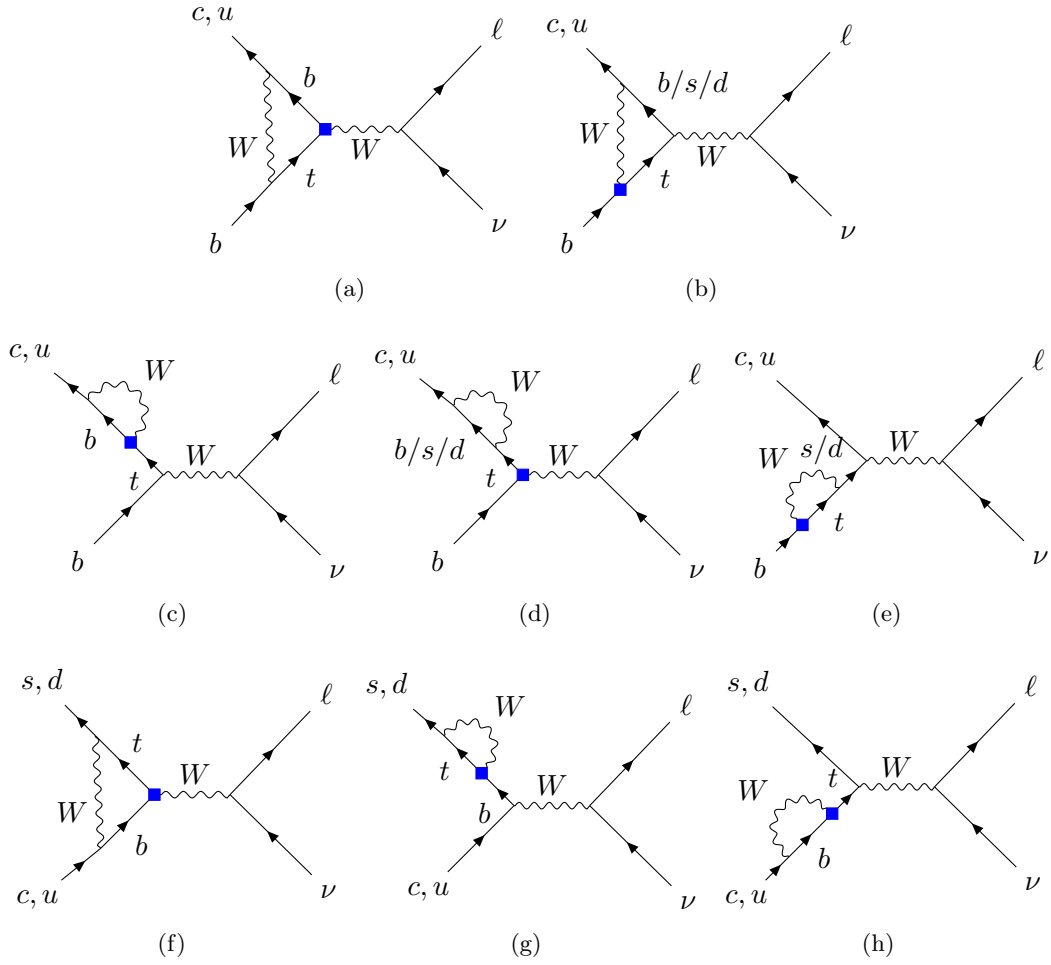
    \begin{figure}
        \centering
        \subfloat[]{ 
        \begin{tikzpicture}
	\begin{feynman}
		\vertex (a1){\(b\)};
		\vertex [square dot,blue,right=1.cm of a1](a2)       {};
		\vertex [square dot,blue,right=1.5cm of a2](a3)       ;
		\vertex [right=1.cm of a3](a4){\(b\)};
				
		\diagram* {
        (a1) --[fermion, arrow size=1pt](a2) --[fermion, , arrow size=1pt,edge label'={\(t\)}](a3) --[fermion, arrow size=1pt](a4),
        (a2) --[boson, half left, looseness=2, edge label={\(W\)}](a3),
	};				
	\end{feynman}
	\end{tikzpicture}}
    \subfloat[]{ \begin{tikzpicture}
	\begin{feynman}
		\vertex (a1){\(b\)};
		\vertex [square dot,blue,right=1.cm of a1](a2)       ;
		\vertex [square dot,blue,right=1.5cm of a2](a3)       {};
		\vertex [right=1.cm of a3](a4){\(b\)};
				
		\diagram* {
        (a1) --[fermion, arrow size=1pt](a2) --[fermion, , arrow size=1pt,edge label'={\(t\)}](a3) --[fermion, arrow size=1pt](a4),
        (a2) --[boson, half left, looseness=2, edge label={\(W\)}](a3),
	};				
	\end{feynman}
	\end{tikzpicture}}
        \caption{Conter-term diagrams corresponding to the process $b \to u_{j} \ell \nu $. }
        \label{fig:FCCC_conter-term}
    \end{figure}
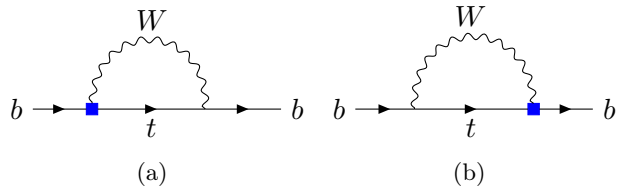

In our EFT framework, we get contributions in FCCC processes at one-loop level in the form of corrections to the SM tree-level diagram. Here we will mainly focus on semileptonic and leptonic decays of the mesons. The effective vertex of $ Wtb $ will contribute to FCCC processes like $d_{i} (u_{i}) \to u_{j} (d_{j}) \ell \nu $ via one-loop diagram. We will have contributions to vertex correction as well as self-energy correction diagrams. The relevant Feynman diagrams are shown in fig.~\ref{fig:FCCC_diagrams}. The first row in fig.~\ref{fig:FCCC_diagrams} denotes the FCCC processes involving $b$-quark. The diagram in fig.~\ref{fig:FCCC_conter-term} is the counter-term diagram for the processes involving $b$-quark. The processes in the 2nd row denote the charge current processes involving quarks other than $b$. 

Fig.~\ref{fig:FCCC_diagrams} shows how the introduction of the effective $Wtb$ vertex modifies only the quark vertex, leaving the lepton vertex unaffected.
Following a matching condition, the diagrams given in fig.~\ref{fig:FCCC_diagrams} will contribute only to to the Wilson coefficients $ C^{\ell}_{V_1} $ and $ C^{\ell}_{V_2} $ in eq.~\eqref{eq:four_ferm_eff_b2c}. Since there are no NP effects in the lepton sector, the scalar and tensor operators, i.e., $ C^{\ell}_{S_1}, C^{\ell}_{S_2}, C^{\ell}_{T} $, do not contribute.
 
Considering all the contributions, We can write
\begin{subequations}\label{eq:FCCC_total_loop}
    \begin{eqnarray}
        C^{\ell}_{V_{1(2)}} = C_{V_{1(2)}}^{\rm vertex} + C_{V_{1(2)}}^{\rm self} + C_{V_{1(2)}}^{\rm WF}\,.  
    \end{eqnarray}
\end{subequations}
The last term in eq.~\eqref{eq:FCCC_total_loop}, corresponding to the wave-function renormalisation, arises only for the processes related to $b$-quark. The relevant loop contributions are discussed in the appendix \ref{appndx:FCCC}. We have also matched our loop contributions with the LEFT Hamiltonian at the EW scale, and to incorporate the low-energy FCCC constraints, we need to further run down the  Wilson coefficients to the $\mu_{b}$ scale following the RGE equation as given below:
\begin{align}\label{eq:FCCC_EW_to_mb}
	\begin{pmatrix}
		C_{V_1}\\
		C_{V_2}\\
		C_{S_1}\\
		C_{S_2}\\
		C_T
	\end{pmatrix}_{{\mu_b}}&=\begin{pmatrix}
		1.00716 &0 &0 & 0& 0\\
		0 & 1.00358 &0 &0 &0\\
		0 & 0 & 1.35728 & 0 &0 \\
		0 & 0 & 0 & 1.35728 & -0.01629\\
		0 & 0 & 0 & -0.0003 & 0.90964
	\end{pmatrix} \begin{pmatrix}
		C_{V_1}\\
		C_{V_2}\\
		C_{S_1}\\
		C_{S_2}\\
		C_T
	\end{pmatrix}_{\mu_{\rm{EW}}}\,,
\end{align}

\noindent In our EFT framework, the NP contributions to the decays $P \to M \ell \nu_\ell$ and $P \to \ell \nu_\ell$ affect the vertex factor, as seen from the rate expressions of eq.~\eqref{eq:brlep} and \eqref{eq:br_semileptonic}. Consequently, these contributions will modify the overall rate normalisation but leave the shape of the $q^2$ distributions unchanged. When extracting the CKM elements from these decay rates, the result will be $|V'_{u_i d_j}| = |V_{u_i d_j}(1 + C_{V_{1}}^{\ell} \pm C_{V_{2}}^{\ell})|$, rather than simply $|V_{u_i d_j}|$. Thus, the CKM elements $|V'_{u_i d_j}|$, which are derived from purely leptonic decays or $P \to M \ell \nu_\ell$ processes, can be used to constrain both the NP parameters and the Wolfenstein parameters $A$, $\lambda$, $\rho$, and $\eta$, which are needed to parametrise $|V_{u_i d_j}|$. 
The list of the inputs and observables used to perform the CKM fit can be found in our previous work \cite{Kolay:2024wns, Biswas:2021pic}. Along with this list, we have also used the updated binned data on the decay width of the $B\to D^{()}\ell\nu$ processes and $R_{D}^{(*)}$ in our analysis as given by Belle-II \cite{Belle-II:2025yjp}.

\section{Impact on Electroweak Precision \& Other Observables} \label{sec:EWPOs}
Due to the high precision of electroweak observables measured at the $W$ and $Z$ poles, it becomes possible to place stringent constraints on potential NP models. In our framework, we get contributions to the $W$ boson self-energy correction as well as correction to the $Z \to b \bar{b}$ vertex. In the following section, we will discuss how the anomalous couplings of the $Wtb$ vertex affect the $W$ mass and $Z$-pole observables. Also, the study of the decay of Higgs boson to bottom-quark pair is also an important channel to constrain NP as the decay width is consistent with SM in $1\sigma$ error range. We will discuss the NP effect in this decay channel in the following section. 

\subsection{Oblique parameters }
Besides low-energy FCNC and FCCC observables, we will also get contributions to the oblique parameters. The S, T, and U parameters are oblique corrections that quantify the effects of new physics on electroweak gauge boson self-energies. The S parameter measures deviations in neutral current interactions, the T parameter reflects custodial symmetry violation and mass splitting between the W and Z bosons, and the U parameter captures additional modifications to the W boson propagator beyond S and T, though it is usually small.
 These parameters provide a model-independent framework to constrain physics beyond the SM.
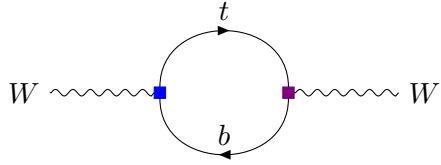
\begin{figure}[htb!]
    \centering
    \begin{tikzpicture}
        \begin{feynman}
            \vertex (a1){\(W\)};
            \vertex[square dot,blue,right=1.8cm of a1](a2){};
            \vertex[square dot,violet,right=1.7cm of a2](a3){};
            \vertex[right=1.8cm of a3](a4){\(W\)};
            
            \diagram*{
                (a1) --[boson](a2) --[fermion, half left, looseness=1.5,arrow size=1pt,edge label={\(t\)}](a3) --[boson](a4),
                (a3) --[fermion, half left, looseness=1.5,arrow size=1.pt,edge label'={\(b\)}](a2)
            }; 
        \end{feynman}
    \end{tikzpicture}
     \caption{$W$ boson mass correction in our framework, different coloured squares represent distinct effective $Wtb$ vertices from SMEFT operators. Each diagram includes one effective vertex at a time, with all other interactions kept as in the SM.
 }
    \label{fig:W_boson_feynman}
\end{figure}

 For our case, the anomalous couplings of the $Wtb$ vertex will modify the self-energy of the W boson. The most general expression for gauge boson self-energy can be written as: 
\begin{eqnarray}\label{eq:self_energy_W}
    \Sigma_{V}(q^2) = \left( g^{\mu \nu} - \frac{q^{\mu}q^{\nu}}{q^2}\right) \Sigma_{V,T}(q^2) + \frac{q^{\mu } q^{\nu}}{q^2} \Sigma_{V,L}(q^2)\,.
\end{eqnarray}

One related observable is the $\rho$ parameter, which in SM can be written as: 
\begin{equation}
    \rho = \frac{G_{NC}}{G_{CC}} = \frac{M_W^2}{c_{W}^2 M_Z^2} = 1\,.
\end{equation}
At SM, tree-level, the above expression is valid. Any deviation from unity denotes the SM loop contribution, possibly accompanied by the presence of BSM. A detailed discussion on this can be found in \cite{Kolay:2024wns}. 
The contribution at one-loop level in $\Delta \rho$, for our case can be expressed as: 
 \begin{eqnarray}
     (\Delta \rho)^{(1)} = -\frac{\Sigma_{W,T}(0)}{M_W^2}\,.
 \end{eqnarray}
The effect of non-standard contributions to the EWPO is also parametrised in terms of the [S, T, U] parameters.  The mass of the $W$ boson  $M_W$, in terms of a oblique parameter $\Delta r $ can be written as: 
\begin{equation}\label{eq:delr_MW}
    M_W^2 \left( 1- \frac{M_W^2}{M_Z^2}\right) = \frac{\pi \alpha_{em}}{\sqrt{2} G_{F}} \frac{1}{1-\Delta r }\,.
\end{equation}
\noindent The NP contribution in $\Delta r $ can be expressed in terms of $[S,T,U]$ as: 
\begin{equation}\label{eq:delr_oblique}
    \Delta r = \frac{\alpha_{em}}{s_{W}^2} \left( - \frac{1}{2} \Delta S + c_{W}^2 \Delta T + \frac{c_{W}^2 - s_{W}^2}{4 s_{W}^2} \Delta U \right)\,.
\end{equation}
We use eq.~\eqref{eq:delr_MW} to obtain the NP value allowed in the observable $\Delta r $, which we denote as: $\delta(\Delta r)$. Eq.~\eqref{eq:delr_oblique} is used to calculate the expression of the theory for the same. The expressions for all oblique parameters, in detail, along with their updated experimental results, can be found in \cite{Kolay:2024wns, Kolay:2025jip}.  
In our case, we get $W$ boson mass correction via the diagram given in fig.~\ref{fig:W_boson_feynman}. Here, we can also see, similar to the previous low-energy processes, the effective $Wtb$ vertex comes twice in the loop. Similar to the previous case, we neglect the higher-order correction and consider one effective vertex at a time (while taking the other effective vertex as simply $V_{tb} P_L $, as in SM) and add the contributions of two diagrams. To calculate the $W$ boson mass correction, we only need the transverse component of the self-energy defined in eq.~\eqref{eq:self_energy_W}, namely $\Sigma_{W, T}$, the expression is given in eq.~\eqref{eq:W-boson_transverse} of appendix \ref{appndx:EWPOs}. We have also considered the contributions from the $W$-boson ratio observables $R_{\ell_i/\ell_j}$, with $\ell_{i,j}=e,\mu,\tau$, which arise from the same Feynman diagram in fig.~\ref{fig:W_boson_feynman}.

\subsection{ \texorpdfstring{$Z$-pole Observables}{Z pole}}
\begin{figure}[htb!]
    \centering
    \subfloat[]{\begin{tikzpicture}
        \begin{feynman}
            \vertex[square dot](a); 
            \vertex[left=1.3cm of a](b){$Z$}; 
            \vertex[square dot,blue,above right=1.cm of a](c){$ $};
            \vertex[above right=1.cm of c](d){$b$};
            \vertex[square dot, violet,below right=1.cm of a](e){$ $};
            \vertex[below right=1.cm of e](f){$b$};
            
            \diagram*{
                (b) --[boson](a),
                (f) --[fermion,arrow size=1pt](e) --[fermion,arrow size=1pt,edge label=\(t\)](a) --[fermion, arrow size=1pt,edge label=\(t\)](c) --[fermion,arrow size=1pt](d),
                (c) --[boson, edge label=W](e)
            }; 
        \end{feynman}
    \end{tikzpicture}}~~
               \subfloat[]{\begin{tikzpicture}
			\begin{feynman}
			\vertex (a1){$Z$};
			\vertex [right=1.3cm of a1](a2);
			\vertex [square dot,blue,above right=1.cm of a2](a3){};
			\vertex [square dot,violet,below right=1.cm of a2](a4){};
			\vertex [ below right=1.cm of a4](a5){\(b\)};
			\vertex [ above right=1.cm of a3](a6){\(b\)};
			
			\diagram* {
				(a1) --[boson](a2),
				(a5) --[fermion, arrow size=1pt](a4) --[fermion, arrow size=1pt,edge label'={\(t\)}](a3) --[fermion, arrow size=1pt](a6),
                (a2) --[boson, edge label=\(W\)](a3), (a2) --[boson, edge label'=\(W\)](a4)
			};	 	
			\end{feynman}
			\end{tikzpicture}}~~
            \subfloat[]{\begin{tikzpicture}
			\begin{feynman}
			\vertex (a1){$Z$};
			\vertex [right=1.3cm of a1](a2);
			\vertex [square dot,violet,above right=0.5cm of a2](a3){};
			\vertex [square dot,blue,above right=0.7cm of a3](a4){};
			\vertex [above right=0.8cm of a4](a5){\(b\)};
			\vertex [below right=1.3cm of a2](a6){\(b\)};
			
			\diagram* {
				(a1) --[boson](a2),
				(a6) --[fermion, arrow size=1pt](a2) --[fermion, arrow size=1pt,edge label'={\(b\)}](a3) --[fermion, arrow size=1pt, edge label'={\(t\)}](a4) --[fermion, arrow size=1pt](a5),
				(a3) --[boson, half left, looseness=2, edge label={\(W\)}](a4),
			};	 	
			\end{feynman}
			\end{tikzpicture}}
     \caption{Feynman diagrams for $Zb\bar{b}$ vertex correction by the $Wtb$ effective vertex. Each diagram includes one colored effective vertex at a time,
while keeping others as SM types.}
    \label{fig:Z_pole_feynman}
\end{figure}
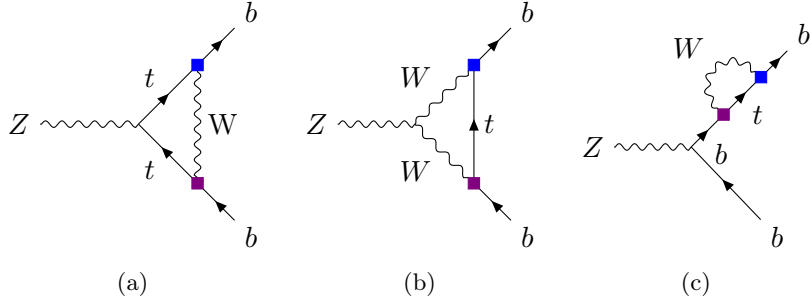
The electroweak $Z$-pole observables are also impacted by the effective anomalous couplings of the $Wtb$ vertex. The corresponding Feynman diagrams are shown in fig.~\ref{fig:Z_pole_feynman}. The branching ratio of the decay $Z \to \bar{f} f $ can be written as follows: 
\begin{align}
\label{eq:decaywidth_Zpole}
\Gamma_{\rm tot}(Z \to f \bar{f}) = \frac{N_c^b}{48} \frac{\alpha}{s_W^2 c_W^2} M_Z \sqrt{1-\mu_{f}^2}& \biggl( |g_{Af}^{\rm tot}|^2 (1-\mu_{f}^2) + |g_{Vf}^{\rm tot}|^2(1+\frac{\mu_{f}^2}{2})\biggr)(1+\delta_{f}^0)(1+\delta_{b}) \nonumber \\ &  (1+\delta_{\rm QCD})(1+\delta_{\rm QED}) (1+\delta^{f}_{\mu}) \,,
\end{align}
with $ \mu_{f}^2 = \frac{4 m_{f}^2}{M_Z^2}$, which is only significant for the decay to $b \bar{b}$. Other terms like $ \delta_{f}^0$, $ \delta_{b}$,  $\delta_{QCD}$,  $\delta_{QED}$,  $\delta_{f}^{\mu} $ are the various corrections to the leading order decay rate, for a detailed discussion see the refs.~\cite{Soni:2010xh,Bernabeu:1990ws} and the references therein. The variables $g_{Af}$ and $g_{Vf}$ are the effective axial and vector couplings of fermion pairs with Z-boson, respectively, which can be understood from the following:
\begin{equation}
    \mathcal{L}_{Zb\bar{b}} = -\frac{g}{2\cos \theta_{w}}\bar{b} \gamma_{\mu} (g_{Vb}^{tot} + g_{Ab}^{tot} \gamma_{5})b Z^{\mu}\,.
\end{equation}
The vertices can be written in terms of the NP correction as: 
\begin{eqnarray}
    g_{Vb}^{\rm tot} & = &  g_{Vb}^{\rm SM} + \delta g_{Vb} \,,\nonumber \\
    g_{Ab}^{\rm tot} & = &  g_{Ab}^{\rm SM} + \delta g_{Ab} \,.
\end{eqnarray}
The normalized NP contribution to the decay width can be written as:
\begin{eqnarray}
    \frac{\Gamma_{\rm tot}- \Gamma_{\rm SM}}{\Gamma_{\rm SM} }  = \delta_{b}^{\rm NP} = 2 \frac{  |g_{Ab}^{\rm tot}| \, \, \delta g_{Vb} + |g_{Vb}^{\rm tot}| \, \,  \delta g_{Ab}}{ |g_{Ab}^{\rm tot}|^2 (1 - \mu_{b}^2) + |g_{Vb}^{\rm tot}|^2 (1 + \frac{\mu_{b}^2}{2}) }\,.
\end{eqnarray}
Here, we have neglected the higher order contributions like $\delta g_{Ab}^2 \,, \delta g_{Vb}^2 \,, \delta g_{Ab} \mu_{b}^2$. The expressions for the NP-modified couplings $\delta g_{Vb}$ and $\delta g_{Ab}$ are provided in eq.~\eqref{eq:loop_Zpole} of appendix~\ref{appndx:Z_Pole}.
Since anomalous couplings of $Wtb$ vertex can only modify the vertex of the process $ Z \to \bar{b} b $, the impacted observables are the ratio observable $R_{b}$ and asymmetric observables $A_{FB}$ and $A_{b}$. As $A_{FB} ( = \frac{3}{4} A_{b} A_{e} ) $ and $A_{b}$ are related, we can only consider one independent observable between the two, since $A_{e}$ remains unmodified in this case. We will take $A_{b}$ into account. $R_{b}$ is defined as following: 
\begin{eqnarray}
    R_{b} = \frac{\Gamma_{b}}{ \Gamma_{\rm had}} = R_{b}^{\rm SM} \frac{1 +\delta_{b}^{\rm NP}}{ 1 + R_{b}^{\rm SM} \delta_{b}^{\rm NP}}\,.
\end{eqnarray}
The observable $A_{b}$ is calculated from the process $e^{+} e^{-} \to \bar{f} f $ scattering process via mediation of $Z$ boson. The expression can be written as:
\begin{eqnarray}
    A_{f} = 2 \frac{g_{Af}^{ \rm tot} g_{Vf}^{\rm tot}}{(1-\mu_{f}^2) |g_{Af}^{\rm tot}|^2 + (1+\frac{\mu_{f}^2}{2}) |g_{Vf}^{\rm tot} |^2}\,.
\end{eqnarray}
\noindent The details can be found in \cite{Kolay:2024wns}.  In our analysis we have considered the observables $R_{b}$ and $A_{b}$, whose SM values are given by \cite{ParticleDataGroup:2024cfk} : 
\begin{eqnarray}
    R_{b}^{\rm SM} &=& 0.9347 \,, \\
    A_{b}^{\rm SM} &=& 0.21582 \pm  0.00002\,.
\end{eqnarray}
The updated experimental values of $W$ and $Z$ pole observables can be found from \cite{Kolay:2024wns, Kolay:2025jip}. 


\subsection{Trilinear vector boson coupling correction}
The general couplings of two charged vector bosons with a neutral vector boson $WWV (V=\gamma, Z)$ are  known as trilinear gauge couplings (TGC) and the effective Lagrangian can be written as \cite{Argyres:1992vv,Hagiwara:1986vm}:

\begin{align}\label{eq:TGC_couplings}
    \mathcal{L}_{WWV} &=i g_{WWV}\Bigg( g_1^V (W^+_{\mu \nu}W^{-\mu} W^{+\nu}-W^+_{\mu}W^-_{\nu}W^{\mu\nu})+ \kappa_V W^+_{\mu} W^-_{\nu} V^{\mu \nu} \nonumber\\
    &+\frac{\lambda_V}{M_W^2} W^+_{\lambda \mu}W^{-\mu}_{\nu} V^{\nu \lambda} + i g_4^V W_{\mu}^+ W^-_{\nu}(\partial^{\mu} V^{\nu}+ \partial^{\nu} V^{\mu}) -i g_5^V \epsilon^{\mu \nu \rho \sigma} (W_{\mu}^+ \stackrel{\leftrightarrow}{\partial}_{\rho} W_{\nu}^-) V_{\rho} \nonumber\\
    &+ \tilde{\kappa}_V W_{\mu}^+ W^-_{\nu} \tilde{V}^{\mu \nu} +\frac{\tilde{\lambda}_V}{M_W^2} W^+_{\lambda \mu} W^{-\mu} \tilde{V}^{\nu \lambda} \Bigg)\,.
\end{align}

Here $W_{\mu \nu}^{\pm}=(\partial_{\mu} W^{\pm}_{\nu}-\partial_{\nu} W^{\pm}_{\mu})$, $V_{\mu \nu}=(\partial_{\mu} V_{\nu}-\partial_{\nu} V_{\mu})$ , $\tilde{V}_{\mu \nu}=\frac{1}{2} \epsilon_{\mu \nu \rho \sigma} V^{\rho \sigma}$ and $(A \stackrel{\leftrightarrow}{\partial}_{\mu} B)=A (\partial_{\mu} B) -(\partial_{\mu} A) B$. In the momentum space, the above equation can be expressed as (fig.~\ref{fig:TGC_feynman_SM})

\begin{equation} \label{eq:TGC_momentum_space}
\mathcal{L}_{WWV}	= i g_{WWV} \Gamma^{\alpha\beta \mu}_V(P, p_1, p_2)\,.
\end{equation}
with 
\begin{align}\label{eq:TGC_SM}
	\Gamma^{\alpha \beta \mu}_V (p_1, p_2, P)&= \Bigg( f_1^V (p_1-p_2)^{\mu} g^{\alpha \beta} -\frac{f_2^V}{M_W^2} (p_1 -p_2)^{\mu} P^{\alpha} P^{\beta}  + f_3^V (P^{\alpha} g^{\mu \beta}
	-P^{\beta} g^{\mu \alpha}) \nonumber\\
	&+ i f_4^V (P^{\alpha} g^{\mu \beta}+P^{\beta} g^{\mu \alpha})
	+i f_5^V \epsilon^{\mu \alpha \beta \rho} (p_1 -p_2)_{\rho}\nonumber\\
	&- f_6^V \epsilon^{\mu \alpha \beta \rho} P_{\rho} -\frac{f_7^V}{M_W^2} (p_1 -p_2)^{\mu} \epsilon^{\alpha \beta \rho \sigma} P_{\rho} (p_1-p_2)_{\sigma} \Bigg) \,.
\end{align}

\begin{figure}[t]
    \centering
    \begin{minipage}[]{0.3\textwidth}
        \begin{tikzpicture}
            \begin{feynman}
                \vertex[blob](a) {};
                \vertex[above right=2.5cm of a](b) {\(W^-_{\alpha }\)};
                \vertex[below right=2.5cm of a](c) {\(W^+_{\beta}\)};
                \vertex[left=2cm of a](d) {\(V_{\mu}\)};
                
                \diagram*{
                    (d) -- [boson, momentum=\(P\)] (a),
                    (a) -- [boson, momentum=\(p_1\)] (b),
                    (a) -- [boson, momentum'=\(p_2\)] (c),
                };
            \end{feynman}
        \end{tikzpicture}
    \end{minipage}
    \hspace{0.01cm}
    \begin{minipage}[]
        {0.65\textwidth}
        \centering
        \begin{equation} 
             = i g_{WWV} \Gamma^{\alpha\beta \mu}_V(P, p_1, p_2)\,.
        \end{equation}
        \vspace{0.5cm}
    \end{minipage}
\caption{Triple gauge boson vertex.}
    \label{fig:TGC_feynman_SM}
\end{figure}

\begin{figure}[t]
	\centering
	\subfloat[]{
		\begin{tikzpicture}
			\begin{feynman}
				\vertex[](a); 
				\vertex[left=1cm of a](b){$V_{\mu}$}; 
				\vertex[square dot,blue, above right=1.cm of a](c){$ $};
				\vertex[above right=1.5 cm of c](d){$W^+_{\alpha}$};
				\vertex[ square dot, violet,below right=1.cm of a](e){$ $};
				\vertex[below right=1.5 cm of e](f){$W^-_{\beta}$};
				
				\diagram*{
					(b) --[boson,momentum=\(P\)](a),
					(c) --[boson, momentum=\(p_1\)](d),
					(e) --[boson,momentum'=\(p_2\)](f),
					(e) --[fermion, arrow size=1pt,edge label=\(t\)](a) --[fermion, arrow size=1pt, edge label=\(t\)](c) --[fermion,  arrow size=1pt, edge label=\(b\)](e)
				}; 
			\end{feynman}
		\end{tikzpicture}
	}
	\subfloat[]{
		\begin{tikzpicture}
			\begin{feynman}
				\vertex[](a); 
				\vertex[left=1. cm of a](b){$V_{\mu}$}; 
				\vertex[square dot,blue, above right=1.cm of a](c){$ $};
				\vertex[above right=1.5 cm of c](d){$W^-_{\alpha}$};
				\vertex[ square dot,violet, below right=1.cm of a](e){$ $};
				\vertex[below right=1.5cm of e](f){$W^+_{\beta}$};
				
				\diagram*{
					(b) --[boson,momentum=\(P\)](a),
					(c) --[boson,momentum=\(p_1\)](d),
					(e) --[boson, momentum'=\(p_2\)](f),
					(e) --[fermion, arrow size=1pt,edge label=\(b\)](a) --[fermion, arrow size=1pt, edge label=\(b\)](c) --[fermion, arrow size=1pt, edge label=\(t\)](e)
				}; 
			\end{feynman}
		\end{tikzpicture}
	}
	\caption{Feynman diagrams contributing to the trilinear gauge coupling $WW\gamma$ and $WWZ$ are modified by the effective $Wtb$ vertex (blue and violet dots). Each diagram includes one colored effective vertex at a time
		while keeping others as SM types.}
	\label{fig:TGC_NP_Feynman}
\end{figure}
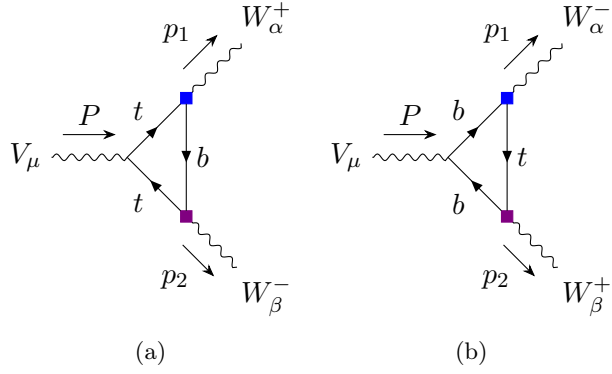

Here, all the form factors $f_i^V$ are dimensionless functions of $P^2$. 
One can draw a connection between TGCs of eq.~\eqref{eq:TGC_couplings} and the form-factors of eq.~\eqref{eq:TGC_momentum_space} as:

\begin{subequations}\label{eq:TGC_coupling_compare}
    \begin{align}
    g_1^V &= f_1^V -\frac{s}{2 M_W^2} f_2^V \,, \\
    \lambda^V &= f_2^V \,,\\
    \kappa^V &= f_3^V -\left(1-\frac{s}{2 M_W^2} \right) f_2^V -f_1^V \,, \\
    g_4^V &= f_4^V \,, \\
    g_5^V &= f_5^V\,, \\
    \tilde{\kappa}^V &= f_6^V -2 f_7^V \,, \\
    \tilde{\lambda}^V &= - 2 f_7^V \,.
\end{align}
\end{subequations}

 In our framework, we get corrections to the TGCs via the diagrams in fig.~\ref{fig:TGC_NP_Feynman}. By calculating the Feynman diagrams, we get beyond the SM contributions to the form factors $f_i^V$ of eq.~\eqref{eq:TGC_SM}, which can further be used to derive the couplings of eq.~\eqref{eq:TGC_couplings} following  the relations of eq.~\eqref{eq:TGC_coupling_compare}. The corresponding SM predictions and experimental results are given in table~\ref{tab:tcg_comparison}.
 The analytical expressions of the various form factors are not included in the main text due to their lengthy and cumbersome nature.

\begin{table}[htb!]
\centering
\rowcolors{1}{cyan!10}{blue!20!green!5}
\renewcommand{\arraystretch}{1.5}
\setlength{\tabcolsep}{15pt}
\begin{tabular}{|c|c|c|}
\hline
    \rowcolor{cyan!20}
        \text{TGC Coupling} & \text{SM Value \cite{ALEPH:2013dgf}} & \text{Experimental Value \cite{ALEPH:2013dgf}}  \\
        \hline
        \hline
        \vspace{0.0cm}
        $g^1_Z$ & $1$ & $0.984^{+0.018}_{-0.020}$   \\
        $\kappa_{\gamma}$ & $1$ & $0.982 \pm 0.042$  \\
        $\lambda_{\gamma}$ & $0$ & $-0.022 \pm 0.019$ \\
        $\kappa_Z$ & $1$ & $0.92 \pm 0.06$  \\
        $\lambda_Z$ & $0$& $-0.09 \pm 0.06$  \\
        $g^5_Z$ & $0$ & $-0.07 \pm 0.09$ \\
        $g^4_Z$ & $0$ & $-0.03 \pm 0.17$ \\
        $\tilde{\kappa}_Z$ & $0$ & $-0.12^{+0.06}_{-0.04}$  \\
        $\tilde{\lambda}_Z$ & $0$ & $-0.09 \pm 0.07$  \\
        \hline
\end{tabular} 
    \caption{The SM and the experimental values of the TGC couplings.}
    \label{tab:tcg_comparison}
\end{table}

\subsection{\texorpdfstring{$H \to b \bar{b}$ Decay}{Higgs decay}}
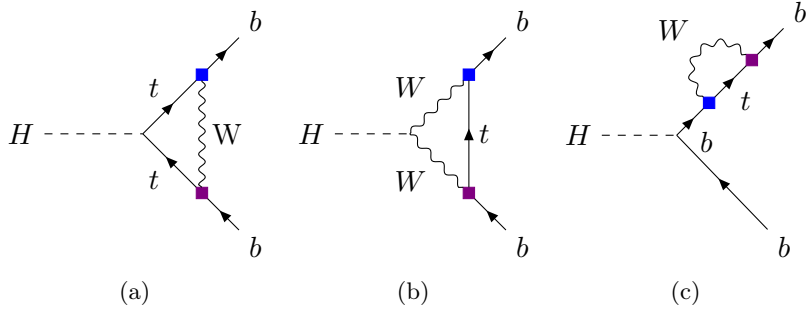
\begin{figure}[htb!]
    \centering
    \subfloat[]{\begin{tikzpicture}
        \begin{feynman}
            \vertex[square dot](a); 
            \vertex[left=1.3cm of a](b){$H$}; 
            \vertex[square dot,blue,above right=1.cm of a](c){$ $};
            \vertex[above right=1.cm of c](d){$b$};
            \vertex[square dot, violet,below right=1.cm of a](e){$ $};
            \vertex[below right=1.cm of e](f){$b$};
            
            \diagram*{
                (b) --[scalar](a),
                (f) --[fermion,arrow size=1pt](e) --[fermion,arrow size=1pt, edge label=\(t\)](a) --[fermion, arrow size=1pt,edge label=\(t\)](c) --[fermion,arrow size=1pt](d),
                (c) --[boson, edge label=W](e)
            }; 
        \end{feynman}
    \end{tikzpicture}} \hspace{0.1cm}
    \subfloat[]{\begin{tikzpicture}
			\begin{feynman}
			\vertex (a1){$H$};
			\vertex [right=1.3cm of a1](a2);
			\vertex [square dot,blue,above right=1.cm of a2](a3){};
			\vertex [square dot,violet,below right=1.cm of a2](a4){};
			\vertex [ below right=1.cm of a4](a5){\(b\)};
			\vertex [ above right=1.cm of a3](a6){\(b\)};
			
			\diagram* {
				(a1) --[scalar](a2),
				(a5) --[fermion, arrow size=1pt](a4) --[fermion, arrow size=1pt,edge label'={\(t\)}](a3) --[fermion, arrow size=1pt](a6),
                (a2) --[boson, edge label=\(W\)](a3), (a2) --[boson, edge label'=\(W\)](a4)
			};	 	
			\end{feynman}
			\end{tikzpicture}}  \hspace{0.1cm}
            \subfloat[]{\begin{tikzpicture}
			\begin{feynman}
			\vertex (a1){$H$};
			\vertex [right=1.3cm of a1](a2);
			\vertex [square dot,blue,above right=0.5cm of a2](a3){};
			\vertex [square dot,violet,above right=0.8cm of a3](a4){};
			\vertex [above right=0.9cm of a4](a5){\(b\)};
			\vertex [below right=1.7cm of a2](a6){\(b\)};
			
			\diagram* {
				(a1) --[scalar](a2),
				(a6) --[fermion, arrow size=1pt](a2) --[fermion, arrow size=1pt,edge label'={\(b\)}](a3) --[fermion, arrow size=1pt, edge label'={\(t\)}](a4) --[fermion, arrow size=1pt](a5),
				(a3) --[boson, half left, looseness=2, edge label={\(W\)}](a4),
			};	 	
			\end{feynman}
			\end{tikzpicture}}
     \caption{Feynman diagrams showing the $Hb\bar{b}$ vertex correction induced by the insertion of an effective $Wtb$ operator. Each diagram includes one colored effective vertex at a time,
while keeping the other as SM types. }
    \label{fig:Htobb}
\end{figure}

 The Higgs boson decay to bottom quarks $(H \to b \bar{b})$ is an important probe of new physics. In SM, we get $H \to b \bar{b}$ at tree level, where the coupling is of purely scalar type. In the presence of NP, we can get both scalar and pseudoscalar-type couplings. In our framework, we get a correction to this tree-level decay via the Feynman diagrams shown in fig.~\ref{fig:Htobb}.
\noindent After including the loop, the effective vertex of $H \to b \bar{b}$ can be written by: 
\begin{equation}\label{eq:Eff_H2bb}
    \mathcal{H}^{\rm eff}_{H \to b\bar{b}}  = C_{HL} (\bar{b} P_{L} b ) + C_{HR} (\bar{b} P_{R} b )\,.
\end{equation}
As mentioned earlier, we get both left- and right-handed scalar currents. 
Here, we write $C_{HL(R)}  = C_{HL(R)}^{SM} + C_{HL(R)}^{NP}$. In SM, we have  $C_{HL}^{SM} = C_{HR}^{SM} = \frac{m_{b}}{v}$, $v$ being the vev, and the expression of $C_{HL(R)}^{NP}$ is provided in eq.~\eqref{eq:loop_Hdecays} of appendix~\ref{appndx:H_decay}.
The contribution of the 3rd diagram of the fig.~\ref{fig:Htobb}, will be added as wavefunction correction \cite{Denner:1991kt}. The decay width by taking the Hamiltonian of eq.~\eqref{eq:Eff_H2bb} is given by : 
\begin{equation}
    \Gamma(H \to b \bar{b}) = \frac{N_{c}}{16 \pi m_{H}^2 }\sqrt{m_{H}^2 - 2 m_{b}^2}  \left[  m_{H}^2 \left( C_{HL}^2 + C_{HR}^2 \right)  - 2 m_{b}^2 \left(C_{HL} + C_{HR} \right)^2 \right]\,.
\end{equation}
 Here, the color factor is $N_{c} =3$ accounts for possible colours of $b\bar{b}$ pairs. Here, $m_{b} = m_{b} (m_{H}) \approx 2.7 $ GeV. The experimental value of the decay width is given by: 
 \begin{equation}
     \Gamma (H \to b \bar{b}) = \left(  1.961 \pm 0.923 \right) \times 10^{-3} \text{ GeV} \,.
 \end{equation}
The above is calculated by using $\Gamma_{H}^{\rm tot} = 3.7^{\, \, + 1.9}_{\, \, -1.4}$ MeV, which is the PDG average \cite{ParticleDataGroup:2024cfk}, calculated using ATLAS \cite{ATLAS:2023dnm} and CMS \cite{CMS:2022ley} values of the decay width.
\subsection{Top chromo-magnetic dipole moment}\label{sec:top_EDM}
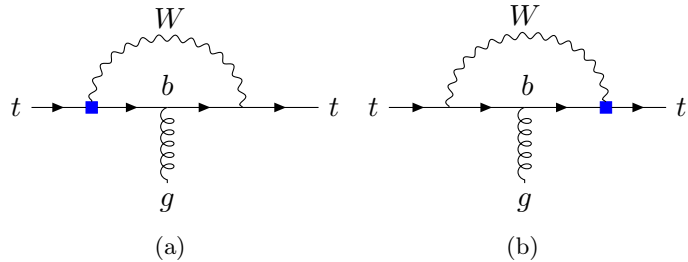
\begin{figure}[htb!]
    \centering
    \subfloat[]{ 
        \begin{tikzpicture}
	\begin{feynman}
		\vertex (a1){\(t\)};
		\vertex [square dot,blue,right=1cm of a1](a2){};
		\vertex [square dot,blue,right=1cm of a2](a3);
            \vertex [right=1cm of a3](a4);
		\vertex [right=1cm of a4](a5){\(t\)};
            \vertex [right=1cm of a2](a6);
            \vertex[below=1cm of a6](a7){\(g\)};
            \node at ($(a2)!0.5!(a4)$) [above=1pt] {\(b\)};
				
		\diagram* {
        (a1) --[fermion, arrow size=1pt](a2) --[fermion, arrow size=1pt](a3) --[fermion, arrow size=1pt](a4) --[fermion, arrow size=1pt](a5),
        (a2) --[boson, half left, looseness=1.5, edge label={\(W\)}](a4),
        (a6) --[gluon](a7),
	};				
	\end{feynman}
	\end{tikzpicture}}
    \subfloat[]{ 
        \begin{tikzpicture}
	\begin{feynman}
		\vertex (a1){\(t\)};
		\vertex [right=1cm of a1](a2);
		\vertex [right=1cm of a2](a3);
            \vertex [square dot,blue,right=1cm of a3](a4){};
		\vertex [right=1cm of a4](a5){\(t\)};
            \vertex [right=1cm of a2](a6);
            \vertex[below=1cm of a6](a7){\(g\)};
            \node at ($(a2)!0.5!(a4)$) [above=1pt] {\(b\)};
				
		\diagram* {
        (a1) --[fermion, arrow size=1pt](a2) --[fermion, arrow size=1pt](a3) --[fermion, arrow size=1pt](a4) --[fermion, arrow size=1pt](a5),
        (a2) --[boson, half left, looseness=1.5, edge label={\(W\)}](a4),
        (a6) --[gluon](a7),
	};				
	\end{feynman}
	\end{tikzpicture}}
    \caption{Feynman diagrams illustrating the top cMDM correction induced by the effective $Wtb$ operator.}

    \label{fig:top_CMDM}
\end{figure}

In the current data-dominated era, numerous measurements are being performed in the top sector. Due to its large mass and strong couplings to the electroweak sector, the top quark provides a favourable platform to probe NP. One of the most precisely measured observable in the top sector is the chromo-magnetic dipole moment (cMDM). In the SM, the cMDM arises from QCD corrections to the $\bar{t}t g$ vertex. The general Lagrangian that describes the effective $t\bar{t}g$ interaction can be written as \cite{CMS:2019kzp}:
\begin{align}\label{eq:top_EDM}
    \mathcal{L}_{t\bar{t}g}=-g_s \left(\bar{t}\gamma^{\mu}G_{\mu}^a T^a t+i \frac{\hat{d}_t}{2 m_t}\bar{t}\sigma^{\mu\nu}\gamma_5 G_{\mu\nu}^a T^a t+\frac{\hat{\mu}_t}{2 m_t}\bar{t}\sigma^{\mu\nu}G_{\mu\nu}^a T^a t\right)\,,
\end{align}
where $G_{\mu}^a$ are the gluon fields, $T^a$ are the generators of $\mathrm{SU}(3)_c$ and the gluon field strength tensor can be written as $G_{\mu\nu}^a \equiv \partial_{\mu}G_{\nu}^a - \partial_{\nu}G_{\mu}^a - g_s f^{abc} G_{\mu}^b G_{\nu}^c$. 
The coefficient $\hat{\mu}_{t}$ of the CP-even operator is called the chromo-magnetic dipole moment, while the coefficient $\hat{d}_{t}$ of the CP-odd operator is called the chromo-electric dipole moment. 
Fig.~\ref{fig:top_CMDM} shows the relevant Feynman diagrams for the dipole moments of the top quark. Similar diagrams can be drawn for the bottom quark as well. By replacing the gluon with a photon, we will have the corresponding Feynman diagrams contributing to electric and magnetic dipole moments. 
For the electric and magnetic dipole moments of the top and bottom quarks, only an indirect bound on them can be derived. The available bounds (or upper limit, whichever is applicable) are very relaxed compared to the SM values. 
In our case, the anomalous $Wtb$ couplings will only affect the chromo-magnetic dipole moment $\mu_t$. Being an amplitude level observable, the NP contributions to the cMDM $\mu_t^{\rm NP}$ will be simply added to the SM contributions $\mu_t^{\rm SM}$. The corresponding loop contributions to the top cMDM are shown in appendix~\ref{appndx:EWPOs}.
At the scale $q^2 = M_Z^2$, the SM prediction is 
$\hat{\mu}_t^{\rm SM}(M_Z^2) = -0.0253$~\cite{Montano-Dominguez:2021zmy}. The experimental value provided by the CMS collaboration~\cite{CMS:2019kzp} is
$\hat{\mu}_t^{\rm Exp}(M_Z^2) = -0.024^{+0.013\,+0.016}_{-0.009\,-0.011}$, which is
in good agreement with the SM prediction.

\section{Results and Analysis}\label{sec:results_analysis}
In the previous sections, we have shown how the anomalous couplings of $Wtb$ vertex can modify FCNC, FCCC, and EWPO processes, including their corresponding SMEFT operators. We have also shown the variation of the Wilson coefficients with the scale of the observables. In our analysis, we have taken all the above-mentioned observables into account and fitted them to their updated experimental values in order to extract the NP parameters. We have done separate and simultaneous analyses by taking FCNC, FCCC, and EWPOs processes. We have performed a chi-squared analysis by taking the above-mentioned observables into account. All the fits are performed using a Mathematica® Package OptEx \cite{sunando_patra_2019_3404311}. In this section, we will first present the results from a separate analysis, followed by the results from the combined analysis, where all the affected observables were considered. Also, for the separate analysis, we will present the results in a one-parameter scenario, while for the combined scenario, we will present in both one-parameter and multi-parameter scenarios and the correlations among the couplings.   
We will also present the results at different energy scales to study the variation of the parameters with the energy scale. 
Further, we will study improvements in the branching ratios of decays that are highly suppressed in the SM and have only upper experimental limits, and will be affected by our chosen SMEFT parameters.

\subsection{Fit Results from EWPOs and Other relevant Observables}
In sec.~\ref{sec:EWPOs}, we have discussed how the EWPOs are affected by the anomalous $Wtb$ couplings and their corresponding SMEFT coefficients. We have considered the $W$ pole observables such as $\Delta r$, $Z$ pole observables such as ratio observable $R_{b}$, asymmetry observable $A_{b}$, $\mathcal{B}(H \to b \bar{b})$, TGCs and top cMDM. Here we will present the fit result we get by taking all these observables into account. Note that, in this case, our observables are at $\mu_{EW}$ scale, so we don't need further running of the effective couplings to a low-energy scale, which we have to do for the other cases (FCCC, FCNC).

\begin{table}[htbp]
\centering
\rowcolors{1}{cyan!10}{blue!20!green!5}
\renewcommand{\arraystretch}{2.0}
\setlength{\tabcolsep}{5pt}
\resizebox{\textwidth}{!}{%
\begin{tabular}{|c|c|c|c|c|}
\hline
\rowcolor{cyan!20}
\multicolumn{5}{|c|}{\textbf{Observables $\rightarrow$  W-mass + Z-pole + TGC + Higgs Decay+ Top cMDM}} \\ 
\hline
\hline
\textbf{Scale}  & 
$\frac{v^2}{\Lambda^2}\mathcal{C}^{\phi q (3)}_{33}$ & 
$\frac{v^2}{\Lambda^2} \mathcal{C}^{\phi u d}_{33}$ & 
$\frac{v^2}{\Lambda^2}\mathcal{C}^{*\, dW}_{33}$ & 
$\frac{v^2}{\Lambda^2}\mathcal{C}^{uW}_{33}$ \\ 
\hline
$\mu_{\rm EW}$  &  $(-0.86\pm 1.29)\times 10^{-2}$  &  $(1.55 \pm 2.45)$  &  $(0.02 \pm 1.73)$  &  $(1.81\pm 3.35)\times 10^{-2}$  \\
\hline
\end{tabular}%
}
\caption{Fit results of the anomalous couplings of $Wtb$ vertex from the EWPOs and other relevant observable.}
\label{tab:fit_res_EWPOs}
\end{table}

Table ~\ref{tab:fit_res_EWPOs} shows the fitted values of the normalized SMEFT couplings from the EWPOs at one parameter scenario (only considering a single NP parameter at a time while other parameters are considered SM type). 
Here, the fit results we get for $\frac{v^2}{\Lambda^2}\mathcal{C}^{\phi q (3)}_{33}$ and $\frac{v^2}{\Lambda^2} \mathcal{C}^{uW}_{33}$ are of the order of $\sim \mathcal{O}(10^{-2})$, whereas $ \frac{v^2}{\Lambda^2}\mathcal{C}^{\phi u d}_{33}$ and $ \frac{v^2}{\Lambda^2} \mathcal{C}^{*\, dW}_{33}$ are less constrained, with values of $\sim \mathcal{O}(10^{-1})$. 
From the fit, we obtain the constraints on the couplings at the electroweak scale ($\mu_{\rm EW}$).  Among all the observables of EWPOs, the most constrained result we are getting from the $\delta(\Delta r)$ as the latest CMS \cite{CMS-PAS-SMP-23-002} result shows a very small deviation from the SM, as can also be seen from \cite{Kolay:2025jip}. On the other hand, the Higgs decay width gives the most relaxed fit in this scenario. 

\subsection{Fit Results from FCCC Observables}
In sec.~\ref{sec:FCCC_process}, we have discussed the FCCC processes that are getting modified in the presence of effective $Wtb$ couplings. Here, we will present the values of the SMEFT couplings we obtain from the CKM fit, with inputs and observables discussed in sec.~\ref{sec:FCCC_process}. 
\begin{table}[htb!]
	\centering
	\rowcolors{1}{cyan!10}{blue!20!green!5}
	\renewcommand{\arraystretch}{1.5}
	\setlength{\tabcolsep}{3pt} 
	\begin{tabular}{|c|c|c|c|c|}
		\hline
		\rowcolor{cyan!20}
		\multicolumn{5}{|c|}{\textbf{Observables $\rightarrow$ FCCC semileptonic + leptonic decays}} \\ 
		\hline
		\hline
		\textbf{Scale}  & 
		$\frac{v^2}{\Lambda^2}\mathcal{C}^{\phi q (3)}_{33}$ & 
		$\frac{v^2}{\Lambda^2} \mathcal{C}^{\phi u d}_{33}$ & 
		$\frac{v^2}{\Lambda^2}\mathcal{C}^{*\, dW}_{33}$ & 
		$\frac{v^2}{\Lambda^2}\mathcal{C}^{uW}_{33}$ \\ 
		\hline
		$\mu_{b}$ & $(0.0 \pm  2.97) \times 10^{-2}$ & $( 0.0\pm 0.94 ) \times 10^{-2} $ & $(0.0 \pm 1.59 ) \times 10^{-2}$ & $(-1.59 \pm 2.57 ) \times 10^{-2}  $ \\
		\hline
	\end{tabular}
	\caption{Value of the fitted parameters from all the FCCC observables at $\mu_{\rm EW}$ and $\mu_{t}$ scale.}\label{tab:FCCC_fit}
\end{table}

Table~\ref{tab:FCCC_fit} presents the fitted values of the SMEFT couplings $\mathcal{C}^{\phi q (3)}_{33}$, $\mathcal{C}^{\phi u d}_{33}$, $\mathcal{C}^{*\, dW}_{33}$, and $\mathcal{C}^{uW}_{33}$ normalised with $v^2 / \Lambda^2$. These are related to the $Wtb$ anomalous couplings according to eq.~\eqref{eq:Wtb_SMEFT_correspondence}.
All the values shown in the table are zero-consistent, as all the data are consistent with the SM. Here we get all the couplings $\sim \mathcal{O}(10^{-2})$. 
As we can see, here we get $ \frac{v^2} {\Lambda^2}\mathcal{C}^{\phi u d}_{33}$ and $ \frac{v^2}{\Lambda^2} \mathcal{C}^{*\, dW}_{33}$ are two order smaller than that we get from EWPOs. Other couplings $\frac{v^2}{\Lambda^2}\mathcal{C}^{\phi q (3)}_{33}$ and $\frac{v^2}{\Lambda^2} \mathcal{C}^{uW}_{33}$ remains as the same order as that we get from EWPOs which is $\sim \mathcal{O}(10^{-2})$, although the magnitude of $\frac{v^2}{\Lambda^2}\mathcal{C}^{\phi q (3)}_{33}$
becomes slightly relaxed. The fit results are obtained on the WCs defined at the energy scale $\mu_b$. The non-negligible correlations among the couplings are also obtained. We will show the correlations among the couplings only for the combined analysis.

\subsection{Fit Results from FCNC Observables}\label{sec:FCNC_fit}
Here we will discuss the constraints on the SMEFT couplings $\frac{v^2}{\Lambda^2}\mathcal{C}^{\phi q (3)}_{33}$, $ \frac{v^2}{\Lambda^2}\mathcal{C}^{\phi u d}_{33}$, $ \frac{v^2}{\Lambda^2} \mathcal{C}^{*\, dW}_{33}$, and $\frac{v^2}{\Lambda^2} \mathcal{C}^{uW}_{33}$ from a fit to all the FCNC observables such as neutral meson mixing, radiative decays, invisible decays, rare decays and differential branching fraction, asymmetry observables, LFUV observables from $b \to s \ell \ell $ transition, that are discussed in sec.~\ref{sec:FCNC_process}.

\begin{table}[htbp]
	\centering
	\rowcolors{1}{cyan!10}{blue!20!green!5}
	\renewcommand{\arraystretch}{2.0}
	\setlength{\tabcolsep}{5pt} 
	\resizebox{\textwidth}{!}{%
		\begin{tabular}{|c|c|c|c|c|}
			\hline
			\rowcolor{cyan!20}
			\multicolumn{5}{|c|}{\textbf{Observables $\rightarrow$ Meson Mixing + Rare + Radiative + $b \to s \ell \ell$ + Invisible }  } \\ 
			\hline
			\hline
			\textbf{Scale}  & 
			$\frac{v^2}{\Lambda^2}\mathcal{C}^{\phi q (3)}_{33}$ & 
			$\frac{v^2}{\Lambda^2} \mathcal{C}^{\phi u d}_{33}$ & 
			$\frac{v^2}{\Lambda^2}\mathcal{C}^{*\, dW}_{33}$ & 
			$\frac{v^2}{\Lambda^2}\mathcal{C}^{uW}_{33}$ \\ 
			\hline
			$\mu_{b}$  &  $(-1.23 \pm 0.89)\times 10^{-2}$  &  $ (0.72 \pm 0.85 )\times 10^{-3}$  &  $(-2.61\pm 3.44 )\times 10^{-4}$  &  $(3.71\pm 2.27)\times 10^{-3}$
			\\
			\hline
		\end{tabular}%
	}
	\caption{Fitted values of the anomalous couplings of $Wtb$ vertex from all the FCNC observables at electroweak and top-mass scales.}
	\label{tab:FCNC_fit}
\end{table} 

Table \ref{tab:FCNC_fit} shows the fitted results of the SMEFT couplings $\frac{v^2}{\Lambda^2}\mathcal{C}^{\phi q (3)}_{33}$, $ \frac{v^2}{\Lambda^2}\mathcal{C}^{\phi u d}_{33}$, $ \frac{v^2}{\Lambda^2} \mathcal{C}^{*\, dW}_{33}$, and $\frac{v^2}{\Lambda^2} \mathcal{C}^{uW}_{33}$ from FCNC observables. Here, we have done the fit in one-parameter scenario, where only one parameter is fitted at a time while setting the other parameters as SM types. The fit results of the couplings are shown at EW scale $\mu_{\rm EW}$ and top scale $\mu_{t}$. We get the value of the coupling $\frac{v^2}{\Lambda^2}\mathcal{C}^{\phi q (3)}_{33}$ is the order of $\sim  \mathcal{O}(10^{-2}) $. The other two couplings $ \frac{v^2}{\Lambda^2}\mathcal{C}^{\phi u d}_{33}$ and $\frac{v^2}{\Lambda^2} \mathcal{C}^{uW}_{33}$, we get $\sim \mathcal{O}(10^{-3})$. The coupling $ \frac{v^2}{\Lambda^2} \mathcal{C}^{*\, dW}_{33}$ is obtained of the order of $\sim \mathcal{O}({10^{-4}})$. We have presented the fitted values at the low energy scale, i.e. $\mu_{b}$. Similar to the previous case, the couplings get a little relaxed with increasing the scale. Note that the most stringent bounds on three of the four couplings are coming from the FCNC processes.  We obtain strongest bounds on the couplings $\frac{v^2}{\Lambda^2}\mathcal{C}^{\phi u d}_{33}$ and $\frac{v^2}{\Lambda^2} \mathcal{C}^{*\, dW}_{33}$, which correspond to $V_{R}$ and $g_{L}$ in anomalous $Wtb$ interactions, primarily come from radiative decays, as discussed in sec.~\ref{sec:radiative}. The observables $R_{K^{(*)}}$ also play an important role as their experimental value does not deviate much from the SM. Similarly, the couplings $\frac{v^2}{\Lambda^2}\mathcal{C}^{\phi q (3)}_{33}$ and $\frac{v^2}{\Lambda^2} \mathcal{C}^{uW}_{33}$, related to $V_{L}$ and $g_{R}$, are mainly constrained by data from neutral meson mixing, as covered in sec.~\ref{sec:meson_mixing}.  

From the FCCC processes and EWPOs, we get the fitted values of the couplings as zero-consistent. Notably, in this case, two of the couplings $\frac{v^2}{\Lambda^2}\mathcal{C}^{\phi q (3)}_{33}$ and $\frac{v^2}{\Lambda^2} \mathcal{C}^{uW}_{33}$ get non-zero values at $1\sigma$, as can be seen from table~\ref{tab:FCNC_fit}. This is due to the inclusion of neutral meson mixing data, specifically $B^{0}-\bar{B}^{0}$ mixing, as reflected in eq.~\eqref{eq:mixing_delta_Exp}. Our observable $\Delta_{d}$ does not agree with SM at the $1\sigma$ error bound, as it is defined by $\left( \Delta M^{NP} / \Delta M^{SM} \right)$, which is discussed in sec.~\ref{sec:meson_mixing}. This slight difference accounts for the nonzero values obtained for $\frac{v^2}{\Lambda^2}\mathcal{C}^{\phi q (3)}_{33}$ and $\frac{v^2}{\Lambda^2} \mathcal{C}^{uW}_{33}$. The allowed solutions will be consistent with zero if we drop $\Delta_d$ from the analysis; we have shown the respective results in the appendix \ref{apndx:fit_without_deltad}.

\subsection{Combined Analysis }
In the previous section, we have shown the fitted results of the SMEFT couplings from separate individual analyses of FCCC, FCNC, and EWPOs. The FCNC and FCCC processes are low-energy processes ($\mu_{b}$), whereas the EWPOs are high-energy observables ($\mu_{EW}$). As mentioned in the text, we have used proper RGEs to connect the NP parameters with the observables at different energy scales.
Here we will present the final fit results from the combined analysis. We have performed the fit for both one-parameter and multi-parameter scenarios.
\begin{table}[t]
	\centering
	\rowcolors{1}{cyan!10}{blue!20!green!5}
	\renewcommand{\arraystretch}{2.0}
	\setlength{\tabcolsep}{5pt}
	\resizebox{\textwidth}{!}{%
		\begin{tabular}{|c|c|c|c|c|}
			\hline
			\rowcolor{cyan!20}
			\multicolumn{5}{|c|}{\textbf{Combined analysis: FCCC + FCNC + EWPOs +Others}} \\ 
			\hline
			\hline
			$\text{Scale}$ & $\frac{v^2}{\Lambda^2} \mathcal{C}^{\phi q (3)}_{33}$  &  $ \frac{v^2}{\Lambda^2} C^{\phi ud}_{33}$ &  $\frac{v^2}{\Lambda^2} C^{*\, dW}_{33}$ &  $ \frac{v^2}{\Lambda^2} C^{uW}_{33}$   \\
			\hline
			$\mu_{b}$  & $( -1.06 \pm 0.69 ) \times 10^{-2}$ &  $( -0.43 \pm 0.86 ) \times 10^{-3}$& $(1.37 \pm 3.49 ) \times 10^{-4} $ & $(2.84 \pm 2.17 ) \times 10^{-3}$ \\ 
			\hline
			$\mu_{\rm EW}$  & $( -1.21 \pm 0.80 ) \times 10^{-2}$ &  $( -0.53 \pm 1.05 ) \times 10^{-3}$& $(1.68 \pm 4.31 ) \times 10^{-4} $ & $(3.53 \pm 2.69 ) \times 10^{-3}$ \\ 
			\hline
			$\mu_t$ & $(-1.24 \pm 0.81 ) \times 10^{-2}$ & $( -0.54 \pm 1.09 ) \times 10^{-3}$ & $(1.73 \pm 4.45 ) \times 10^{-4}$  & $(3.65 \pm  2.78 ) \times 10^{-3}$ \\
			\hline
		\end{tabular}%
	}
	\caption{Fit results of the SMEFT couplings from the combined analysis, considering all FCCC and FCNC observables, as well as EWPOs. }
	\label{tab:combined_one_param_SMEFT}
\end{table}

Table~\ref{tab:combined_one_param_SMEFT} shows the fitted values from the combined analysis, where we have taken all the observables discussed in this article. This table presents the fitted values of the SMEFT couplings in the one-parameter scenario. As we have seen in the previous section, the most constrained values on $\frac{v^2}{\Lambda^2} \mathcal{C}^{\phi q (3)}_{33}$, $ \frac{v^2}{\Lambda^2} C^{uW}_{33}$ are coming from neutral meson mixing whereas $ \frac{v^2}{\Lambda^2} C^{\phi ud}_{33}$ and $\frac{v^2}{\Lambda^2} C^{*\, dW}_{33}$ is mostly constrained from radiative decays. In the combined analysis, we get values of $\frac{v^2}{\Lambda^2} \mathcal{C}^{\phi q (3)}_{33}$ of order $\sim \mathcal{O}(10^{-2})$. The couplings $ \frac{v^2}{\Lambda^2} C^{uW}_{33}$ and $ \frac{v^2}{\Lambda^2} C^{\phi ud}_{33}$ are of the order of $\sim \mathcal{O}(10^{-3})$, whereas $\frac{v^2}{\Lambda^2} C^{*\, dW}_{33}$ we get is one order smaller i.e., $ \sim \mathcal{O}(10^{-4})$. Recalling the results obtained from data on different sectors, we easily understand which data or data sets play an important role in the constraints of the different couplings.    

The first row of table~\ref{tab:combined_one_param_SMEFT} shows the fitted result at scale $\mu_{b}$, whereas the second and third rows show the results at scale $\mu_{EW}$ and $\mu_{t}$, respectively, which we have obtained following RGEs defined earlier. We note a slight variation ($\approx 20\% \to 25\%)$ of the maximum allowed values of these couplings with the changes of the scale from $\mu_b\to \mu_{EW}$ or $\mu_t$.

As can be seen from FCNC fit results, the couplings $\frac{v^2}{\Lambda^2} \mathcal{C}^{\phi q (3)}_{33}$, $ \frac{v^2}{\Lambda^2} C^{uW}_{33}$ are not zero consistent because of the data from neutral meson mixing. 
If we do not include the observable $\Delta_{d}$ in the analysis, then all the couplings' values will be zero-consistent. The fit results corresponding to this scenario are shown in the appendix \ref{apndx:fit_without_deltad}.


\begin{table}[htb!]
\centering
\rowcolors{1}{cyan!10}{blue!20!green!5}
\renewcommand{\arraystretch}{3.0}
\setlength{\tabcolsep}{10pt}
\begin{tabular}{|c|c|c|}
\hline
\rowcolor{cyan!20}
\textbf{Scale} & \textbf{Scenario} & \textbf{Values} \\
\hline
\hline
 \cellcolor{cyan!15} & $\left(\frac{v^2}{\Lambda^2}\mathcal{C}^{\phi q(3)}_{33}\,, \frac{v^2}{\Lambda^2}\mathcal{C}^{\phi u d}_{33}\right)$ & 
\shortstack[l]{
    $\left( (-1.20\pm 0.80 )\times 10^{-2} \,,
   (-0.44\pm 1.04)\times 10^{-3} \right)$
} \\

\cellcolor{cyan!15}& $\left(\frac{v^2}{\Lambda^2}\mathcal{C}^{\phi q (3)}_{33}\,, \frac{v^2}{\Lambda^2}\mathcal{C}^{*\, dW}_{33}\right)$ & 
\shortstack[l]{
    $\left((-1.12\pm 0.80 )\times 10^{-2}\,,(1.73 \pm 4.30)\times 10^{-4}\right)$
} \\

\cellcolor{cyan!15}& $\left(\frac{v^2}{\Lambda^2}\mathcal{C}^{\phi q(3)}_{33}\,, \frac{v^2}{\Lambda^2}\mathcal{C}^{uW}_{33}\right)$ & 
\shortstack[l]{
    $\left((-0.99 \pm 1.05)\times 10^{-2}\,,(1.17 \pm 3.55) \times 10^{-3}\right)$
} \\

\cellcolor{cyan!15}\cellcolor{cyan!15}& $\left(\frac{v^2}{\Lambda^2}\mathcal{C}^{\phi u d}_{33}\,, \frac{v^2}{\Lambda^2}\mathcal{C}^{*\, dW}_{33}\right)$ & 
\shortstack[l]{
    $\left((-0.68\pm 4.88)\times 10^{-3} \,, (0.06 \pm 1.97)\times 10^{-3}\right)$
} \\

\cellcolor{cyan!15}& $\left(\frac{v^2}{\Lambda^2}\mathcal{C}^{\phi u d}_{33}\,, \frac{v^2}{\Lambda^2}\mathcal{C}^{ uW}_{33}\right)$& 
\shortstack[l]{
    $\left((0.5 \pm 1.04)\times 10^{-3}\,, (3.37  \pm 2.69 )\times 10^{-3}\right)$
} \\

\multirow{-6}{*}{\cellcolor{cyan!15} $\mu_{\rm EW}$} & $\left(\frac{v^2}{\Lambda^2}\mathcal{C}^{*\,dW}_{33}\,, \frac{v^2}{\Lambda^2}\mathcal{C}^{ uW}_{33}\right)$ & 
\shortstack[l]{
    $\left((1.62 \pm 4.31)\times 10^{-4}\,,(3.36 \pm 2.69)\times 10^{-3}\right)$
} \\
\hline
\end{tabular}
\caption{Fit results of SMEFT couplings from the combined analysis in two-parameter scenario. }
\label{tab:combined_two_param}
\end{table}
\begin{table}[htb!]
\centering
\rowcolors{1}{cyan!10}{blue!20!green!5}
\renewcommand{\arraystretch}{3.0}
\setlength{\tabcolsep}{10pt}
\begin{tabular}{|c|c|c|}
\hline
\rowcolor{cyan!20}
\textbf{Scale} & \textbf{Scenario} & \textbf{Values} \\
\hline
\hline
$\mu_{\rm EW}$ & \shortstack[l]{$\frac{v^2}{\Lambda^2}\mathcal{C}^{\phi q(3)}_{33}, \frac{v^2}{\Lambda^2}\mathcal{C}^{\phi u d}_{33},$ \\ $\frac{v^2}{\Lambda^2}\mathcal{C}^{* dW}_{33}, \frac{v^2}{\Lambda^2}\mathcal{C}^{ uW}_{33}$} & 
\shortstack[l]{
    $(-0.97 \pm 1.09) \times 10^{-2}\,, (-0.72\pm 4.89)\times 10^{-3}\,, $\\
    $(-0.01\pm 1.97)\times 10^{-3}\,, (1.19 \pm 3.71)\times 10^{-3}$
} \\
\hline
\end{tabular}
\caption{Fit results of the SMEFT couplings from the combined analysis in the four-parameter scenario. }
\label{tab:combined_four_param}
\end{table}

\paragraph{Fit in multi-operator scenarios:}
Table~\ref{tab:combined_two_param} shows the values of the fitted SMEFT couplings from a combined analysis of all the impacted observables in a two-parameter scenario, i.e., in the analyses, we consider the contributions of two parameters at a time while setting the other two parameters at their SM values. The allowed ranges of values of the WCs remain almost the same as those in the one-parameter scenario, as different parameters are constrained by different processes, as mentioned. In the one parameter scenario, we obtained non-zero allowed solutions for $\mathcal{C}^{\phi q(3)}_{33}$ and $\mathcal{C}^{uW}_{33}$ when considered at their respective 1$\sigma$ confidence interval (CI). However, when both the operators are considered simultaneously, the allowed regions of both the coefficients are consistent with zero even at 1$\sigma$ CI. 

Furthermore, we have done an analysis that considers the contribution of all four operators simultaneously. In table~\ref{tab:combined_four_param}, we present the corresponding allowed values of the SMEFT couplings obtained from a combined analysis. In this scenario, the allowed regions of all four coefficients are consistent with those obtained in the one- or two-operator scenarios. All the allowed solutions are zero-consistent. Additionally, the maximum allowed range of $\frac{v^2}{\Lambda^2}\mathcal{C}^{* dW}_{33}$ could be of order $\mathcal{O}(10^{-3})$ which is relatively relaxed as compared to one or two operator scenarios.

\begin{figure}[t]
    \centering
    \includegraphics[scale=0.92]{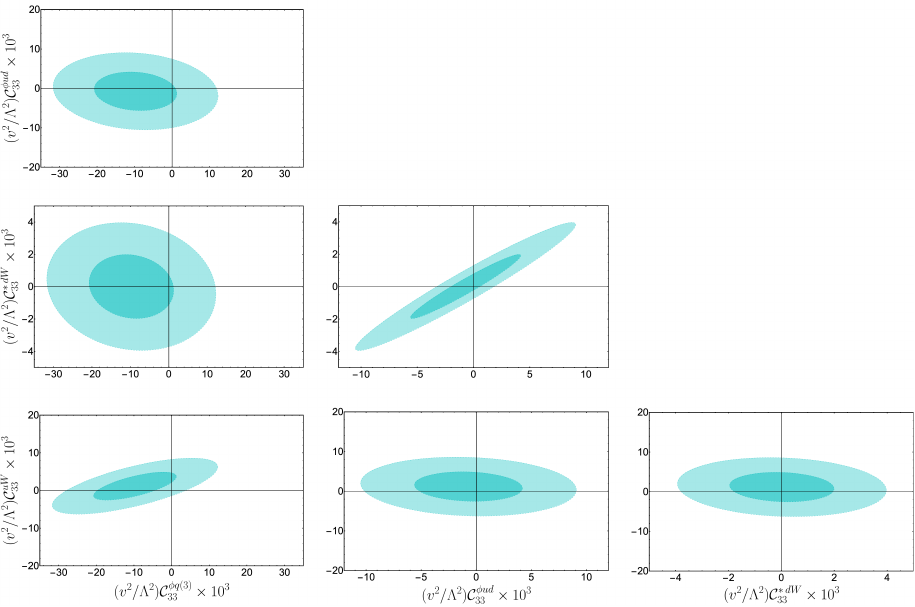}
     \caption{Correlations among the SMEFT Wilson coefficients are shown. The darker and lighter cyan-blue ellipses represent $1\sigma$ and  $2\sigma$ errors, respectively. Wilson coefficients are defined at the reference scale $\mu_{\rm EW}$.}
     \label{fig:correlations_SMEFT}
\end{figure}

In the multi-parameter analysis, it will be important to study the correlations between the respective parameters. The six plots in fig.~\ref{fig:correlations_SMEFT} show the correlations among the four SMEFT couplings $ \mathcal{C}^{\phi q(3)}_{33}, \mathcal{C}^{\phi u d}_{33},  \mathcal{C}^{* dW}_{33}  $, and $ \mathcal{C}^{uW}_{33}$, scaled with $\frac{v^2}{\Lambda^2} $, obtained from the fit of the four-parameter scenario. We have shown the correlations considering the errors in the parameters at their $2\sigma$ CI. The darker ellipse indicates the $1\sigma$ error bands, while the lighter ones represent the $2\sigma$  band, respectively. 

Strong positive correlations are observed between the couplings \( ( \mathcal{C}^{*\,dW}_{33}-\mathcal{C}^{\phi u d}_{33}) \) and \( ( \mathcal{C}^{\phi q(3)}_{33}-\mathcal{C}^{uW}_{33}) \). Similarly, mild negative correlations are seen among the pairs \( ( \mathcal{C}^{\phi q(3)}_{33}-\mathcal{C}^{*\,dW}_{33})\), \((\mathcal{C}^{\phi q(3)}_{33}-\mathcal{C}^{\phi u d}_{33})\), \((\mathcal{C}^{uW}_{33}\)--\(\mathcal{C}^{*\,dW}_{33})\), and \((\mathcal{C}^{uW}_{33}\)--\(\mathcal{C}^{\phi u d}_{33})\), as illustrated in fig.~\ref{fig:correlations_SMEFT}. These correlation trends are also reflected in the two-parameter fit results presented in table~\ref{tab:combined_two_param}.
The strong positive correlation between $(\mathcal{C}^{*\,dW}_{33}-\mathcal{C}^{\phi u d}_{33})$ is coming from the FCCC observables, whereas the correlation among $(\mathcal{C}^{\phi q(3)}_{33}-\mathcal{C}^{uW}_{33})$ is due to the FCNC observables that we have discussed earlier. 
From the correlation plots, it is evident that all the contours include the point \((0,0)\) within the \(1\sigma\) uncertainty band, except for the \((\mathcal{C}^{\phi q(3)}_{33},\,\mathcal{C}^{uW}_{33})\) contour. In particular, the contours involving \((\mathcal{C}^{\phi q(3)}_{33},\,\mathcal{C}^{uW}_{33})\) intersect the origin only within the \(1\sigma\) to \(2\sigma\) uncertainty region. This deviation is mainly driven by the positive correlation between the two SMEFT couplings, which are individually consistent with zero, but not simultaneously within the correlated fit.

\begin{table}[t]
	\centering
	\rowcolors{1}{cyan!10}{blue!20!green!5}
	\renewcommand{\arraystretch}{2.0}
	\setlength{\tabcolsep}{5pt}
	\resizebox{0.9\textwidth}{!}{%
		\begin{tabular}{|cccc|}
			\hline
			\rowcolor{cyan!20}
			Coupling & ATLAS \cite{CMS:2020ezf}& CMS \cite{CMS:2020ezf}& This work \\
			\hline
			\hline
			$\text{Re}(V_R)$ & $[-0.17, 0.25]$ & $[-0.12, 0.16]$ &   $(-0.26\pm 1.05)\times 10^{-3}$ \\
			$\text{Re}(g_L)$ & $[-0.11, 0.08]$ & $[-0.09, 0.06]$ &  $(0.24 \pm 1.22)\times 10^{-3}$ \\
			$\text{Re}(g_R)$ & $[-0.03, 0.06]$ & $[-0.06, 0.01]$ &  $(4.99\pm 7.62)\times 
			10^{-3}$  \\
			$\text{Re}(V_{L}) $ \cite{ParticleDataGroup:2024cfk,CMS:2020vac}  & \multicolumn{2}{c}{$ (-0.012 \pm 0.036) $}  & $(-1.21 \pm 0.80)\times 10^{-2}$ \\
			\hline
		\end{tabular}
	}
	\caption{Comparison of the allowed ranges for the anomalous couplings $V_R$, $g_L$, and $g_R$ at 95\% CL obtained from our analysis with the most recent bounds from \cite{CMS:2020ezf}. The extreme right column gives our result with $2\sigma$ error except $V_L$, which is given with $1\sigma$ error. The couplings are assumed to be real.}
	\label{tab:compare_AnomalousCouplings}
\end{table}

\paragraph{\underline{Bounds on $Wtb$ anomalous couplings:}}
Using the results on the SMEFT coefficients, we have obtained the predictions for the $Wtb$ anomalous couplings defined in eq.~\eqref{eq:Wtb_SMEFT_correspondence}.
We present the respective results in table~\ref{tab:compare_AnomalousCouplings}, which we have obtained using the results presented in table~\ref{tab:combined_one_param_SMEFT} for the scale $\mu_t$. For comparison, in the same table, we have presented the constraints on the anomalous \( Wtb \) couplings obtained by the analysis done by ATLAS and CMS~\cite{CMS:2020ezf} using several experimental data. The experimental constraints are available for \( V_{R} \), \( g_{L} \), and \( g_{R} \) at $95\%$  confidence level. The last column in table~\ref{tab:compare_AnomalousCouplings} shows our fitted results in one-parameter scenario, with \( 2\sigma \) error. Our bounds are two orders of magnitude stronger than those provided by ATLAS and CMS.
In the CMS results, no bound has been provided for the coupling \( V_L \) or \( \mathcal{C}^{\phi q (3)}_{33} \). However, CMS \cite{CMS:2020vac} has provided the data on $|f_{LV} V_{tb} |$, which in our parameterization is equivalent to $V_{L}'  $, from which $V_{L}$ can be calculated using eq.~\eqref{eq:VL_from_Vtb} i.e., $V_{L}^{\prime} = V_{tb} + V_L $. Using $V_{tb}$ as given by PDG \cite{ParticleDataGroup:2024cfk}, we get experimental equivalent $V_{L} = (-0.012 \pm 0.036) $. If we use $V_{tb}$ as given by the CKMFitter group \cite{ckmfitter}, the extracted value of $V_{L}$ would be $V_{L} = (-0.001 \pm 0.024)$. Another study has determined the value of the coupling \( V_{L}^{\prime} \) (defined in our parametrization as \( V_{L}^{\prime} = V_{tb} + V_L \)) to be in the range \([0.85, 1.08]\) at 95\% confidence level~\cite{Tonero:2021jhe}, based on data collected from the LHC and Tevatron~\cite{Deliot:2017byp}.
In our work, we extract the value of \( V_L \) as given in the last row of the table~\ref{tab:compare_AnomalousCouplings}. The values of $V_L$, both from experimentally equivalent data and our extraction, are quoted with $1\sigma$ uncertainties, whereas the remaining three parameters are provided at the $2\sigma$ uncertainty level.

\paragraph{\underline{Scale dependency of the couplings:}}
In the above paragraphs, we have shown the constraints on the SMEFT couplings along with their corresponding $Wtb$ anomalous couplings separately and collectively for FCNC, FCCC, and EWPOs. In the tables, we have presented our results at three different scales, i.e. at $\mu_b$, $\mu_{\rm  EW} $ and $\mu_{t}$, respectively. The following will discuss the running of the relevant coupling above $\mu_{\rm  EW} $ or $\mu_{t}$ up to the scale $\Lambda$. In the running, we have considered the values of $\Lambda$ up to 10 TeV. 

\begin{figure}[t]
	\centering
	\subfloat[]{\includegraphics[width=7.4cm,height=5cm]{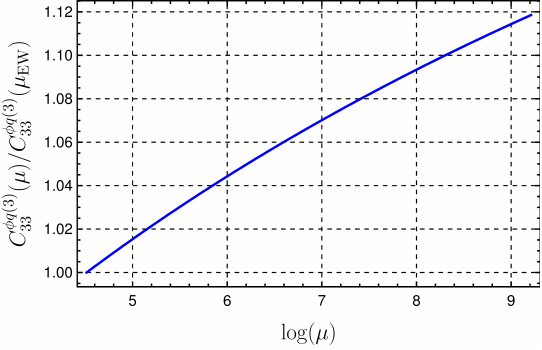}\label{fig:Cphiq3}} \hspace{0.01cm}
	\subfloat[]{\includegraphics[width=7.4cm,height=5cm]{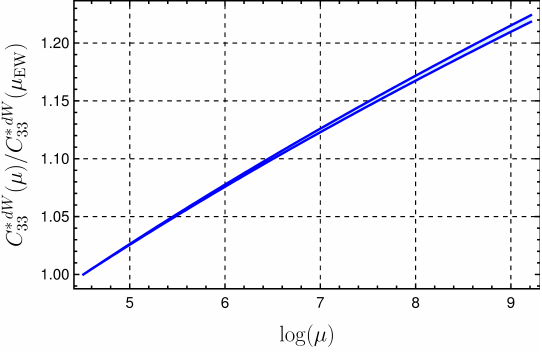} \label{fig:cdW}}\hspace{0.01cm}
	\subfloat[]{\includegraphics[width=7.4cm,height=5cm]{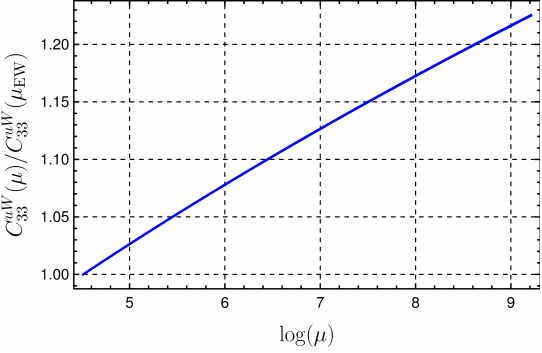}\label{fig:CuW}}\hspace{0.01cm}
	\subfloat[]{\includegraphics[width=7.4cm,height=5cm]{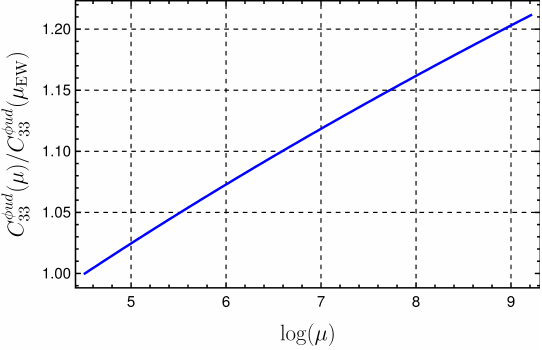}\label{fig:Cphiud}}\hspace{0.01cm}
	\caption{Evolution of the extracted SMEFT couplings over the energy range up to $10$ TeV.}
	\label{fig:scale_variation_SMEFT}
\end{figure}

To obtain the SMEFT coefficients for different values of $\Lambda$, one can directly solve the RGEs given in eq.~\eqref{eq: RGE}. However, we have followed the equation as given below
\begin{align}\label{eq:RGEratios}
	\frac{\mathcal{C}_i(\mu_2)}{\mathcal{C}_i(\mu_{EW})}=1+\left(\sum_i\frac{\gamma_{ii}(\mu_1)}{16\pi^2}+\sum_{i\neq j}\frac{\gamma_{ij}(\mu_1)}{16\pi^2}\frac{\mathcal{C}_j(\mu_{EW})}{\mathcal{C}_i(\mu_{EW})}\right)\log\left(\frac{\mu_2}{\mu_{EW}}\right)\,.
\end{align}
Here, the $\gamma$ matrices are part of the $\beta$ functions, with 
\begin{equation}
\beta_i = \sum_i \gamma_{ii} \mathcal{C}_i + \sum_{i \neq j} \gamma_{ij} \mathcal{C}_j.
\end{equation}
The expressions for $\beta_i$'$s$ are provided in appendix~\ref{appndx:RGE_beta} from where one could extract the required $\gamma_{ij}$'$s$. Note that the SMEFT coefficients are normalised by the scale $\Lambda$. Hence, to study a variation of the coefficients with varying $\Lambda$, we define the ratio $\mathcal{C}_i (\mu_2)/\mathcal{C}_i (\mu_{EW})$ in which the normalisation factor $\Lambda^2$ cancels. Earlier, we have extracted the values of $\mathcal{C}_i (\mu_{EW})$ in table~\ref{tab:combined_one_param_SMEFT}, which we use in the above equation, and obtain the variations of $\mathcal{C}_i (\Lambda)$. The respective variations are shown in fig.~\ref{fig:scale_variation_SMEFT}. It is evident from the figure that the RGEs play a crucial role. The $\mathcal{C}^{\phi q (3)}_{33}(\Lambda)$ increases up to 12\% from it's value $\mathcal{C}^{\phi q (3)}_{33}(\mu_{\rm EW})$ when we vary $\Lambda$ up to 10 TeV. More or less, all the other SMEFT couplings vary approximately up to 20\% from their respective values at $\mu_{\rm EW}$ when we vary $\Lambda$ up to 10 TeV. 

Earlier, we extracted the values of anomalous $Wtb$ couplings at the scale $\mu_t$. In the following, we have extracted these couplings at different higher NP scales, like $\Lambda = 1,$ 5, and 10 TeV, respectively. We present the respective numbers in table~\ref{tab:combined_one_param}. We obtain these numbers using the SMEFT couplings at the respective scales following eq.~\eqref{eq:RGE} or eq.~\eqref{eq:RGEratios}. Depending on the values of $\Lambda$, the maximum allowed values of these couplings could vary up to 10\% to 20\%.


\begin{table}[t]
	\centering
	\rowcolors{1}{cyan!10}{blue!20!green!5}
	\renewcommand{\arraystretch}{2.0}
	\setlength{\tabcolsep}{5pt}
	\resizebox{\textwidth}{!}{%
		\begin{tabular}{|c|c|c|c|c|}
			\hline
			\rowcolor{cyan!20}
			\multicolumn{5}{|c|}{\textbf{Observables $\rightarrow$  All Combined }} \\ 
			\hline
			\hline
			$\text{Scale}$  &  $V_L$  &  $V_R$  &  $g_L$  &  $g_R$  \\
			\hline
			$\mu_t$ & $(-1.23 \pm 0.81)\times10^{-2}$ & $(-2.72 \pm 5.43) \times 10^{-4}$ & $(2.45 \pm 6.30)\times 10^{-4}$ & $(5.16 \pm 3.94)\times 10^{-3}$\\
			\hline
			$\mu = \rm 1\, TeV $ & $(-1.29 \pm 0.85)\times10^{-2}$ & $(-2.94 \pm 5.86) \times 10^{-4}$ & $(2.65 \pm 6.82)\times 10^{-4}$ & $(5.60 \pm 4.27)\times 10^{-3}$\\
			\hline
			$\mu=\rm 5\, TeV$ & $(-1.34 \pm 0.88)\times10^{-2}$ & $(-3.12 \pm 6.23) \times 10^{-4}$ & $(2.82 \pm 7.26)\times 10^{-4}$ & $(5.96 \pm 4.55)\times 10^{-3}$\\
			\hline
			$\mu = \rm 10\, TeV $ & $(-1.36 \pm 0.90)\times10^{-2}$ & $(-3.19 \pm 6.37) \times 10^{-4}$ & $(2.88 \pm 7.44)\times 10^{-4}$ & $(6.11 \pm 4.67)\times 10^{-3}$\\
			\hline
		\end{tabular}%
	}
	\caption{Fit results of the anomalous couplings of $Wtb$ vertex from the combined analysis taking all the available observables that are discussed in the text. Predictions at different scales. }
	\label{tab:combined_one_param}
\end{table}

\subsection{Predictions for Branching Ratios of Top-Quark FCNC Decays}
\begin{figure}[t]
    \centering
    \subfloat[]{
    \begin{tikzpicture}
        \begin{feynman}
            \vertex[](a);
            \vertex[below left=1.7cm of a](c){$t$};
            \vertex[square dot, blue, below left=1.cm of a](d){};
            \vertex[above left=1.5cm of a](e){$u,c$};
            \vertex[above left=1.cm of a](f);
            \vertex[right=1.cm of a](g){$\gamma,Z,g$};
            \diagram*{
                (c) --[fermion,arrow size=1pt](d) --[fermion,arrow size=1pt, edge label'=\(b\)] (a) --[fermion,arrow size=1pt, edge label'=\(b\)] (f) --[fermion,arrow size=1pt] (e),
                (d) --[boson, edge label=\(W\)] (f),
                (a) --[boson] (g)
            };
        \end{feynman}
    \end{tikzpicture}}
    \hspace{0.001cm}
    \subfloat[]{\begin{tikzpicture}
        \begin{feynman}
            \vertex[](a);
            \vertex[below left=1.7cm of a](c){$t$};
            \vertex[square dot,blue,below left=1.cm of a](d){};
            \vertex[above left=1.5cm of a](e){$u,c$};
            \vertex[above left=1.cm of a](f);
            \vertex[right=1.cm of a](g){$\gamma,Z$};
            \diagram*{
                (c) --[fermion,arrow size=1pt](d) --[fermion,arrow size=1pt, edge label=\(b\)](f) --[fermion,arrow size=1pt](e),
                (a) --[boson](g),
                (a) --[boson, edge label=\(W\)](d), 
                (a) --[boson, edge label'=\(W\)](f)
            };
        \end{feynman}
    \end{tikzpicture}}
    \hspace{0.001cm}
    \subfloat[]{\begin{tikzpicture}
        \begin{feynman}
            \vertex[](a);
            \vertex[below left=1.7cm of a](c){$t$};
            \vertex[square dot,blue, below left=1.cm of a](d){};
            \vertex[above left=1.5cm of a](e){$u,c$};
            \vertex[below left=.4cm of a](f);
            \vertex[right=1.cm of a](g){$\gamma,Z,g$};
            \diagram*{
                (c) --[fermion,arrow size=1pt](d) --[fermion,arrow size=1pt, edge label'=\(b\)](f) --[fermion,arrow size=1pt, edge label'=\(u/c\)](a) --[fermion,arrow size=1pt](e),
                (a) --[boson](g),
                (d) --[boson, half left, looseness=2, edge label=\(W\)](f)
            };
        \end{feynman}
    \end{tikzpicture}}
    \\[1em] 
    \subfloat[]{
    \begin{tikzpicture}
        \begin{feynman}
            \vertex[](a);
            \vertex[below left=1.7cm of a](c){$t$};
            \vertex[square dot,blue,below left=1.cm of a](d){};
            \vertex[above left=1.6cm of a](e){$u,c$};
            \vertex[above left=1.cm of a](f);
            \vertex[right=1.cm of a](g){$H$};
            \diagram*{
                (c) --[fermion,arrow size=1pt](d) --[fermion,arrow size=1pt, edge label'=\(b\)] (a) --[fermion,arrow size=1pt, edge label'=\(b\)] (f) --[fermion,arrow size=1pt] (e),
                (d) --[boson, edge label=\(W\)] (f),
                (a) --[scalar] (g)
            };
        \end{feynman}
    \end{tikzpicture}}~~
    \hspace{0.001cm}
    \subfloat[]{\begin{tikzpicture}
        \begin{feynman}
            \vertex[](a);
            \vertex[below left=1.7cm of a](c){$t$};
            \vertex[square dot,blue, below left=1.cm of a](d){};
            \vertex[above left=1.6cm of a](e){$u,c$};
            \vertex[above left=1.cm of a](f);
            \vertex[right=1.cm of a](g){$H$};
            \diagram*{
                (c) --[fermion,arrow size=1pt](d) --[fermion,arrow size=1pt, edge label=\(b\)](f) --[fermion,arrow size=1pt](e),
                (a) --[scalar](g),
                (a) --[boson, edge label=\(W\)](d), 
                (a) --[boson, edge label'=\(W\)](f)
            };
        \end{feynman}
    \end{tikzpicture}}~~
    \hspace{0.001cm}
    \subfloat[]{\begin{tikzpicture}
        \begin{feynman}
            \vertex[](a);
            \vertex[below left=1.7cm of a](c){$t$};
            \vertex[square dot,blue,below left=1.cm of a](d){};
            \vertex[above left=1.4cm of a](e){$u,c$};
            \vertex[below left=.4cm of a](f);
            \vertex[right=1.0cm of a](g){$H$};
            \diagram*{
                (c) --[fermion,arrow size=1pt](d) --[fermion,arrow size=1pt, edge label'=\(b\)](f) --[fermion,arrow size=1pt, edge label'=\(u/c\)](a) --[fermion,arrow size=1pt](e),
                (a) --[scalar](g),
                (d) --[boson, half left, looseness=2, edge label=\(W\)](f)
            };
        \end{feynman}
    \end{tikzpicture}}
    \caption{Top-quark FCNC processes are modified by the SMEFT operators corresponding to effective $Wtb$ vertex.}
    \label{fig:Top_FCNC}
\end{figure}
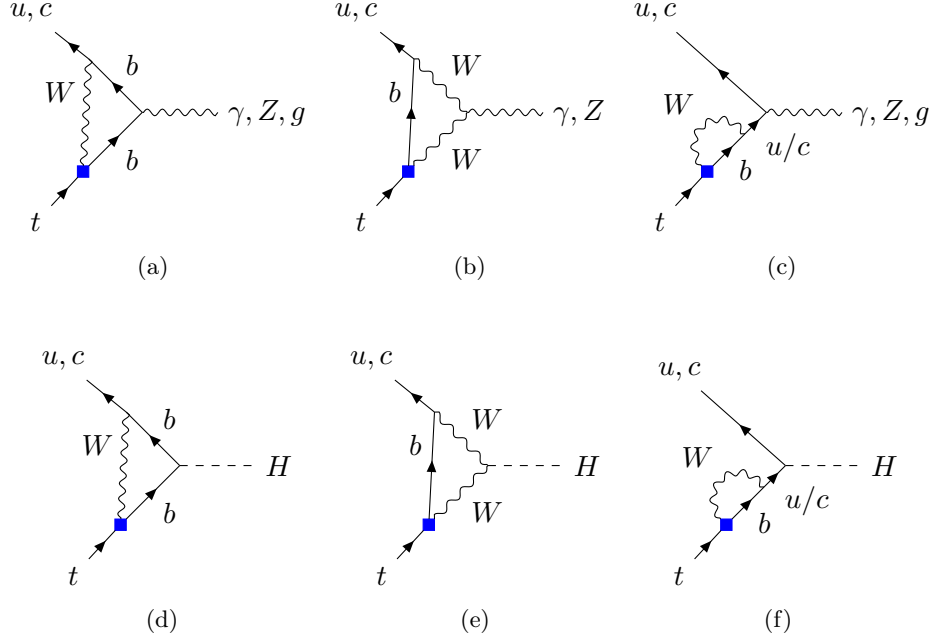

In the previous sections, we presented the bounds on the $Wtb$ anomalous couplings and their corresponding SMEFT couplings, derived separately and collectively from FCNC, FCCC, and EWPO constraints. Our results indicate that $V_{R}$ and $g_{L}$ are of the order $\mathcal{O}(10^{-4})$, while $V_{L}$ and $g_{R}$ are the order of $\mathcal{O}(10^{-2})$ and $\mathcal{O}(10^{-3})$ respectively.  FCNC decays of the top quark, such as $t \to u_{j} \gamma$, $t \to u_{j} Z$, $t \to u_{j} H$, and $t \to u_{j} g$ (where $u_{j} = u, c$), occur via one-loop diagrams in the SM and are highly suppressed due to the GIM mechanism. The experimental upper limits on these decay channels are several orders of magnitude higher than the SM predictions, as shown in table~\ref{tab:top_FCNC_prediction}. Consequently, these decay channels could be sensitive to new interactions beyond the SM \cite{Bhattacharya:2025mlg,Bhattacharya:2023beo}.
The $Wtb$ effective interaction we discuss above will contribute to these  FCNC decays of the top quark at one-loop level. We have shown the respective diagrams in fig.~\ref{fig:Top_FCNC}. In this section, we provide predictions for the branching fractions of all these channels, considering the bounds we have obtained on $Wtb$ couplings. 

 
The most general effective Lagrangian contribute to $t \to u_j X (=\gamma,g,H,Z)$ processes can be written as
\begin{align} \label{eq:top_FCNC_lag}
\mathcal{L}_{t\to u_j X} &=-\frac{1}{m_t}\bar{u}_j i\sigma_{\mu\nu}\epsilon^{\mu\,*}_{\gamma}k^{\nu}\left(A^{\gamma}_L P_L +A^{\gamma}_R P_R\right)t -  \frac{1}{m_t}\bar{u}_j i\sigma_{\mu\nu}T^a \epsilon^{\mu\,*}_{g}k^{\nu}\left(A^{g}_L P_L +A^{g}_R P_R\right)t \nonumber\\
& - \frac{1}{m_t}\bar{u}_j i\sigma_{\mu\nu}\epsilon^{\mu\,*}_{Z}k^{\nu}\left(A^{Z}_L P_L +A^{Z}_R P_R\right)t+ \bar{u}_j \gamma^{\mu} \left(B_L^Z P_L+B_R^Z P_R\right)t \,Z^{\mu} \\
& + \bar{u}_j\left(C_L^H P_L +C_R^H P_R\right)t\, H \,. \nonumber
\end{align}
Here $A, B  \, \,\text{and}\,\,C$ are the different tensor, vector, and scalar type form factors, respectively. Here, $\epsilon_V$ denotes the polarisation of a vector gauge boson. Using eq.~\eqref{eq:top_FCNC_lag} the branching ratio of $t \to u_i X $ with $X=(\gamma,g,H,Z)$ are then given by:
    \begin{align}
    \mathcal{B}(t\to u_j \gamma) &= \frac{m_t}{16 \pi \Gamma_t}\left(|A_L^{\gamma}|^2 +|A_R^{\gamma}|^2\right) \,,   \nonumber \\
    \mathcal{B}(t\to u_j g) &=\frac{m_t C_F}{16 \pi \Gamma_t}\left(|A_L^{g}|^2 +|A_R^{g}|^2\right) \,,  \nonumber  \\
    \mathcal{B}(t\to u_j Z) &= \frac{1}{32\pi m_t M_Z^2 \Gamma_t}\left(1-\frac{M_Z^2}{m_t^2}\right)^2 \Bigg[M_Z^2(2m_t^2+M_Z^2)\left(|A_L^Z|^2+|A_R^Z|^2\right)+(m_t^2+2 M_Z^2)\times  \,  \nonumber \\
    & m_t^2\left(|B_L^Z|^2+|B_R^Z|^2\right) +3 m_t^2 M_Z^2\left((A_L^Z B_R^{*\, Z}+ A_L^{*\, Z} B_R^Z)+(A_R^Z B_L^{*\, Z}+ A_R^{Z \,*} B_L^Z)\right)\Bigg] \,, \nonumber  \\
    \mathcal{B}(t\to u_j H) &= \frac{m_t}{32 \pi \Gamma_t}\left(1-\frac{m_H^2}{m_t^2}\right)^2 \left(|C_L^H|^2+|C_R^H|^2\right)  \,. 
\end{align}
We have used $\Gamma_t=1.42^{+0.19}_{-0.15}$ GeV \cite{ParticleDataGroup:2024cfk}, which is the width of the top quark and $C_F = 4/3 $, is the color factor. The diagrams given in fig.~\ref{fig:Top_FCNC} contribute to all the form factors $A_{L,R}^{\gamma,Z,g}$, $B_{L,R}^{Z}$ and $C_{L,R}^{H}$, respectively. Although we have given the most general form of the branching ratios, we have taken all the form factors to be real during the calculation.
The branching ratio for $t \to u_j \ell^+ \ell^-$ can also be extracted from the decay $t \to u_j X\,(\gamma, Z, H)$, where the intermediate boson $X$ subsequently decays into a lepton pair. The effective Lagrangian of $t \to u_{j} \ell \ell $ decay is given by for our case as:
\begin{align}\label{eq:tqll_pred}
    \mathcal{L}_{t\to u_j \ell^+ \ell^-}&= -\left(\bar{u}_j \gamma_{\mu}\left(D^V_L P_L +D^V_R P_R\right)t\right)\left(\bar{\ell}\gamma^{\mu}\ell\right)-\left(\bar{u}_j \gamma_{\mu}\left(D^A_L P_L +D^A_R P_R\right)t\right)\left(\bar{\ell}\gamma^{\mu}\gamma^5 \ell\right)\nonumber\\
&-\left(\bar{u}_j\left(E^S_L P_L+E^S_R P_R\right)t\right)\left(\bar{\ell}\,  \ell\right)\,.
\end{align}
Here, $D_{L(R)}^{V(A)}$ and $E^{S}_{L(R)}$ are the effective couplings of vector and scalar type. 
Since we don't get any new physics contribution to the lepton sector, for scalar interaction, the lepton sector remains purely scalar type.
For this effective basis, the differential decay rate can be written as (by neglecting the lepton mass)\footnote{In our numerical analysis, we have taken into account the effect of both the final state quark and the lepton masses.}:
\begin{align}
    \frac{d \Gamma(q^2)}{d q^2}&=\frac{q^2(m_t^2-q^2)}{256 \pi^3 m_t^3}\left(|E_L^S|^2+|E_R^S|^2+4\left(|D_L^V|^2+|D_R^V|^2+|D_L^A|^2+|D_R^A|^2\right)\right)\,.
\end{align}

\begin{table}[htb!]
\centering
\rowcolors{1}{cyan!10}{blue!20!green!5}
\renewcommand{\arraystretch}{1.4}
\setlength{\tabcolsep}{10pt}
\begin{tabular}{|l|c|c|c|}
\hline
\rowcolor{cyan!20}
\textbf{Branching Ratio} & \textbf{SM \cite{TopQuarkWorkingGroup:2013hxj}} & \textbf{Experimental bound}&\textbf{Our predictions} \\ 
\hline
\hline
BR$(t \to \gamma c)$  & $5 \times 10^{-14}$    & \shortstack{$4.16 \times 10^{-5}$ LH \\ $4.16 \times 10^{-5}$ RH} \cite{ATLAS:2022per}  & $(1.98 \pm 4.14) \times 10^{-11}$               \\ \hline

BR$(t \to \gamma u)$        & $4 \times 10^{-16}$     & \shortstack{$0.85 \times 10^{-5}$ LH \\ $1.2 \times 10^{-5}$ RH} \cite{ATLAS:2022per}  &    $(1.59 \pm 3.33) \times 10^{-13}$         \\ \hline

BR$(t \to Z c)$        & $1 \times 10^{-14}$     & \shortstack{$1.3 \times 10^{-4}$ LH \\ $1.2 \times 10^{-4}$ RH} \cite{ATLAS:2023qzr}   &  $(0.65 \pm 1.16)\times 10^{-10} $         \\ \hline

BR$(t \to Z u)$        & $7 \times 10^{-17}$     & \shortstack{$6.2 \times 10^{-5}$ LH \\ $6.6 \times 10^{-5}$ RH} \cite{ATLAS:2023qzr}    &  $(5.24 \pm 9.35) \times 10^{-13}$          \\ \hline

BR$(t \to H c)$        & $3 \times 10^{-15}$     & $4.3 \times 10^{-4}$ \cite{ATLAS:2023ujo}     & $(0.70\pm 1.26) \times 10^{-11}$           \\ \hline

BR$(t \to H u)$        & $2 \times 10^{-17}$     & $4.0 \times 10^{-4}$ \cite{ATLAS:2023ujo}    &  $(0.56 \pm 1.01) \times 
 10^{-13}$          \\ \hline
 
BR$(t \to g c)$        & $5 \times 10^{-12}$    & $3.7 \times 10^{-4}$ \cite{ATLAS:2021amo}    & $(1.17 \pm 2.32) \times 10^{-11}$           \\ \hline

BR$(t \to g u)$        & $4 \times 10^{-14}$    & $6.1 \times 10^{-5}$  \cite{ATLAS:2021amo}    &      $(0.95\pm 1.87)\times 10^{-13}$      \\ \hline
BR$(t \to c e^+ e^-)$ & $8.48 \times 10^{-15}$\,\,\cite{Frank:2006ku} & - & $(0.80 \pm 1.61)\times 10^{-12} $\\

BR$(t \to c \mu^+ \mu^-)$ & $9.55 \times 10^{-15}$ \,\cite{Frank:2006ku} & - & $(1.07 \pm 2.24)\times 10^{-12} $\\

BR$(t \to u e^+ e^-)$\tablefootnote{For SM, branching ratio for $t \to u \ell \ell $ can be calculated from $t \to c \ell \ell$, by just considering the CKM effect, please refer to the text.} &$6.81 \times 10^{-17}$\,\, & -  & $(0.64\pm 1.30)\times 10^{-14}$\\

BR$(t \to u \mu^+ \mu^-)$ & $7.68 \times 10^{-17}$ \, \, & - & $(3.17\pm 6.44)\times 10^{-14}$\\
\hline

BR$(t \to c \nu \bar{\nu})$ & $2.99 \times 10^{-14}$\,\,\cite{Frank:2006ku}& - & $(3.64 \pm 7.40)\times 10^{-12}$\\
BR$(t \to u \nu \bar{\nu})$ &  $2.40 \times 10^{-16}$ & - &$(2.93 \pm 5.95)\times 10^{-14}$\\
\hline
\end{tabular}
\caption{Comparison of SM and experimental bounds and our predictions of top-FCNC decays.}
\label{tab:top_FCNC_prediction}
\end{table}

Table~\ref{tab:top_FCNC_prediction} presents our predictions for various top FCNC decay channels alongside the corresponding SM values and experimental bounds. We calculated these values using our fit results of four-parameter scenarios. 
Experimental upper limits on these processes are of the order of $\mathcal{O}(10^{-5})$ to $\mathcal{O}(10^{-4})$, while the SM predictions are several orders smaller, ranging from $\mathcal{O}(10^{-12})$ to $\mathcal{O}(10^{-17})$. Our predictions are approximately $\mathcal{O}(10^3)$ times larger than the SM values; however, the experimental upper limits still remain several orders of magnitude higher than both. These limits are set due to the absence of observed events for such processes at the detector. With improved detector precision, the upper bounds are expected to decrease further. If future experiments detect such decays with branching ratios close to our predictions, it would indicate that the NP contribution originates from the anomalous $Wtb$ vertex. 

We have also predicted the branching ratios for the three-body top decays: $t \to u_{j} \ell \ell$. The experimental upper bounds of these decay processes are not available. 
The SM predictions for the branching ratio BR($t \to c \ell \ell$) are available in ref.~\cite{Frank:2006ku}. Although they have not provided the branching for $t \to u \ell \ell $, we can extract that from $t \to c \ell \ell$ by taking into account the CKM elements for $t \to u $ transition, as the two are related through CKM matrix elements. The ratio of decay width will be given by: $\Gamma_{t \to u \ell \ell }/ \Gamma_{t \to c \ell \ell } \propto  (V_{ub} / V_{cb} )^2$, while neglecting mass of the final state particles. We also have predicted the decay $t \to u_{j} \nu \bar{\nu}$. No experimental upper limit is available for these decay processes, and the SM prediction is of the order of $\sim \mathcal{O} (10^{-14})$. In our scenario, we get the branching ratio two orders higher than SM for both $t \to c \nu \bar{\nu}$ and $t \to u \nu \bar{\nu}$.

\paragraph{\underline{$Z(H) \to d_{i} \bar{d}_{j}$ Decays}:}
The anomalous $Wtb$ couplings can contribute to FCNC decays of $Z$ and the Higgs boson. 
In the SM, the decays $Z(H) \to d_{i} \bar{d}_{i}$ occur at tree level, while the flavour-changing decays $Z(H) \to d_{i} d_{j}$ (for $i \neq j$) proceed via one-loop diagrams, similar to the one given in fig.~\ref{fig:Top_FCNC}. Although no bounds from experiment are currently available, the SM predictions for $Z(H) \to b \bar{s} (\bar{d})$ are of the order $\mathcal{O}(10^{-7}) \left(\mathcal{O}(10^{-9})\right)$ \cite{Aranda:2020tqw, Eilam:2002as, PhysRevLett.57.1514}. While no events have been observed in colliders, the best upper limit from LEP and SLD experiments \cite{Fuster:1999, Atwood:2002ke} is:  
\begin{equation*}
 \mathcal{B} (Z \to b \bar{s} (\bar{d}) ) \leq 1.8 \times 10^{-3} \quad (\text{at 90\% CL})\,.
\end{equation*}
Using the fit results from table~\ref{tab:combined_one_param}, we predicted the branching ratios for $Z(H) \to b \bar{s}(\bar{d})$. However, our predictions for new physics contributions are very small and indistinguishable from the SM expectations; thus, we do not present the results here.

\section{Summary}\label{sec:summary}
In this work, we consider the effect of heavy NP contributions on the $Wtb$ vertex through the effective couplings $V_{L}$, $V_{R}$, $g_{L}$, and $g_{R}$. While the SM includes only the left-handed vector-type coupling, the extended Lagrangian introduces right-handed vector-type and tensor-type currents for both chiralities. In the dim-6 SMEFT basis, these couplings arise at tree level from the operators $ \mathcal{O}^{\phi q(3)}_{33}$, and $\mathcal{O}^{\phi u d}_{33}\,, \mathcal{O}^{*dW}_{33} $ and $ \mathcal{O}^{uW}_{33}$.

These SMEFT operators contribute to various FCNC, FCCC, and electroweak observables at the one-loop level. Processes involving top and bottom quarks, either as external or loop particles, are affected by this effective vertex. For FCNCs, we consider processes such as $B^{0}_{(s)}-\bar{B}^{0}_{(s)}$ meson mixing, rare decays like $B^{0}_{(s)} \to \mu^{+} \mu^{-}$, radiative decays with the quark-level transition $b \to s(d) \gamma(g)$, and all observables related to $b \to s \ell \ell$, including branching ratios, angular observables, and ratio observables like $R_{K^{(*)}}$. Invisible processes such as $B \to K^{(\ast)} \nu \bar{\nu}$ are also considered.
For FCCCs, we include semileptonic and leptonic decays of the type $P \to M \ell \nu$ and $P \to \ell \nu$. We perform a CKM fit considering all available observables. Beyond low-energy FCNC and FCCC observables, we analyse electroweak observables such as $W$ and $Z$-pole observables. The anomalous $Wtb$ couplings also contribute to TGCs and $H \to b \bar{b}$ decays.

We have obtained bounds on the SMEFT couplings defined at the scale $\mu_b$ using the available low-energy data on FCNC and FCCC processes. Furthermore, we have obtained bounds on similar couplings defined at the electroweak scale $\mu_{EW }$ using the data on electroweak precision observables. A RGE connects the couplings at different scales. We have discussed the role of different sets of observables in constraining the SMEFT couplings and finally obtained the bounds on these couplings from the combined analysis of all the data. We obtain the most up-to-date tight constraints on the relevant SMEFT couplings. Also, we have shown the variations of those couplings with the scale of new physics $\Lambda$ up to 10 TeV. Using the bounds on the SMEFT couplings, we obtain the bounds on the anomalous $Wtb$ anomalous couplings, which are orders of magnitude tighter than those obtained by the ATLAS and CMS collaborations.

Using the extracted values of the SMEFT couplings $\frac{v^2}{\Lambda^2}\mathcal{C}^{\phi q(3)}_{33}$, $\frac{v^2}{\Lambda^2}\mathcal{C}^{\phi u d}_{33}$,  $\frac{v^2}{\Lambda^2}\mathcal{C}^{* dW}_{33} $ and $\frac{v^2}{\Lambda^2}\mathcal{C}^{ uW}_{33}$, we have predicted the branching ratios for top-FCNC processes such as $t \to u(c) \gamma$, $t \to u(c) Z$, and $t \to u(c) H$. Notably, the SM predictions and experimental bounds for these processes differ by several orders of magnitude. Our predictions exceed the SM values by a few orders but remain within the reach of future colliders.
We have also calculated the branching ratios for $t \to c(u) \ell \ell$ and $t \to c(u) \nu \bar{\nu}$ processes, which can be derived from the decays $t \to u(c) Z$, $t \to u(c) \gamma$, and $t \to u(c) H$, as no NP effects are introduced at the lepton vertex.
We also examine NP contributions to Higgs and $Z$ boson FCNCs, such as $H \to b s(d)$ and $Z \to b s(d)$. However, the NP effects are too small to be distinguished from the SM predictions.
\appendix
\section{RGEs : Beta Functions}\label{appndx:RGE_beta}
In sec.~\ref{sec:methodology}, we discussed the scale evolution methodology for SMEFT and LEFT Wilson coefficients. The evolution begins from $\mu_{\Lambda} $ to $\mu_{\rm EW}$, followed by further evolution to $\mu_{b}$. The beta functions corresponding to the SMEFT Wilson coefficients are given below:
\begin{subequations}\label{eq:beta_lambda2EW}
\begin{align}
    \Bigg[\beta^{uW}\Bigg]_{rs}=&-\frac{1}{36}\left(33 g_2^2+19g_1^2-96g_s^2\right)\Bigg[\frac{\mathcal{C}^{uW}}{\Lambda^2}\Bigg]_{rs}+2\Bigg[\Gamma_d\Gamma_d^{\dagger}\frac{\mathcal{C}^{uW}}{\Lambda^2}\Bigg]_{rs}+\Bigg[\frac{\mathcal{C}^{uW}}{\Lambda^2}\Gamma_u^{\dagger}\Gamma_u \Bigg]_{rs}  \nonumber\\
&+\gamma_{H}^{(Y)}\Bigg[\frac{\mathcal{C}^{uW}}{\Lambda^2}\Bigg]_{rs}+\Bigg[\gamma_q^{(Y)}\frac{\mathcal{C}^{uW}}{\Lambda^2}\Bigg]_{rs}+\Bigg[\frac{\mathcal{C}^{uW}}{\Lambda^2} \gamma_{u}^{(Y)}\Bigg]_{rs}-\Bigg[\frac{\mathcal{C}^{dW}}{\Lambda^2} \Gamma_d^{\dagger} \Gamma_{u}\Bigg]_{rs} \,, 
\end{align}
\begin{align}
    \Bigg[\beta^{dW}\Bigg]_{rs}=&-\frac{1}{36}\left(33 g_2^2+31g_1^2-96g_s^2\right)\Bigg[\frac{\mathcal{C}^{dW}}{\Lambda^2}\Bigg]_{rs}+2\Bigg[\Gamma_u\Gamma_u^{\dagger}\frac{\mathcal{C}^{dW}}{\Lambda^2}\Bigg]_{rs}+\Bigg[\frac{\mathcal{C}^{dW}}{\Lambda^2}\Gamma_d^{\dagger}\Gamma_d \Bigg]_{rs}\nonumber\\
&+\gamma_{H}^{(Y)}\Bigg[\frac{\mathcal{C}^{dW}}{\Lambda^2}\Bigg]_{rs}+\Bigg[\gamma_q^{(Y)}\frac{\mathcal{C}^{dW}}{\Lambda^2}\Bigg]_{rs}+\Bigg[\frac{\mathcal{C}^{dW}}{\Lambda^2} \gamma_{d}^{(Y)}\Bigg]_{rs}-\Bigg[\frac{\mathcal{C}^{uW}}{\Lambda^2} \Gamma_u^{\dagger} \Gamma_{d}\Bigg]_{rs} \,, \nonumber \\
\\
    \Bigg[\beta^{\phi u d}\Bigg]_{rs} =& -3 g_1^2\Bigg[\frac{\mathcal{C}^{\phi u d}}{\Lambda^2}\Bigg]_{rs}+ 2\Bigg[\Gamma_u^{\dagger} \Gamma_u \frac{\mathcal{C}^{\phi u d}}{\Lambda^2}\Bigg]_{rs}+ 2 \Bigg[\frac{\mathcal{C}^{\phi u d}}{\Lambda^2} \Gamma_d^{\dagger} \Gamma_d \Bigg]_{rs}+ 2\gamma_H^{(Y)}\Bigg[\frac{\mathcal{C}^{\phi u d}}{\Lambda^2}\Bigg]_{rs}\nonumber\\
    & +\Bigg[\gamma_u^{(Y)} \frac{\mathcal{C}^{\phi u d}}{\Lambda^2}\Bigg]_{rs}+\Bigg[\frac{\mathcal{C}^{\phi u d}}{\Lambda^2}\gamma_d^{(Y)}\Bigg]_{rs}\,, \nonumber \\
    \\
    \Bigg[\beta^{\phi q (3)}\Bigg]_{rs} =& +\frac{1}{2}\left(\Bigg[\Gamma_d \Gamma_d^{\dagger} \frac{\mathcal{C}^{\phi q (3)}}{\Lambda^2}\Bigg]_{rs}+\Bigg[\Gamma_u \Gamma_u^{\dagger} \frac{\mathcal{C}^{\phi q (3)}}{\Lambda^2}\Bigg]_{rs}+ \Bigg[\frac{\mathcal{C}^{\phi q(3)}}{\Lambda^2} \Gamma_d \Gamma_d^{\dagger}\Bigg]_{rs}
    + \Bigg[\frac{\mathcal{C}^{\phi q (3)}}{\Lambda^2} \Gamma_u \Gamma_u^{\dagger}\Bigg]_{rs} \right)\nonumber\\
    &-\frac{17}{3} g_2^2 \Bigg[\frac{\mathcal{C}^{\phi q (3)}}{\Lambda^2}\Bigg]_{rs}+ 2\gamma_H^{(Y)}\Bigg[\frac{\mathcal{C}^{\phi q (3)}}{\Lambda^2}\Bigg]_{rs}+\Bigg[\gamma_q^{(Y)}\frac{\mathcal{C}^{\phi q (3)}}{\Lambda^2}\Bigg]_{rs}+\Bigg[\frac{\mathcal{C}^{\phi q (3)}}{\Lambda^2} \gamma_q^{(Y)}\Bigg]_{rs} \,.
\end{align}
\end{subequations}
However, in SMEFT, there exist operators that can mix with our chosen set; we have verified that their contributions are subleading and can therefore be safely neglected.
 In eq.~\eqref{eq:beta_lambda2EW} wavefunction renormalisation terms are
\begin{eqnarray}
    &\gamma_H^{(Y)}=\rm{Tr}\left(3\Gamma_u\Gamma_u ^{\dagger}+3\Gamma_d \Gamma_d^{\dagger}+\Gamma_e\Gamma_e^{\dagger}\right) \,,& \nonumber\\
    &\Bigg[\gamma_q^{(Y)}\Bigg]_{rs} =\frac{1}{2}\Bigg[\Gamma_u \Gamma_u^{\dagger}+\Gamma_d \Gamma_d^{\dagger}\Bigg] \,,& \nonumber\\
    &\Bigg[\gamma_u^{(Y)}\Bigg]_{rs} =\Bigg[\Gamma_u^{\dagger} \Gamma_u\Bigg]_{rs}\,, & \\
   & \Bigg[\gamma_d^{(Y)}\Bigg]_{rs} =\Bigg[\Gamma_d^{\dagger} \Gamma_d\Bigg]_{rs}\,,& \nonumber\\
    &\Bigg[\gamma_e^{(Y)}\Bigg]_{rs} = \Bigg[ \Gamma_e^{\dagger} \Gamma_e\Bigg]_{rs}\,.& \nonumber
\end{eqnarray}
Here, $\Gamma_u$ and $\Gamma_d$ represent the Yukawa matrices; in our notation, all right-handed fields are in the mass basis, and left-handed lepton doublet is in the mass basis of charged leptons, whereas left-handed quark doublet is defined in arbitrary flavour basis. In this basis, we can express the Yukawa matrices as:
\begin{align}
    \Gamma_u &=U_{u_L}\Gamma_u^{\rm diag},&  \Gamma_d&=U_{d_L}\Gamma^{\rm diag}_d & \Gamma_q&=\begin{pmatrix}
        U_{u_L}^{\dagger} u_L\\
        U_{d_L}^{\dagger} d_L
    \end{pmatrix}
    \label{eq:yukawa_arbitary}
\end{align}
 We can perform a unitary transformation in the flavour space of the quark doublet that turns the doublet component into a mass basis. Performing a rotation in the down quark basis of the eq.~\eqref{eq:yukawa_arbitary}. 
 \begin{align}
     q\to U_{d_L} q \,,
 \end{align}
 we end up in with, so-called $down\, aligned \,basis$,
 \begin{align}
     \Gamma_u&=V_{\rm CKM}^{\dagger} \Gamma_u^{\rm diag} & \Gamma_d &=\Gamma_d^{\rm diag} & \Gamma_q&=\begin{pmatrix}
         V_{\rm CKM}^{\dagger} u_L\\
         d_L
     \end{pmatrix}
 \end{align}
 where $V_{\rm CKM}= U_{u_L}^{\dagger}U_{d_L}$ is the CKM matrix. In our work, we have used this down-aligned basis, where the down-type Yukawa matrix is diagonal, and components of the down-type quark doublet are written in the mass basis and as mentioned earlier, no mixing in the lepton sector is considered, $\Gamma_e=\Gamma_e^{\rm diag}$. The values of the following SM parameters are used for running at the EW scale: $g_2=0.6515$, $g_1=0.3576$, $g_s=1.220$, and $\lambda=0.2813$.
We also have taken into account the scale dependency of the SM couplings $g_{1}$, $g_{2}$ and $g_{s}$. Corresponding $\beta$ functions are given below:

\begin{eqnarray}
        &\left[\beta_{g_2}\right] =-\frac{19}{6}g_2^3\,,& \nonumber\\ 
        & \left[\beta_{g_1}\right] = \frac{41}{6}g_1^3\,,&\nonumber\\ 
    &\left[\beta_{g_s}\right]= -7 g_s^3 &\\
&\Bigg[\beta_{\Gamma_d}\Bigg]_{rs} =\frac{3}{2}\left(\Bigg[\Gamma_d \Gamma_d^{\dagger} \Gamma_d\Bigg]_{rs}-\Bigg[\Gamma_u \Gamma_u^{\dagger} \Gamma_d\Bigg]_{rs}\right)+\left(\gamma_H^{(Y)}-\frac{9}{4}g_2^2-\frac{5}{12}g_1^2-8g_s^2\right)\Bigg[\Gamma_d\Bigg]_{rs}-\Bigg[\Gamma_u C_{\phi u d}\Bigg]_{rs} \,, &\nonumber\\
&\Bigg[\beta_{\Gamma_u}\Bigg]_{rs} =\frac{3}{2}\left(\Bigg[\Gamma_u \Gamma_u^{\dagger} \Gamma_u\Bigg]_{rs}-\Bigg[\Gamma_d \Gamma_d^{\dagger} \Gamma_u\Bigg]_{rs}\right)+\left(\gamma_H^{(Y)}-\frac{9}{4}g_2^2-\frac{17}{12}g_1^2-8g_s^2\right)\Bigg[\Gamma_u\Bigg]_{rs}-\Bigg[\Gamma_d C_{\phi u d}^{\dagger}\Bigg]_{rs}\,.\nonumber&
\end{eqnarray}

\section{Loop Contributions for FCNC processes}
In this section, we provide all the loop amplitudes associated with the FCNC processes, along with their matching expressions in the traditional LEFT basis. The loop factors are given as a function of the scale $\mu$. They can be subsequently evolved to connect with low-energy observables via RGEs, as discussed earlier.
\paragraph{\underline{Loop Factor of Meson mixing}:} \label{appndx:mixing_loop}
The loop factor obtained from fig.~\ref{fig:meson_mixing} can be given by:
{\footnotesize 
\begin{align}\label{eq:meson_mixing_loop}
    C_{V_{L}} (\mu)&=\sum_{d_j}\frac{g_2^4 V_{tb} V_{td_j}^{*\,2}}{128\pi^2 M_W^2}\Bigg[\left(6\frac{m_t}{M_W}g_R(\mu)+4(x_t-3)V_L(\mu)\right)\log\left(\frac{\mu^2}{m_t^2}\right) +\frac{6\left(\frac{m_t}{M_W}(3 x_t-1)g_R(\mu)-2(2x_t-1)V_L(\mu)\right)}{(x_t-1)^3}\log x_t\nonumber\\
    &+\frac{\left(-\frac{m_t}{M_W}(x_t^2+10 x_t+1)g_R(\mu)+2(2 x_t^3-3 x_t^2+9 x_t-2)V_L(\mu)\right)}{(x_t-1)^2}\Bigg]\nonumber\\
    &+\sum_{u_i,d_j}\frac{g^4 V_{td_j} V_{u_i d_j}^* V_{u_i b}^*}{128 \pi^2 M_W^2}\Bigg[+2\left(3 \frac{m_t}{M_W}g_R(\mu)+(x_u+x_t-6)V_L(\mu)\right)\log\left(\frac{\mu^2}{M_W^2}\right)\nonumber\\
    &+2\frac{\left(3 x_t^3 \frac{m_t}{M_W}g_R(\mu)+\left(x_t^2(x_t^2-8 x_t+4)\right)V_L(\mu)\right)}{(x_{u_i}-x_t)(x_t-1)^2}\log x_t-2\frac{\left(3 x_{u_i}^3 \frac{m_t}{M_W}g_R(\mu)+\left(x_{u_i}^2(x_{u_i}^2-8 x_{u_i}+4)\right)V_L(\mu)\right)}{(x_{u_i}-x_t)(x_t-1)^2}\log x_{u_i}\nonumber\\
    &-\frac{\frac{m_t}{M_W}\left(5 x_t-5 x_{u_i}(x_t-1)+1\right)g_R(\mu)+\left(3 x_t^2-3 x_{u_i}^2(x_t-1)+x_{u_i}(-3x_t^2+16 x_t-13)-13 x_t+4\right)V_L(\mu)}{(x_{u_i}-1)(x_t-1)}\Bigg]\,.
\end{align}}
Here, $d_j = (s,d)$ depending on whether we are calculating for $B_{s}^{0} - \bar{B}_{s}^{0}$ or $B^{0} - \bar{B}^{0}$ process, respectively. 
The fields $u_i$ denote the up-type quarks ($u$, $c$), respectively.
The quantities $x_t$ and $x_{u_i}$ are defined as $x_t = \frac{m_t^2}{M_W^2}$ and $x_{u_i} = \frac{m_{u_i}^2}{M_W^2}$. Contribution to the right-handed current will vanish upon neglecting the external particles' mass. Hence, we haven't given their expression explicitly. 
\paragraph{\underline{Loop Factors for $b \to s \ell \ell $ processes and Radiative decays}:}\label{appndx:Loop_b2sll}
In secs.~\ref{sec:radiative} and \ref{sec:b2sll}, we have shown the contributions of the diagrams in fig.~\ref{fig:FCNC_feynman}, in terms of the loop functions $C_{T_{I}}\,, C_{V_{IJ}} $, and $C_{S_{IJ}}$, with $\{I,J\} = \{L,R\}$, corresponding to the left and right chirality. For simplicity, we redefine some of the previously mentioned couplings:
\begin{align}
C_{V_{I,J}}&=C_{V_{I,J}}^{\gamma}+C_{V_{I,J}}^Z +C_{V_{I,J}}^{\rm box}\,.
\end{align}
The couplings $C_{V_{I,J}}^\gamma$ and $C_{V_{I,J}}^Z$ represent vector-type couplings that appear in semileptonic decays, with the superscripts $\gamma$ and $Z$ referring to the mediators of the semileptonic decay. Since the scalar-type coupling $C_{S_{I,J}}$ is associated with semileptonic decays mediated exclusively by the Higgs boson, we adhere to the previously used notation.

{\footnotesize 
\begin{align}\label{eq:radiative_gamma_CTR}
    C_{T_{R}}^{\gamma}(\mu) & = \frac{i g_1 g_2^3 V_{ts}^*}{16\pi^2 \sqrt{g_1^2+g_2^2}M_W} \Bigg[-\frac{(3 x_t +10) g_L (\mu)+24 \frac{m_t}{M_W}V_R(\mu)}{24 }\log \left(\frac{\mu^2}{M_W^2} \right)\nonumber\\
    &+\frac{\left(15 x_t^3 + 26 x_t^2+ 25 x_t+6\right)g_L(\mu)+\frac{m_t}{M_W}\left(60 x_t^2 -28 x_t+104\right)V_R(\mu)}{48(x_t-1)^2}\nonumber\\
    &+ \frac{x_t^2(3 x_t^2 +x_t +32) g_L(\mu)+ 4\frac{m_t}{M_W}(6 x_t^2 -9 x_t +20)V_R(\mu)}{24(x_t-1)^3}\log x_t
    \Bigg]\,,
    \end{align}
    \begin{align}\label{eq:radiative_gluon_CTR}
    C_{T_{R}}^g(\mu) &= \frac{i g_2^2 g_s T^a V_{ts}^*}{16\pi^2 M_W}\Bigg[-
    \frac{1}{4}\log\left(\frac{\mu^2}{M_W^2}\right)g_L(\mu)+\frac{\left(x_t^2-4 x_t+3\right)g_L(\mu)-8\frac{m_t}{M_W}\left(x_t+1\right)V_R(\mu)}{8(x_t-1)^2} \nonumber\\
    &+\frac{\left(x_t^2(x_t-1)\right)g_L(\mu)+ 8\frac{m_t}{M_W}x_t V_R(\mu)}{4(x_t-1)^3}\log{x_t}\Bigg] \,,
    \end{align}
    \begin{align}\label{eq:loop_b2sll_gamma_med}
    C_{V_{L(L,R)}}^{\gamma}(\mu) &=\frac{1}{2} \frac{g_1 g_2^3 e V_{ts}^*}{16 \pi^2 M_W^2}\Bigg[-\frac{\frac{m_t}{M_W}\left(g_R(\mu) -\frac{m_t}{M_W}V_L(\mu)\right)}{8 }\log\left(\frac{\mu^2}{M_W^2}\right)\nonumber\\
    & + \frac{3\frac{m_t}{M_W}\left(3 x_t^4+11 x_t^2 -16 x_t +8\right)g_R(\mu)-\left(9 x_t^5-36 x_t^4 +19 x_t^3 +36 x_t^2-72 x_t +16\right)V_L(\mu)}{72(x_t-1)^4}\log x_t\nonumber\\
    &+\frac{9 \frac{m_t}{M_W}(9 x_t^3+17 x_t^2-31 x_t +17)g_R(\mu)+\left(-81 x_t^4 + 255 x_t^3 -173 x_t^2 +(112+43 x_t)\right)V_L(\mu)}{432 (x_t-1)^3}\Bigg]\,,
    \end{align}
    \begin{align}\label{eq:loop_b2sll_Z_med}
    C_{V_{L(L,R)}}^Z(\mu)& =\frac{1}{2}\frac{g_2^2 a_{L,(R)}^f V_{ts}^*}{16 \pi^2 M_Z^2 \sqrt{g_1^2+g_2^2}}\Bigg[\frac{\frac{m_t}{M_W}\left(3 g_1^2- 9 (x_t-2) g_2^2\right)g_R(\mu)-\left((g_1^2-3g_2^2)(2-3x_t)\right)V_L(\mu)}{24}\log\left(\frac{\mu^2}{M_W^2}\right)\nonumber\\&+\frac{\frac{m_t}{M_W}\left(3 g_2^2 (5 x_t^2 -9 x_t -2)+g_1^2 (x_t+5)\right)g_R(\mu) 
    + \left(3 g_2^2 (5 x_t^2 +x_t-8)+g_1^2(-5 x_t^2+31 x_t+8)\right)V_L(\mu)}{48 (x_t-1)} \nonumber\\
    &+\frac{3x_t\frac{m_t}{M_W}\left((2-x_t^2)g_1^2+3(x_t^2-4x_t+2)g_2^2\right)g_R(\mu)-x_t\left((g_1^2-3 g_2^2)(3x_t^2-8 x_t-12)\right)V_L(\mu)}{24(x_t-1)^2}\log x_t
    \Bigg] \,,
    \end{align}
    \begin{align}\label{eq:loop_b2sll_box}
      C_{V_{LL}}^{\rm box}(\mu) &= \frac{g_2^4 V_{ts}^*}{16 \pi^2 M_W^2}\Bigg[\frac{-3\frac{m_t}{M_W}g_R(\mu)+(6-x_t)V_L(\mu)}{16}\log\left(\frac{\mu^2}{M_W^2}\right)-\frac{\frac{m_t}{M_W}(5 x_t^2+1)g_R(\mu)+(3x_t^2-13 x_t+4)V_L(\mu)}{32 (x_t-1)}\nonumber\\
    &+\frac{3\frac{m_t}{M_W} x_t^2 g_R(\mu)+ (x_t^2 -8 x_t +4)x_t V_L(\mu)}{16 (x_t-1)^2} \log x_t\Bigg] \,,
\end{align}
\begin{align}\label{eq:loop_b2sll_scalar_left}
    C_{S_{R(L,R)}}(\mu) &= \frac{1}{2}\frac{g_2^3 V_{ts}^*}{16\pi^2 }\frac{m_{\ell}}{v m_H^2}\Bigg[\frac{-12 g_L(\mu)-4\frac{m_t}{M_W} (x_t-2)V_R(\mu)}{8}\log\left(\frac{\mu^2}{M_W^2}\right)-\frac{\frac{m_t}{M_W}(x_t^2+2 x_t+4)V_R(\mu) + (2+x_t)g_L(\mu)}{2(x_t-1)}\nonumber\\
    &+\frac{\frac{m_t}{M_W}x_t(x_t^2-4x_t+10)V_R(\mu)+3 x_t^2 g_L(\mu)}{2(x_t-1)^2}\log x_t-\frac{\frac{m_t}{m_b}\left((x_t+1)g_L(\mu)+\frac{m_t}{M_W}(x_t-1)V_R(\mu)\right)}{4}\nonumber\\
    &-\frac{\frac{m_t}{m_b}\left(3(x_t+1)g_L(\mu)+\frac{m_t}{M_W}(x_t-3)V_R(\mu)\right)}{4}\log\left(\frac{\mu^2}{M_W^2}\right)+\frac{\frac{m_t}{m_b}\left(x_t(x_t-4)V_R(\mu)+3 x_t^2 g_L(\mu)\right)}{4(x_t-1)}\log x_t\Bigg] \,, 
    \end{align}
    \begin{align}\label{eq:loop_b2sll_scalar_right}
    C_{S_{L(L,R)}}(\mu) &= \frac{1}{2}\frac{g_2^3 V_{ts}^*}{16\pi^2} \frac{m_{\ell}}{v \, m_H^2}\Bigg[\frac{3 \left(g_R(\mu) +\frac{m_t}{M_W}V_L(\mu)\right)}{8} \log \left(\frac{\mu^2}{M_W^2}\right)+\frac{\left(x_t(x_t+5)g_R(\mu)+\frac{m_t}{M_W}(-5 x_t^2 +5 x_t+6)V_L(\mu)\right)}{16(x_t-1)}\nonumber\\
    &+\frac{3x_t^2(x_t-2)\left(g_R(\mu)+\frac{m_t}{M_W}V_L(\mu)\right)}{8(x_t-1)^2}\log x_t-\frac{\frac{m_t}{m_b}\left((x_t+1)g_L(\mu)+\frac{m_t}{M_W}(1-x_t)V_R(\mu)\right)}{4}\nonumber\\
    &-\frac{\frac{m_t}{m_b}\left(3(x_t+1)g_L(\mu)+\frac{m_t}{M_W}(1-3x_t)V_R(\mu)\right)}{4}\log\left(\frac{\mu^2}{M_W^2}\right) + \frac{\frac{m_t}{m_b}\left(-x_t(x_t-4)V_R(\mu)+ 3 x_t^2 g_L(\mu)\right)}{4(x_t-1)}\log x_t\Bigg] \,.
\end{align}}
The couplings of $Z$ boson with fermions are defined as $a^f_L = \frac{g_2}{\cos\theta}(Q_f \sin^2\theta - I^3)$ and $a^f_R = \frac{g_2}{\cos\theta} Q_f \sin^2\theta$, where $I^3$ and $Q_f$ represent the third component of the isospin and the charge of the fermion, respectively. Additionally, we define $x_t = \frac{m_t^2}{M_W^2}$. $m_{\ell} $ is corresponding to the mass of the lepton $\ell$ for the decay $b \to s \ell \ell$. 
 
\paragraph{\underline{Loop Factors of the invisible decay $b \to s(d) \nu \bar{\nu}$}:}\label{appndx:invisible}
The effect of $Wtb$ anomalous coupling on invisible decay is discussed in sec.~\ref{sec:invisible_decays}. 
The loop factor obtained from fig.~\ref{fig:invisible} will contribute to both left- and right-handed currents. Upon neglecting the mass of the external legs, we only have contributions to the left-handed operator, which is given below: 
{\footnotesize
\begin{align}\label{eq:loop_invisible}
    C_{L}^{\nu, \rm NP} (\mu) &=-\frac{1}{g_1^2 V_{tb}}\Bigg[\frac{\frac{m_t}{M_W}\left(3 g_1^2- 9 (x_t-2) g_2^2\right)g_R(\mu)-\left((g_1^2-3g_2^2)(2-3x_t)\right)V_L(\mu)}{24}\log\left(\frac{\mu^2}{M_W^2}\right)\nonumber\\&+\frac{\frac{m_t}{M_W}\left(3 g_2^2 (5 x_t^2 -9 x_t -2)+g_1^2 (x_t+5)\right)g_R(\mu) 
    + \left(3 g_2^2 (5 x_t^2 +x_t-8)+g_1^2(-5 x_t^2+31 x_t+8)\right)V_L(\mu)}{48 (x_t-1)} \nonumber\\
    &+\frac{3x_t\frac{m_t}{M_W}\left((2-x_t^2)g_1^2+3(x_t^2-4x_t+2)g_2^2\right)g_R(\mu)-x_t\left((g_1^2-3 g_2^2)(3x_t^2-8 x_t-12)\right)V_L(\mu)}{24(x_t-1)^2}\log x_t
    \Bigg] \nonumber\\
    &+ \frac{1}{V_{tb}}\left(1+\frac{g_2^2}{g_1^2}\right)\Bigg[\frac{-3\frac{m_t}{M_W}g_R(\mu)+(6-x_t)V_L(\mu)}{8}\log\left(\frac{\mu^2}{M_W^2}\right)-\frac{\frac{m_t}{M_W}(5 x_t^2+1)g_R(\mu)+(3x_t^2-13 x_t+4)V_L(\mu)}{16 (x_t-1)}\nonumber\\
    &+\frac{3\frac{m_t}{M_W} x_t^2 g_R+ (x_t^2 -8 x_t +4)x_t V_L(\mu)}{8 (x_t-1)^2} \log x_t\Bigg] \,, \\
    C_{R}^{\nu, \rm NP} (\mu) &= 0 \,. 
\end{align}}

\section{Loop Contributions for FCCC Processes:}\label{appndx:FCCC}
\paragraph{\underline{Wavefunction Renormalisation of the $b$ quark}:}
Unlike the FCNC process, in FCCC (and EWPOs), the processes which include $b$ quark as an external leg will be modified by the $Wtb$ effective couplings according to fig.~\ref{fig:FCCC_conter-term}. 
To compute the renormalisation matrix for the one-loop correction to the $b$-quark leg, we follow the renormalisation procedure described in ref.~\cite{Denner:1991kt}.
\begin{align}\label{eq:b_WF}
    b_{\rm ren}&=Z_{ii}^{1/2\,,L(R)} b=\left(\mathbb{1}+\frac{1}{2}\delta Z_{ii}^{L(R)}\right)b \,, \\
    \delta Z_{ii}^{L(R)}&=-\text{Re}\Sigma_{ii}^{L(R)}(m_b^2)-m_b^2\frac{\partial^2}{\partial p^2} \text{Re}\Bigg[\Sigma_{ii}^{L}(p^2)+\Sigma_{ii}^{R}(p^2)+2\Sigma_{ii}^{S}(p^2)\Bigg]\Bigg|_{p^2=m_b^2} \,. 
\end{align}
Here, $\Sigma_{ii}^{L,R,S}$ denote the different renormalisation matrices that arise from the correction of the $b$ leg. Further details can be found in ref.~\cite{Denner:1991kt}.
{\footnotesize
\begin{align}
    \Sigma_{ii}^L(p^2,\mu) &=\frac{g_2^2 V_{tb}}{32 \pi^2 M_W^2}\Bigg[-\left(3 m_t M_W g_R(\mu)+(p^2-3 m_t^2)V_L(\mu)\right)\log\left(\frac{\mu^2}{m_t^2}\right)\nonumber\\
     &+\frac{\left(m_t M_W(3 (m_t^2-M_W^2)-4 p^2)g_R(\mu)-(2 p^4-M_W^2 p^2(1+5 x_t)+M_W^4(x_t^2+x_t-2)\right)}{p^2}\nonumber\\
    &-\frac{\left(3 m_t M_W(m_t^4+M_W^4+p^4-2 m_t^2(M_W^2+p^2)\right)g_R(\mu)}{2 p^2} \log\left(x_t\right)\\
    &-\frac{\left(-m_t^6+p^6-3x_t M_W^2 p^2+M_W^6(3 x_t-2)+3 M_W^4 p^2(1+x_t^2)\right)V_L(\mu)}{2 p^2}\log\left(x_t\right)\nonumber\\
    &-\frac{\left(3 m_t M_W(M_W^2-m_t^2+p^2)g_R(\mu)+(p^4+M_W^2 p^2(1-x_t)+M_W^4(x_t^2+x_t-2)\right)}{2p^2}\text{DiscB}[p^2,m_t,M_W]\nonumber\Bigg]\\
    \Sigma_{ii}^R(p^2,\mu)&=0
    \end{align}
    
    \begin{align}\Sigma_{ii}^S(p^2,\mu)&=\frac{g_2^2 V_{tb}}{64 m_b \pi^2 M_W}\Bigg[-\left(3 (-p^2+2M_W^2(x_t+1))g_L(\mu)+2 m_t M_W(x_t-3)V_R(\mu)\right)\log\left(\frac{\mu^2}{m_t^2}\right)\nonumber\\
    &-\left((5 M_W^2(x_t+1)-4p^2)g_L(\mu)+2 m_t M_W( x_t-4)V_R(\mu)\right)\\
    &+\frac{3 \left(2m_t M_W(-m_t^2+M_W^2+p^2)V_R(\mu)+(p^4-2 M_W^2 p^2(x_t+1)+M_W^4(x_t^2-1))g_L(\mu)\right)}{2 p^2}\log x_t\nonumber\\
    &-3 \left(( M_W^2(1+x_t)-p^2)g_L(\mu)- 2 m_t M_W V_R(\mu)\right)\text{DiscB}[p^2,m_t,M_W]\Bigg] \,. \nonumber
\end{align}}

{\small \paragraph{\underline{Loop contributions from vertex and external leg corrections}:}
Here, we will write down the loop contribution of the FCCC processes shown in the diagrams of fig.~\ref{fig:FCCC_diagrams}. Here, the contributions will be written in terms of the matched Wilson coefficients of eq.~\eqref{eq:four_ferm_eff_b2c}. 
For the processes involving $b$ quark, we have considered the effect of both quarks of the external legs. For the processes involving lighter quarks, we have ignored their effect. Also, the operator $C_{V_2}$ depends entirely on the mass of the final state quark. The loop contributions are written in terms of $x_{p} = m_{p}^2/M_W^2$. Also, for processes involving $b$ quark, $C_{V_2}$ is written in terms of $m_{2}$, with $m_{2} = m_{c (u)}$ for the $b \to c (u) \ell \nu$ decay. Also, since the NP only modifies the hadronic part of the process and the leptonic part is the same as the SM, the WCs will be the same for all leptons. }
{\footnotesize
\begin{align}
    C_{V_{1}}^{\ref{fig:FCCC_a}}(\mu) = &  \frac{G_F \, M_W^2 \, V_{tb}}{4 \sqrt{2} \, \pi^2 }
 \bigg[ \left( \frac{m_b^2 \, m_t}{M_W^3} \, g_R(\mu) + \frac{m_b \, m_t}{2 M_W^2} \, V_R(\mu) - x_t \, V_L(\mu) \right) \log\left( \frac{\mu^2}{m_t^2} \right) \nonumber \\ 
    &  + \left( \frac{m_b^2 \, x_t \, (x_t - 1 + 2x_t)}{3 M_W^2 (x_t - 1)^3} \, V_L(\mu)
    - \frac{m_b \, m_t \left( m_b^2 (x_t - 3) - 9 M_W^2 (1 - x_t)^2 \right)}{6 M_W^4 (x_t - 1)^3} \, V_R(\mu)
 \right. \nonumber \\ 
    &  \left. - \frac{m_b^3 \, (3x_t - 2)}{2 M_W^3 (x_t - 1)^2} \, g_L(\mu)
    - \frac{m_b^2 \, m_t \, x_t (x_t - 1)}{M_W^3 (x_t - 1)^3} \, g_R(\mu) \right) \log x_{t} \\
    & - \frac{\left( 6 x_t^2 M_W^2+ m_b^2  (-4 + 8x_t) - 3 M_W^2 (-3 + 4x_t + x_t^2) \right)}{6 M_W^2 (x_t-1)^2} \, V_L(\mu) \nonumber\\
    & + \frac{m_b^3 (3x_t - 4)}{2 M_W^3 (x_t - 1)} \, g_L(\mu)
    + \frac{m_b \, m_t \left( 4 m_b^2 (x_t - 2) + 3 M_W^2 (1 - x_t)^2 \right)}{12 M_W^4 (x_t - 1)^2} \, V_R(\mu) + \frac{m_b^2 \, m_t (1 + x_t)}{2 M_W^3 (x_t - 1)} \, g_R(\mu) \bigg]\nonumber \,,
\end{align}
\begin{align}
    C_{V_{2}}^{\ref{fig:FCCC_a}}(\mu) = &  \frac{G_F \, M_W^2 \, V_{tb}}{4 \sqrt{2} \, \pi^2 }
 \bigg[ \frac{m_2\, m_b\,  \, }{2 M_W^2} V_L(\mu) \log\left(\frac{\mu^2}{m_t^2}\right) \nonumber \\ 
    & + \left( \frac{m_2\, m_b\, ( -3 + 6 x_t + x_t^2 )}{6\, M_W^2\, (x_t - 1)^3} \, V_L(\mu) 
    + \frac{m_2\, m_b^2\, m_t\, (3 - 3 x_t + x_t^2)}{3\, M_W^4\, (x_t - 1)^3} \, V_R(\mu) \right. \nonumber \\
    & \left. + \frac{m_2\, m_b^4\, (2 x_t - 3)}{6\, M_W^5\, (x_t - 1)^2} \, g_L(\mu) 
    + \frac{m_2\, m_b^3\, m_t\, (x_t - 3)}{6\, M_W^5\, (x_t - 1)^3} \, g_R(\mu) \right) \log x_t + \frac{m_2\, m_b\, (7 - 18 x_t + 3 x_t^2)}{12\, M_W^2 (x_t - 1)^2} \, V_L(\mu)\nonumber\\
    & + \frac{m_2\, m_b^2\, m_t\, (3 x_t - 5)}{6 M_W^4 \, (x_t - 1)^2} \, V_R(\mu) + \frac{m_2\, m_b^4}{6\, M_W^5\, (x_t - 1)} \, g_L(\mu) 
    - \frac{m_2\, m_b^3\, m_t\, (x_t - 2)}{3\, (x_t - 1)^2\, M_W^5} \, g_R(\mu) \bigg] \,,
\end{align}
\begin{align}
    C_{V_{1}}^{\ref{fig:FCCC_b}}(\mu) = & \frac{G_F M_W^2 V_{c d_i} V_{t d_i}}{4 \sqrt{2} \pi^2 V_{cb}} 
\bigg[ \left( \frac{3 m_b}{2 M_W} \, g_L(\mu) + \frac{m_b\, m_t}{2 M_W^2} \, V_R(\mu) - x_t \, V_L(\mu) \right) \log\left( \frac{\mu^2}{m_t^2} \right) \nonumber \\
    & + \left( \frac{m_b^2\, x_t\, (3 x_t - 1)}{3\, M_W^2\, (x_t - 1)^3} \, V_L(\mu) 
    - \frac{m_b\, m_t\, (2 x_t - 1)}{2\, M_W^2\, (x_t - 1)^2} \, V_R(\mu)- \frac{m_b\, (4 x_t - 3)}{2\, M_W\, (x_t - 1)^2} \, g_L(\mu) \right. \nonumber \\
    & \left.  
    + \frac{2 m_b^2\, m_t\, x_t}{3\, M_W^3\, (x_t - 1)^3} \, g_R(\mu) \right) \log x_t + \frac{m_b\, (1 + x_t)}{4 M_W \left(x_t-1 \right)} \, g_L(\mu)\nonumber \\
    & - \frac{ \left( -4 m_b^2 + 9 M_W^2 + 8 m_b^2 x_t - 12 M_W^2 x_t - 3 M_W^2 x_t^2 + 6 M_W^2 x_t^3 \right)}{6 M_W^2 \left( x_t - 1 \right)^2} \, V_L(\mu) \\
    & + \frac{m_b\, m_t\, (1 + x_t)}{4 M_W^2 \left( x_t -1 \right)} \, V_R(\mu) - \frac{m_t \left( 2 m_b^2 + 9 M_W^2 + 2 m_b^2 x_t - 18 M_W^2 x_t + 9 M_W^2 x_t^2 \right)}{6 M_W^3 \left( x_t - 1 \right)^2} \, g_R(\mu) \bigg] \,, \nonumber
\end{align}
\begin{align}
    C_{V_{2}}^{\ref{fig:FCCC_b}}(\mu) = & \frac{G_F M_W^2 V_{c d_i} V_{t d_i}}{4 \sqrt{2} \pi^2 V_{cb}} \bigg[ 
    \left( \frac{3 m_2}{2 M_W} \, g_L(\mu) + \frac{m_2\, m_b}{2 M_W^2} \, V_L(\mu) - \frac{m_2\, m_t}{2 M_W^2} \, V_R(\mu) \right) \log\left( \frac{\mu^2}{m_t^2} \right) \nonumber \\
    & + \left( \frac{m_2\, m_b\, (x_t^2 + 6 x_t - 3)}{6\, M_W^2\, (x_t - 1)^3} \, V_L(\mu) 
    - \frac{m_2\, m_t\, \left( m_2^2 x_t - m_b^2 x_t + 3 M_W^2 x_t^2 - 3 m_2^2 + 3 m_b^2 - 3 M_W^2 \right)}{6\, M_W^4\, (x_t - 1)^3} \, V_R(\mu) \right. \nonumber \\
    & \left. + \frac{m_2\, \left( 2 m_2^2 x_t - 2 m_b^2 x_t - 15 M_W^2 x_t^2 + 24 M_W^2 x_t - 9 M_W^2 \right)}{6\, M_W^3\, (x_t - 1)^3} \, g_L(\mu) 
    - \frac{m_2\, m_b\, m_t\, x_t\, (x_t - 3)}{3\, M_W^3\, (x_t - 1)^3} \, g_R(\mu) \right) \log x_t \nonumber \\
    & + \frac{m_2\, m_b\, (3 x_t^2 - 18 x_t + 7)}{12\, M_W^2 (x_t-1)^2} \, V_L(\mu) - \frac{2\, m_2\, m_b\, m_t}{3\,M_W ^3 (x_t-1)^2} \, g_R(\mu) \bigg] \nonumber \\
    & + \frac{m_2\, m_t\, \left( 4 m_2^2 x_t - 4 m_b^2 x_t - 3 M_W^2 x_t^2 + 18 M_W^2 x_t - 8 m_2^2 + 8 m_b^2 - 15 M_W^2 \right)}{12\, (m_t - M_W)^2\, (m_t + M_W)^2} \, V_R(\mu) \nonumber \\
    & - \frac{m_2\, M_W\, \left( 2 m_2^2 x_t - 2 m_b^2 x_t - 9 M_W^2 x_t^2 + 6 M_W^2 x_t + 2 m_2^2 - 2 m_b^2 + 3 M_W^2 \right)}{12\, (m_t - M_W)^2\, (m_t + M_W)^2} \, g_L(\mu) \,,
\end{align}
\begin{align}
    C_{V_{1}}^{\ref{fig:FCCC_c}} (\mu) = & \frac{G_F\, M_W^2\, V_{tb}}{4 \sqrt{2}\, \pi^2} \bigg[
    \left( \frac{7\, M_W}{8\, m_t} \, g_R(\mu) - \frac{m_b}{4\, m_t} \, V_R(\mu) \right) 
    - \left( \frac{3\, (M_W\, g_R(\mu) - 2\, m_b\, V_R(\mu))}{4\, m_t} \right) \log\left( \frac{\mu^2}{M_W^2} \right) \bigg] \,, 
\end{align}
\begin{align}
    C_{V_{1}}^{\ref{fig:FCCC_d}}(\mu) = & \frac{G_F\, M_W^2\, V_{cd_i}\, V_{td_i}}{4 \sqrt{2}\, \pi^2\, V_{cb}} \bigg[
    \left( \frac{g_L(\mu)\, m_b^3}{2\, m_t^2\, M_W} + \frac{g_R(\mu)\, m_2^2}{m_t\, M_W} - \frac{g_R(\mu)\, m_b^2}{2\, m_t\, M_W} - \frac{m_b\, V_R(\mu)}{m_t} \right) \log \left( \frac{\mu^2}{M_W^2} \right) \\
    & - \frac{m_b\, V_R(\mu)}{2\, m_t} + \frac{g_L(\mu)\, m_b^3}{4\, m_t^2\, M_W} + \left( \frac{g_R(\mu)\, m_2^2}{2\, m_t\, M_W} - \frac{g_R(\mu)\, m_b^2}{4\, m_t\, M_W} \right) \bigg] \,, \nonumber
\end{align}
\begin{align}
    C_{V_{1}}^{\ref{fig:FCCC_e}}(\mu) = & \frac{G_F M_W^2 V_{cd_i} V_{td_i}}{4 \sqrt{2} \pi^2 V_{cb}} \bigg[ 
    \left( \frac{3 V_L (\mu) x_t}{2}
    - \frac{m_t V_R (\mu) \left(x_t^4 - 6 x_t^3 + 12 x_t^2 - 10 x_t + 3\right) \log(x_t)}{m_b (x_t - 1)^3} \right. \nonumber \\
    & \left. + \frac{3 g_L (\mu) m_t}{2 M_W} - g_R (\mu) \left( \frac{3 m_b}{2 M_W} + \frac{3 M_W (1 + x_t) \log(x_t)}{m_b} \right) \right) \log\left( \frac{\mu^2}{M_W^2} \right) \nonumber  \\
    & + \left( \frac{3 V_L (\mu) (2 - x_t) x_t^2}{2 (1 - x_t)^2} 
    - \frac{3 m_t V_R (\mu) \left( M_W^2 (x_t - 1)^2 + m_b^2 x_t \right)}{m_b M_W^2 (x_t - 1)^3} \right. \nonumber \\
    & \left. + \frac{3 g_L (\mu) m_t (2 - x_t) x_t}{2 M_W (x_t - 1)^2} 
    + g_R (\mu) \left( \frac{3 M_W}{m_b (1 - x_t)} - \frac{3 m_b (3 - x_t) x_t^2}{2 M_W (x_t - 1)^3} \right) \right) \log x_t  \\
    & + \frac{V_L (\mu) (5 x_t^2 - 5 x_t - 6)}{4 ( -1 + x_t)} 
    + V_R (\mu) \left( \frac{3 m_b m_t (x_t^2 + 10 x_t + 1)}{2 M_W^2 (-1 + x_t)^2} - \frac{m_t (-1 + x_t)}{m_b} \right) \nonumber \\
    & - \frac{g_L (\mu) m_t (5 + x_t)}{4 M_W (-1 + x_t)} 
    + g_R (\mu) \left( \frac{M_W (1 + x_t)}{m_b} + \frac{m_b (x_t^2 + 10 x_t + 1)}{4 M_W (-1 + x_t)^2} \right) \bigg] \,, \nonumber
\end{align}
\begin{align}
    C_{V_{1}}^{\ref{fig:FCCC_f}}(\mu) = & \frac{G_F M_W^2 V_{ib} V_{tj}}{4 \sqrt{2} \pi^2 V_{cb}}
 \bigg[ \left( - \frac{V_L (\mu) (m_b^2 + m_t^2)}{M_W^2} + \frac{m_b m_t V_R (\mu) }{2 M_W^2} \right) \log \frac{\mu^2}{M_W^2} \nonumber  \\
    & + \left( - \frac{m_b^2 V_L (\mu) x_b}{M_W^2 (x_t - x_b)} + \frac{m_b m_t V_R (\mu) (x_b - 4)\, x_b}{2 M_W^2 (x_b - 1)(x_t - x_b)}\right) \log x_b \\
    & + \left( \frac{m_t^2 V_L (\mu) x_t }{M_W^2 (x_t - x_b)} + \frac{m_b m_t V_R (\mu) (4 - x_t)\, x_t }{2 M_W^2 (x_b - x_t)(x_t - 1)}    \right) \log x_t  - V_L (\mu) \left( \frac{3}{2} + (x_b + x_t) \right) + \frac{m_b m_t V_R (\mu)}{4 M_W^2} \bigg] \,, \nonumber 
\end{align}
\begin{align}
    C_{V_{1}}^{\ref{fig:FCCC_g}}(\mu) = & \frac{G_F M_W^2 V_{ib} V_{tj}}{4 \sqrt{2} \pi^2 V_{cb}}
 \bigg[ \left( \frac{1}{m_b} \left( m_t V_R (\mu) (-3 + x_t) - 3 g_L (\mu) M_W (1 + x_t) \right)  \right) \log\frac{\mu^2}{M_W^2}  \hspace{2cm}   \\
    & + \left(\frac{m_t}{m_b M_W^2 (x_t - 1)} \left( 3 g_L (\mu) m_t x_t + M_W^2 V_R  (\mu)x_t (4 - x_t) \right) \right) \log x_t  + \frac{m_t V_R (\mu) (x_t-1) - g_L (\mu) M_W (1 + x_t)}{m_b}  \bigg] \,, \nonumber
\end{align}
\begin{align}
    C_{V_{1}}^{\ref{fig:FCCC_h}}(\mu) = & \frac{G_F M_W^2 V_{ib} V_{tj}}{4 \sqrt{2} \pi^2 V_{cb}} \bigg[ \left( \frac{m_b V_R (\mu) (-3 + x_t)}{m_t} + \frac{3 g_R (\mu) M_W (1 + x_t)}{m_t} \right) \log \frac{\mu^2}{m_t^2} \nonumber  \\
    & + \left(\frac{3 (g_R (\mu) M_W - m_b V_R (\mu))}{m_t (1 - x_t)}  \right) \log x_t  + \frac{m_b V_R (\mu) (-1 + x_t)}{m_t} + \frac{g_R (\mu) M_W (1 + x_t)}{m_t} \bigg] \,.
\end{align}    
}
\section{Loop Contributions of EWPOs and other relevant observables}\label{appndx:EWPOs}
The impacts of anomalous $Wtb$ couplings on the EWPOs are discussed in the sec.~\ref{sec:EWPOs}, in terms of the loop contributions coming from respective Feynman diagrams. The contribution from the diagram of fig.~\ref{fig:W_boson_feynman} is given in eq.~\eqref{eq:W-boson_transverse}. The other contributions are given below. 

\paragraph{\underline{W-Pole Observables}:}\label{appndx:W_Pole}
The self-energy correction of the SM gauge bosons can be divided into two parts: transverse and longitudinal. The $S, T$, and $U$ parameters are expressed in terms of the transverse component of the self-energy correction, which has an expression:
\begin{eqnarray} \label{eq:W-boson_transverse}
    \Sigma_{W,T}(q^2) = \Sigma_{W,T}^{\rm Pole}(q^2) + \Sigma_{W,T}^{\rm Fin} (q^2)\,, 
\end{eqnarray}
with 
{\footnotesize
\begin{eqnarray}
 & \Sigma_{W,T}^{\rm Pole} (q^2, \mu)& = \frac{G_F M_W }{6 \sqrt{2} \pi^2}\left( -3 g_L(\mu) m_b q^2 + 3 g_R(\mu) m_t q^2 + (3 m_t^2 M_W  - 2 M_W q^2 )V_L(\mu) + 6 m_b m_t M_W V_R (\mu)\right) \log \left( \frac{\mu^2}{m_b^2} \right) \,,\nonumber \\
\end{eqnarray}}
{\footnotesize
\begin{eqnarray}
  \Sigma_{W,T}^{\rm Fin}(q^2, \mu) & = &
\frac{ G_F M_W}{12 \sqrt{2} q^4 \pi^2} \Bigg[ 
- 3 g_R(\mu) m_t q^2 \left( m_t^4 - 2 m_t^2 q^2 - q^4 \right) 
- 3 g_L(\mu) m_b q^2 \left( m_t^4 + q^4 - 2 m_b^2 (m_t^2 + q^2) \right) \nonumber \\
&+ M_W &\left( 3 m_b^4 m_t^2 V_L(\mu) + 3 m_b^2 (- m_t^4) V_L(\mu) + \left( m_t^6 + 3 m_t^2 q^4 - 2 q^6 \right) V_L(\mu) \right. \nonumber \\
& & \left. - 6 m_b^3 m_t q^2 V_R(\mu)+ 6 m_b m_t q^2 \left( m_t^2 + q^2 \right) V_R(\mu) \right) 
\Bigg] \log \left( \frac{m_b^2}{m_t^2} \right) \nonumber \\
& +V_R(\mu) &\left( \frac{\sqrt{2} G_F m_b m_t M_W^2}{\pi^2} + \frac{G_F m_b m_t M_W^2 \, \text{DiscB}(q^2, m_b, m_t)}{\sqrt{2} \pi^2} \right) \nonumber \\
&+ g_L(\mu) & \left( - \frac{G_F m_b}{\sqrt{2} \pi^2} \left( m_t^2 M_W + M_W q^2 + m_t^2 M_W \, \text{DiscB}(q^2, m_b, m_t) \right) \right) \nonumber \\
& + g_R(\mu) &\left( - \frac{G_F m_t M_W}{\sqrt{2} \pi^2} \left( m_t^2 + m_t^2 \, \text{DiscB}(q^2, m_b, m_t) \right) + \frac{G_F m_t M_W q^2}{\pi^2} \right) \nonumber \\
& + V_L(\mu) &\left( \frac{G_F M_W^2}{\sqrt{2} \pi^2} \left( \frac{m_t^2}{3} + \frac{m_t^4}{6 q^2} - \frac{5 q^2}{9} + \frac{m_t^2 \, \text{DiscB}(q^2, m_b, m_t)}{6} - \frac{m_b^2 m_t^2 \, \text{DiscB}(q^2, m_b, m_t)}{3 q^2} \right. \right. \nonumber \\
&& \left. \left. + \frac{m_t^4 \, \text{DiscB}(q^2, m_b, m_t)}{6 q^2} \right) \right)\,.
\end{eqnarray}}
\paragraph{\underline{Z-Pole Observables}:}\label{appndx:Z_Pole}
The effect of NP in the $Z \to b\bar{b}$ process is expressed through the coupling corrections $\delta g_{V(A)}$, which can be further decomposed into vertex and leg corrections. The expressions for these contributions are provided below.
\begin{align}\label{eq:loop_Zpole}
    \delta g_V&=g_{ZL}^{NP}+g_{ZR}^{NP}=\left(g_{ZL}^{\rm Vertex}+ g_{ZL}^{\rm WF}\right)+\left(g_{ZR}^{\rm Vertex}+ g_{ZR}^{\rm WF}\right) \,, \\
    \delta g_A&=g_{ZL}^{NP}- g_{ZR}^{NP}=\left(g_{ZL}^{\rm Vertex}+ g_{ZL}^{\rm WF}\right)-\left(g_{ZR}^{\rm Vertex}+ g_{ZR}^{\rm WF}\right) \,.
\end{align}
{\footnotesize
\begin{align}
    g_{ZL}^{\rm Vertex}&= 
\frac{g_2 V_{tb} \,\,\text{cos}\theta}{32 \pi^2}\Bigg[\left(-4 a_L^t+a_R^t\right)x_t V_L(\mu)\log\left(\frac{\mu^2}{m_t^2}\right)+\frac{\left(a_R^tx_t(-2+2 x_t+x_z)-2a_L^t(2+2 x_t(x_z-1)+5 x_z\right)V_L(\mu)}{x_z}\nonumber\\
&+\frac{2\left(a_R^t x_t(x_t-1)+a_L^t(-2-3x_z+x_t(2+x_z)\right)}{(x_t-1)x_z}\log x_t\nonumber\\
&+ \frac{\left(a_L^t(-4+4 x_t-6 x_z)
+a_R^t x_t(2 x_t +x_z-2)\right)V_L(\mu)}{x_z}\text{DiscB}[M_Z^2,m_t,m_t]\nonumber\\
&-\frac{2M_W^2\left(a_R^tx_t((x_t-1)^2+2 x_z)+a_L^t(-4x_t(1+x_z)+2(1+x_z)^2+x_t^2(2+x_z))\right)}{x_z}\, \, C_{0}(0,0,M_Z^2,m_t,M_W,m_t)\Bigg]\nonumber\\
&-\frac{g_2^4 V_{tb}}{32 \pi^2 \sqrt{g_1^2+g_2^2}}\Bigg[\frac{1}{6}\left(-3 x_t(x_z-6)+(x_z-4)x_z\right)V_L(\mu)\log\left(\frac{\mu^2}{m_t^2}\right)\nonumber\\
&+\frac{\left(-72 +18 x_t^2(2+x_z)-2x_z(x_z+2)(4 x_z-21)+9x_t (4+x_z(3 x_z-4))\right)V_L(\mu)}{18 x_z}\nonumber\\
&+\frac{ \left( - (x_z - 4) x_z^2 - 6 x_t^3 (x_z + 2) - 3 x_t^2 (4 + x_z^2) + x_t (x_z - 4)(x_z (x_z + 3) - 6) \right)V_L(\mu)}{6 x_z (x_t - 1)}\log x_t\nonumber\\
&- \frac{ \left( -24 + 6 x_t^2 (x_z + 2) + x_z \left( 20 - (x_z - 2) x_z \right) + 3 x_t (4 + x_z^2) \right)V_L(\mu)}{6 x_z}\text{DiscB}[M_Z^2,M_W,M_W]\\
&- \frac{M_W^2  (4 + x_t (x_z + 2)) \left(1 + x_t (x_t - 2 + x_z)\right)V_L(\mu)}{x_z}\, \, C_{0}(0,0,M_Z^2,M_W,m_t,M_W) \Bigg] \,, \nonumber
\end{align}
\begin{align}
&g_{ZR}^{\rm Vertex}=0 \,, \\
&g_{ZL}^{\rm WF}=2\left(\frac{\text{cos}\theta}{g_2} a_L^b\right)\left(\frac{1}{2} \delta Z_{ii}^L\right) \,, \\
&g_{ZR}^{\rm WF}=2 \left(\frac{\text{cos}\theta}{g_2} a_R^b\right)\left(\frac{1}{2} \delta Z_{ii}^R\right) \,.
\end{align}}
Here, $a_{L(R)}^t$ and $a_{L(R)}^b$ represent the chiral couplings of the $Z$ boson with the top and bottom quarks (defined earlier), respectively, $x_{i}$'s are defined earlier. The expression for the renormalisation matrix $\delta Z_{ii}^{L(R)}$ is given in eq.~\eqref{eq:b_WF}. The functions $\rm DiscB(a,b,c)$ and $C_{0}(a,b,c,d,e,f)$ are the two-point and three-point loop functions and can be found from \cite{Patel:2016fam}. 

\paragraph{\underline{Observables related to Higgs boson}:}\label{appndx:H_decay}
The expression for the Higgs couplings modified by the NP contribution, arising from the anomalous $Wtb$ effective operator, fig.~\ref{fig:Htobb}, can be further classified into vertex corrections and leg corrections. We present the individual contributions separately for better readability.

\begin{align}\label{eq:loop_Hdecays}
    C_{HL}^{NP}&=C_{HL}^{\rm Vertex}+C_{HL}^{\rm WF}\,,\\
    C_{HR}^{NP}&=C_{HR}^{\rm Vertex}+C_{HR}^{\rm WF}\,.
\end{align}

{\footnotesize
\begin{align}
    C_{HL}^{\rm Vertex}(\mu)&=\frac{g_2^2 V_{tb}}{32 \pi^2 v}\Bigg[\left(m_t (x_h+x_t-3)V_R(\mu)+6M_W(x_t-1)g_L(\mu)\right)\log\left(\frac{\mu^2}{m_t^2}\right)\nonumber\\
    &+\left(3M_W(x_t-2)g_L(\mu)+m_t(2(x_t+x_h)-3)V_R(\mu)\right)\nonumber\\
    &-\left(\frac{M_W(2 x_t^2+7 x_t -6)g_L(\mu)}{(x_t-1)}-m_t(x_h-3)V_R(\mu)\right)\log x_t+x_t\left(3 M_W g_L(\mu)+m_t V_R(\mu) \right)\text{DiscB}[m_H^2,m_t,m_t]\nonumber\\
    &+\left( m_t(x_h-2)V_R(\mu)-2 M_W(x_t+2)g_L(\mu)\right)\text{DiscB}[m_H^2,M_W,M_W]\\
    &+ m_t M_W \left( 3 M_W x_t(x_t+1)g_L(\mu)+ m_t (x_t^2+ 2 x_t x_h -7 x_t)V_R(\mu)\right)\,\,C_{0}(0,0,m_H^2,m_t,M_W,m_t)\nonumber\\
    &-M_W^2 \left( M_W(2 x_t^2+x_t x_h +2 (x_t+1))g_L(\mu)-m_t(x_t x_h-2(2+x_t))V_R(\mu)\right)\,\,C_{0}(0,0,m_H^2,M_W,m_t,M_W)\bigg]\nonumber
    \end{align}

    \begin{align}
    C_{HR}^{\rm Vertex}(\mu)&=\frac{g_2^2 V_{tb}}{32 \pi^2 v}\Bigg[\left(m_t (x_h+x_t-3)V_R(\mu)+6M_W(x_t-1)g_L(\mu)\right)\log\left(\frac{\mu^2}{m_t^2}\right)\nonumber\\
    &-\left(3M_W(x_t-2)g_L(\mu)-m_t(2(x_t+x_h)-3)V_R(\mu)\right)\nonumber\\
    &+\left(\frac{M_W(2 x_t^2+7 x_t -6)g_L(\mu)}{(x_t-1)}+m_t(x_h-3)V_R(\mu)\right)\log x_t-x_t(3 M_W g_L(\mu)- m_t V_R(\mu))\text{DiscB}[m_H^2,m_t,m_t]\nonumber\\
    &+\left( m_t(x_h-2)V_R(\mu)+2 M_W(x_t+2)g_L(\mu)\right)\text{DiscB}[m_H^2,M_W,M_W]\\
    &+m_t M_W \left( -3 M_W x_t(x_t+1)g_L(\mu)+ m_t (x_t^2+ 2 x_t x_h -7 x_t)V_R(\mu)\right)\,\,C_{0}(0,0,m_H^2,m_t,M_W,m_t)\nonumber\\
    &+M_W^2 \left( M_W(2 x_t^2+x_t x_h +2 (x_t+1))g_L(\mu)+m_t(x_t x_h-2(2+x_t))V_R(\mu)\right)\,\,C_{0}([0,0,m_H^2,M_W,m_t,M_W)\bigg]\nonumber
    \end{align}

    \begin{align}
    C_{HL}^{\rm WF}(\mu)&= -2\frac{m_b}{v}\left(\frac{1}{2}\delta Z_{ii}^L\right)\\
    C_{HR}^{\rm WF}(\mu)&= -2\frac{m_b}{v}\left(\frac{1}{2}\delta Z_{ii}^R\right)
\end{align}}
Here, $x_h = \frac{m_H^2}{M_W^2}$, and the contribution from the $b$-leg correction is discussed in Appendix~\ref{appndx:FCCC}. The expression for the renormalisation matrix $\delta Z_{ii}^{L(R)}$ is given in eq.~\eqref{eq:b_WF}.

\paragraph{\underline{Observable related to Top cMDM}:}\label{appndx:top_cMDM}
The expression for the top cMDM modified by the NP contribution, arising from the anomalous $Wtb$ effective operator in fig.~\ref{fig:top_CMDM}, can be written as:
\begin{align}
    \hat{\mu}_t^{\rm NP}&=\frac{g_2^2 V_{tb}}{16\pi^2}\left(\frac{\frac{m_t}{M_W}(3 x_t+2)g_L(\mu)+2(2-3 x_t)V_L(\mu)}{4 x_t}+\frac{m_t}{M_W}g_L(\mu)\log\frac{\mu^2}{M_W^2}\right.\nonumber\\ &\left.  -\frac{\frac{m_t}{M_W}(2 x_t^2-1)  g_L(\mu)-2(1-2 x_t)V_L(\mu)}{4 x_t^2}\log\frac{m_b^2}{M_W^2}\right.\nonumber\\ &\left.+ \frac{\frac{m_t}{M_W}(2 x_t^2-1)g_L(\mu)-2(1-2 x_t)V_L(\mu)}{2 x_t^2}\log\frac{m_b}{M_W(1-x_t)}\right)
\end{align}


\section{\texorpdfstring{Combined Fit Results without $\Delta_{d}$}{without deltad}}\label{apndx:fit_without_deltad}
\begin{table}[htb!]
\centering
\rowcolors{1}{cyan!10}{blue!20!green!5}
\renewcommand{\arraystretch}{2.0}
\setlength{\tabcolsep}{5pt}
\resizebox{\textwidth}{!}{%
\begin{tabular}{|c|c|c|c|c|}
\hline
\rowcolor{cyan!20}
\multicolumn{5}{|c|}{\textbf{Combined analysis: FCCC + FCNC + EWPOs (without $\Delta_d$)}} \\ 
\hline
\hline
$\text{Scale}$ & $\frac{v^2}{\Lambda^2} \mathcal{C}^{\phi q (3)}_{33}$  &  $ \frac{v^2}{\Lambda^2} C^{\phi ud}_{33}$ &  $\frac{v^2}{\Lambda^2} C^{*\, dW}_{33}$ &  $ \frac{v^2}{\Lambda^2} C^{uW}_{33}$   \\
\hline
 $\mu_{\rm EW}$  &  $(-0.85 \pm 1.26 )\times 10^{-2}$  &  $  (-0.78\pm 4.89 )\times 10^{-3}$  &  $ (0.0 \pm 1.98  )\times 10^{-3}$  &  $ (0.69 \pm 2.49 )\times 
10^{-3}$  \\
\hline
\end{tabular}%
}
\caption{Fit results of the SMEFT couplings from the combined analysis, considering all available observables, except for $\Delta_{d}$. }
\label{tab:combined_one_param_SMEFT_without_Deltad}
\end{table}
\begin{table}[htb!]
\centering
\rowcolors{1}{cyan!10}{blue!20!green!5}
\renewcommand{\arraystretch}{2.0}
\setlength{\tabcolsep}{5pt}
\resizebox{\textwidth}{!}{%
\begin{tabular}{|c|c|c|c|c|}
\hline
\rowcolor{cyan!20}
\multicolumn{5}{|c|}{\textbf{Observables $\rightarrow$  All Combined (without $\Delta_d$)}} \\ 
\hline
\hline
$\text{Scale}$  &  $V_L$  &  $V_R$  &  $g_L$  &  $g_R$  \\
\hline
  $\mu_{\rm EW}$  &  $(-0.85 \pm 1.26)\times 10^{-2}$  &  $(- 0.39 \pm 2.44 )\times 10^{-3}$  &  $(0.02 \pm 2.79 )\times 10^{-3}$  &  $(0.98 \pm 3.51 )\times  10^{-3}$  \\
\hline
\end{tabular}%
}
\caption{Fit results of the anomalous couplings of $Wtb$ vertex from the combined analysis taking all the available observables, except for $\Delta_{d}$. }
\label{tab:combined_one_paraM_Without_Deltad}
\end{table}
Tables~\ref{tab:combined_one_param_SMEFT_without_Deltad} and \ref{tab:combined_one_paraM_Without_Deltad} are identical to the tables~\ref{tab:combined_one_param_SMEFT} and \ref{tab:combined_one_param}, respectively, except that in the former, data for $\Delta_{d}$ has been excluded due to its disagreement with the SM at the $1\sigma$ level. As discussed in the text, this exclusion is based on the neutral meson mixing of $B^{0} - \bar{B}^{0}$, which shows a deviation from the SM. Consequently, we obtain better-constrained fitted values for the couplings $\left(\frac{v^2}{\Lambda^2}\mathcal{C}^{\phi q(3)}_{33}, \frac{v^2}{\Lambda^2}\mathcal{C}^{uW}_{33}\right)$. Table~\ref{tab:combined_one_param_SMEFT_without_Deltad} presents these fitted results in terms of SMEFT couplings, while table~\ref{tab:combined_one_param} presents them in terms of the anomalous couplings of the $Wtb$ vertex.

\bibliographystyle{JHEP} 
\bibliography{refs}

\end{document}